\documentclass[12pt,a4]{article}
\usepackage{amsthm,amsmath,latexsym,amssymb,amsfonts,amscd}
\usepackage{graphics,lscape,fancyhdr,array,stmaryrd,euscript}
\usepackage{verbatim}
\usepackage{array}
\usepackage[all]{xy}

\usepackage[numbers,sort&compress]{natbib}
\usepackage[nottoc]{tocbibind}

\newcommand{\pl}{\partial}
\newcommand{\momega}{{\boldsymbol{\omega}}}
\newcommand{\mxi}{{\boldsymbol{\xi}}}
\newcommand{\mC}{{\boldsymbol{C}}}
\newcommand{\me}{{\boldsymbol{e}}}
\newcommand{\mvarepsilon}{{\boldsymbol{\varepsilon}}}
\newcommand{\mw}{{\boldsymbol{w}}}
\newcommand{\mvarphi}{{\boldsymbol{\varphi}}}
\newcommand{\mpi}{{\boldsymbol{\pi}}}
\newcommand{\mA}{{\boldsymbol{A}}}
\newcommand{\mF}{{\boldsymbol{F}}}
\newcommand{\mR}{{\boldsymbol{R}}}
\newcommand{\mN}{{\boldsymbol{\mathcal N}}}
\newcommand{\mQ}{{\boldsymbol{\mathcal Q}}}
\newcommand{\mG}{{\boldsymbol{\mathcal G}}}
\newcommand{\mOmega}{{\boldsymbol{\Omega}}}
\newcommand{\mTheta}{{\boldsymbol{\Theta}}}
\newcommand{\mX}{{\boldsymbol{X}}}
\newcommand{\mJ}{{\boldsymbol{J}}}
\newcommand{\mf}{{\boldsymbol{f}}}

\newcommand{\mE}{{\boldsymbol{E}}}
\newcommand{\mB}{{\boldsymbol{B}}}

\newcommand{\ga}{\alpha}
\newcommand{\gb}{\beta}

\newcommand{\gad}{{\dot{\alpha}}}
\newcommand{\gbd}{{\dot{\beta}}}
\newcommand{\gdd}{{\dot{\gamma}}}

\newcommand{\bry}{{{\bar{y}}}}
\newcommand{\Vcan}{\mathcal{V}^{\text{can}}}
\newcommand{\Tr}[1]{{\left\langle #1 \right\rangle}}

\newcommand{\fud}[2]{{}^{#1}{}_{#2}\,}
\newcommand{\fdu}[2]{{}_{#1}{}^{#2}\,}

\newcolumntype{w}[1]{%
>{\raggedright\hspace{0pt}}m{#1}}%
\newcolumntype{z}[1]{%
>{\raggedleft\hspace{0pt}}m{#1}}%

\newcommand{\besubeqs}{\begin{subequations}}
\newcommand{\esubeqs}{\end{subequations}}

\usepackage{amsthm,amsmath,latexsym,amssymb,amsfonts,amscd,hyperref}
\usepackage{color,tikz,tikz-cd}
\usepackage[all]{xy}
\newtheorem{theorem}{Theorem}[section]

\newtheorem{proposition}[theorem]{Proposition}

\theoremstyle{definition}

\newtheorem{example}[theorem]{Example}

\theoremstyle{remark}
\newtheorem{remark}[theorem]{Remark}

\numberwithin{equation}{section}

\begin{document}
\pagenumbering{gobble}
\hfill
\vskip 0.01\textheight
\begin{center}
{\Large\bfseries 
Higher Spin Gravities and Presymplectic AKSZ Models
\vspace{0.4cm}}

\vskip 0.03\textheight
\renewcommand{\thefootnote}{\fnsymbol{footnote}}
Alexey \textsc{Sharapov}${}^{a}$ \&  Evgeny \textsc{Skvortsov}\footnote{Research Associate of the Fund for Scientific Research -- FNRS, Belgium}${}^{b,c}$
\renewcommand{\thefootnote}{\arabic{footnote}}
\vskip 0.03\textheight

{\em ${}^{a}$Physics Faculty, Tomsk State University, \\Lenin ave. 36, Tomsk 634050, Russia}\\
\vspace*{5pt}
{\em ${}^{b}$ Service de Physique de l'Univers, Champs et Gravitation, \\ Universit\'e de Mons, 20 place du Parc, 7000 Mons, 
Belgium}\\
\vspace*{5pt}
{\em ${}^{c}$ Lebedev Institute of Physics, \\
Leninsky ave. 53, 119991 Moscow, Russia}\\

\end{center}

\vskip 0.02\textheight

\begin{abstract}
As a step towards quantization of Higher Spin Gravities we construct the presymplectic AKSZ sigma-model for  $4d$ Higher Spin Gravity which is AdS/CFT dual of Chern--Simons vector models. It is shown that the presymplectic structure leads to the correct quantum commutator of higher spin fields and to the correct algebra of the global higher spin symmetry currents. The presymplectic AKSZ model is proved to be unique, it depends on two coupling constants in accordance with the AdS/CFT duality, and it passes some simple checks of interactions. 
\end{abstract}
\newpage
\tableofcontents
\newpage
\section{Introduction}
\pagenumbering{arabic}
\setcounter{page}{2}
Quantization of Higher Spin Gravities (HSGRA) is an important open problem together with the problem of constructing more viable HSGRA. Concrete quantum checks of HSGRA require concrete models. Topological theories in $3d$, which are higher spin extensions of Chern--Simons formulation of $3d$ gravity \cite{Blencowe:1988gj,Bergshoeff:1989ns,Campoleoni:2010zq,Henneaux:2010xg} or $3d$ conformal gravity \cite{Pope:1989vj,Fradkin:1989xt,Grigoriev:2019xmp}, are well-behaved perturbatively due to the lack of any propagating degrees of freedom, but their non-perturbative definition requires further study, see e.g. \cite{Alday:2020qkm}. Amplitudes of $4d$ conformal HSGRA \cite{Segal:2002gd,Tseytlin:2002gz,Bekaert:2010ky} were studied in \cite{Joung:2015eny,Beccaria:2016syk,Adamo:2018srx}. The most elaborate checks have been performed for Chiral HSGRA \cite{Metsaev:1991mt,Metsaev:1991nb,Ponomarev:2016lrm,Skvortsov:2018jea,Skvortsov:2020wtf}, which was shown to be one-loop finite \cite{Skvortsov:2018jea,Skvortsov:2020wtf,Skvortsov:2020gpn}. In addition, there is a number of interesting computations of vacuum one-loop corrections \cite{Gopakumar:2011qs,Tseytlin:2013jya,Giombi:2013fka,Giombi:2014yra,Beccaria:2014jxa,Beccaria:2014xda,Beccaria:2015vaa,Gunaydin:2016amv,Bae:2016rgm,Skvortsov:2017ldz}. Finally, the one-loop correction to the four-point function in the bulk dual of the free vector model can be reduced to a CFT computation and seems to be consistent with the vacuum one-loop results \cite{Ponomarev:2019ltz}.\footnote{In this regard it is worth mentioning the collective dipole approach, see e.g. \cite{deMelloKoch:2018ivk,Aharony:2020omh}, which is bound to reproduce the correct holographic correlation functions and all the other observables of vector models. However, its relation to the field theory approach is to be clarified \cite{Aharony:2020omh}.}

The only class of perturbatively local HSGRAs with propagating massless fields and an action seems to be given by Chiral HSGRA and its various truncations.  This was proved for flat space in \cite{Ponomarev:2016lrm} and it is likely to be the case for $AdS_4$ as well. However, Chiral Theory does not have a covariant action at the moment and this hampers the study of quantum corrections.  Furthermore, the chiral theory constitutes \cite{Skvortsov:2018uru} only a subsector of the HSGRA that is dual to Chern--Simons Matter theories  \cite{Giombi:2011kc, Maldacena:2012sf, Aharony:2012nh,Aharony:2015mjs,Karch:2016sxi,Seiberg:2016gmd}. The latter HSGRA is known to feature non-localities beyond what is admissible by the field theory methods \cite{Dempster:2012vw,Bekaert:2015tva,Maldacena:2015iua,Sleight:2017pcz,Ponomarev:2017qab}. 

Nevertheless, the bulk dual of vector models  \cite{Klebanov:2002ja,Sezgin:2003pt} can be approached from the opposite end: abandoning perturbative locality one can get a very explicit description of this HSGRA at the level of formal algebraic structures; hence the name ``Formal HSGRA''. In practice, this means that one can construct an $L_\infty$-algebra that extends a given higher spin algebra. Equivalently, one can think of the corresponding $Q$-manifold, where $Q$ --- an odd, nilpotent $Q^2=0$ vector field --- can be expanded as $Q=f_{BC}^A w^B w^C \pl/\pl w^A+...$ with $f_{BC}^A$ being the structure constants of a higher spin algebra. The same idea can be encoded in the formal dynamical system \cite{Vasiliev:1988sa}
\begin{equation}\label{dWQA}\tag{$\ast$}
    E^A\equiv d\mw^A-Q^A(\mw)=0\,,
\end{equation}
which is a sigma-model with the $Q$-manifold as a target and the space-time manifold as a base. It is rather remarkable that the $Q$'s of HSGRAs remain non-trivial despite non-localities and non-local redefinitions hidden in the very definition of $Q$. With the proviso that the equations are only formally consistent and should not be taken at face value as PDE's or as field's theory equations, we can continue studying the formal structures restricted by the higher spin symmetries with the hope that we either find a `canonical' $Q$ with certain prescriptions that would lead to a reasonable field theory or the formal approach itself will allow one to define the theory as to be able to compute various physical observables. 

The classical equations of motion \eqref{dWQA} are non-Lagrangian as they stand. There is no theorem prohibiting the action principle for HSGRA extended with suitable auxiliary fields; and yet no one has succeeded in formulating it. This prevents applying the standard quantization methods (canonical or path-integral) to this class of field-theoretical models. Therefore, it seems reasonable to try other approaches to the quantization problem that are not so closely related to the Lagrangian or Hamiltonian form of classical dynamics. Let us stress that all holographic HSGRAs are Lagrangian theories, more or less by definition. The actions, however non-local they may be, can be reconstructed \cite{Bekaert:2015tva} from the CFT correlators. At the free level such actions reduce to the Fronsdal action \cite{Fronsdal:1978rb} for an appropriate set of fields.

A possible alternative to the conventional quantization methods has been proposed long ago in \cite{Kazinski:2005eb}. It is based on the concept of a Lagrange anchor, which can be regarded as a covariant counterpart of  (degenerate) Poisson brackets. In the non-degenerate case, the Lagrange anchor is just inverse to the integrating factor in the inverse problem of the calculus of variations. Degeneracy manifests itself in the fact that not all classical degrees of freedom  may fluctuate in quantum theory; some of them remain purely classical.   
Nonetheless,  it is possible to define a full-fledged path integral  for transition amplitudes and quantum averages. 
With the help of Lagrange anchor every (non-)Lagrangian field theory in $d$ dimensions can equivalently be reformulated as a topological {\it Lagrangian} theory in $d+1$ dimensions. Applying then the standard BV quantization to the latter induces a quantization of the former (non-)Lagrangian theory.  It should be noted that the extension of a (non-)Lagrangian theory on the boundary to a topological Lagrangian theory in the bulk depends crucially on the choice of Lagrange anchor and different choices may result in different quantizations of one and the same classical theory. 

In the HSGRA context, the quantization method above was first considered in \cite{Boulanger:2011dd}. However, the Lagrange anchor behind the topological model of \cite{Boulanger:2011dd} does not reproduce the standard propagators for free higher spin fields in Fronsdal's theory. Therefore, the relation between the two theories is unclear beyond the classical equations of motion. In \cite{Kaparulin:2011zz}, the canonical Lagrange anchor was constructed for the subsystem of \eqref{dWQA} that describes a free scalar field. Contrary to the Lagrange anchor of the work \cite{Boulanger:2011dd}, it involves an infinite number of spacetime derivatives, and it seems unlikely that one can remove them all by the inclusion of higher spin fields and interactions. As shown in \cite{Kaparulin:2011aa}, the Lagrange anchors for higher spin fields described by the  Bargmann--Wigner equations also contain higher derivatives, the number of which increases with spin. In the two most recent papers \cite{MISUNA2019134956}, \cite{misuna2020offshell} a similar quantization method was implemented to reproduce propagators for free higher spin fields.

In the present paper, we consider another approach to quantizing (non-)Lagrangian theories. It employs the concept of a covariant presymplectic structure introduced in \cite{1987thyg.book..676C}, \cite{Zuckerman:1989cx}. The notion of presymplectic structure is, in a sense, dual to that of Lagrange anchor; both coincide for Lagrangian theories, being non-degenerate and inverse to each other. The degeneracy of a presymplectic structure reduces the algebra of physical observables admitting quantization. More precisely, the kernel distribution of the presymplectic structure must annihilate quantizable observables. Hence, the bigger the kernel, the smaller the algebra of quantum observables. 
Since all the gauge symmetry generators of the classical equations of motion belong to the kernel distribution, the quantizable observables are automatically gauge-invariant. As with the Lagrange anchor, classical equations of motion do not specify a compatible presymplectic structure uniquely, and different choices may lead to different algebras of quantum observables. While the Lagrange anchor aims at the path-integral quantization of (non-)Lagrangian dynamics, the concept of presymplectic structure is more adapted to the deformation quantization.

As was proved in \cite{Khavkine2013PresymplecticCA}, each covariant presymplectic structure defines and is defined by some Lagrangian. The corresponding Euler--Lagrange equations, however, are weaker than the original ones whereby admitting more solutions. That is why we call such Lagrangians `weak'. One may regard a weak Lagrangian as a solution to the inverse problem of the calculus of variations where the integrating factor is not necessarily invertible.  Finding a covariant presymplectic structure is thus fully equivalent to constructing a weak Lagrangian.

We will show that for  $4d$ HSGRA the corresponding weak Lagrangian has the form of an AKSZ sigma-model\footnote{Historically, the importance of the AKSZ construction for the development of action principles for HSGRA was pointed out in \cite{Boulanger:2011dd}, see also \cite{Boulanger:2015kfa,Bonezzi:2016ttk}, which have been an important reference frame for our work even though we deviate significantly from these papers by considering the presymplectic case, where the groundwork has been laid in \cite{Alkalaev_2014,grigoriev2016presymplectic}. } \cite{Alexandrov:1995kv},  or more precisely, its presymplectic counterpart \cite{Alkalaev_2014}, \cite{grigoriev2016presymplectic}. For free higher spin fields such a weak Lagrangian was proposed in \cite{Sharapov:2016qne}. There is a number of immediate advantages of the presymplectic approach. Among these are $(i)$ minimality: one does not have to introduce any auxiliary fields on top of what are already present in Eq. \eqref{dWQA}; $(ii)$ background independence and gauge invariance: we do not have to pick any particular vacuum, like $AdS_4$ and the gauge symmetry is fully taken into account to every order in the weak curvature expansion; $(iii)$ relation to the canonical quantization and to the Lagrangian formulation that HSGRAs must have in principle; $(iv)$ a complete classification and an explicit description of admissible presymplectic structures in terms of the Chevalley--Eilenberg cohomology of the underlying  higher spin algebra, which can be reduced to a much simpler Hochschild cohomology and, finally, computed with the help of the techniques of \cite{Sharapov:2020quq}.

\vspace{0.1cm}
\noindent The main results of this paper can be summarized as follows:
\vspace{-0.2cm}
\begin{itemize}\setlength\itemsep{-0.2em}
    \item we prove the existence of the weak action -- presymplectic AKSZ sigma-model -- for $4d$ formal HSGRA and construct the first few terms in the weak-field expansion. Higher order corrections are proved to be unobstructed and can be found with the help of general techniques from \cite{Sharapov:2018hnl,Sharapov:2018ioy,Sharapov:2018kjz,Sharapov:2019vyd}. The action depends on one additional coupling constant in accordance with conjectured duality to Chern--Simons vector models \cite{Giombi:2011kc};
    
    \item In arbitrary dimension $d$ the presymplectic AKSZ action for HSGRA begins with
    \begin{align}\notag
        S=\int \big\langle  \mathcal{V}(\momega,...,\momega, \mC)\star (d\momega-\momega \star \momega)\rangle +\mathcal{O}(\mC^2)\,,
    \end{align}
    where $\momega$ and $\mC$ are a connection and a matter field valued in a given higher spin algebra, $\mathcal{V}(\momega,\ldots ,\momega, \mC)$ is a cocycle of this higher spin algebra and $\langle - \rangle$ is the invariant trace; 
    
    \item for free fields we show that the corresponding presymplectic structure gives the correct quantization for fields with $s\geq1$ (the scalar field's presymplectic structure starts to contribute only at the next order). It also gives the correct commutation relations for the global higher spin symmetry currents. The action correctly reproduces the free field dynamics, i.e., on $AdS_4$ it turns to be a genuine action, rather than a weak action;
    
    \item  it is shown that the presymplectic structure obtained from the Fradkin--Vasiliev part \cite{Vasiliev:1986bq} of the cubic action\footnote{This action contains a number of non-abelian cubic interactions that are consistent at the cubic level with the gauged higher spin algebra. However, it is not a complete action of the HSGRA up to the cubic level. The complete cubic action was found in \cite{Sleight:2016dba} and with one more parameter available only in $4d$ in \cite{Skvortsov:2018uru}. } agrees with the one resulting from the presymplectic AKSZ model, as well as the cubic vertex does. Therefore, the perturbative analysis performed up to the cubic level over the anti-de Sitter background is consistent with the background independent considerations of this paper. We also clarify the origin of the action.
\end{itemize}
\vspace{-0.1cm}
\noindent The paper is organized as follows. In the first sections we, to a large extent, review all the required material: formal dynamical systems are defined in Sec. \ref{sec:formal}; the details specific to HSGRA in four dimensions are introduced in Sec. \ref{sec:HSGRA}; we discuss presymplectic AKSZ sigma-models in Sec. \ref{sec:pre-AKSZ} with an example of gravity elaborated on. The covariant phase space of presymplectic models is discussed in Sec. \ref{sec:CPhS}. The central section, to which a skilled reader can immediately scroll down, is Sec. \ref{sec:PSHSGRA}, where we present the presymplectic AKSZ action for $4d$ HSGRA and discuss its properties and relations to other results in the literature. Sec. \ref{sec:waves} elaborates on the general properties of the presymplectic AKSZ models over the maximally symmetric higher spin backgrounds. The main statements of Sec. \ref{sec:PSHSGRA} are supported by technical Appendices A - D, where all necessary cohomology groups are computed. 

\section{Formal dynamical systems }
\label{sec:formal}

The formal dynamical systems we are going to discuss are defined in terms of $Q$-manifolds, see e.g. \cite{Roytenberg:2006qz, Cattaneo_2006, Voronov2019}. By a $Q$-manifold we understand a $\mathbb{Z}$-graded manifold endowed with an integrable  vector field $Q$ of grade\footnote{Here prefer the term {\it grade} to a more conventional {\it degree}; the latter is reserved to the ``degree of a differential form''. The grade of a homogeneous element $a$ is denoted by $|a|\in\mathbb{Z}$, e.g. $|Q|=1$.} one.  Since $Q$ is  odd, the integrability condition 
\begin{equation}\label{Q2}
[Q,Q]=2Q^2=0
\end{equation}
is a non-trivial restriction on $Q$. In mathematics, such vector fields $Q$ are usually called {\it homological} or {\it $Q$-structure} \cite{schwarz1993, Alexandrov:1995kv}. 
 Let us present some typical constructions of $Q$-manifolds that we will need later for reference. 

\begin{example}\label{E21}
A classical example of a $Q$-manifold is the `shifted' tangent bundle $T[1]M$ of an ordinary manifold $M$;  here, the tangent space's coordinates $\theta^\mu$ are assigned the grade one, while the local coordinates $x^\mu$ on the base manifold $M$ have grade  zero. Let $\mathcal{M}$ denote the total space of $T[1]M$ considered as a graded manifold. Then the algebra of `smooth functions' on $\mathcal{M}$ is clearly isomorphic to  the exterior algebra of differential forms $\Lambda^\bullet (M)$. The~isomorphism is established by the relation
\begin{equation}\label{id}
C^{\infty}(\mathcal{M})\ni f(x,\theta)\qquad \Longleftrightarrow \qquad f(x,dx)\in \Lambda^\bullet(M)\,.
\end{equation} 
Upon this identification the de Rham differential on $\Lambda^\bullet (M)$ passes to the canonical homological vector field \begin{equation}\label{rm-d}
\mathrm{d}=\theta^\mu\frac{\partial}{\partial x^\mu}
\end{equation}
on $\mathcal{M}$. The cohomology of the operator $\mathrm{d}: C^\infty (\mathcal{M})\rightarrow C^\infty(\mathcal{M})$ is obviously isomorphic to the de Rham cohomology of $M$. 
\end{example}

\begin{example}\label{E22}
 Let $\mathcal{G}=\bigoplus_{n\in \mathbb{Z}}\mathcal{G}_n$ be a graded Lie algebra with a homogeneous basis $\{e_A\}$ and the 
commutation relations $[e_A,e_B]=f_{AB}^C e_C$. Then one can endow the vector space $\mathcal{G}[1]$, viewed as a $\mathbb{Z}$-graded manifold, with the quadratic homological vector field \begin{equation}\label{EQ}
   Q=-\frac12(-1)^{|e_A|(|e_B|-1)}\xi^A\xi^Bf_{AB}^C\frac{\partial}{\partial \xi^C}\,.
   \end{equation}
Here $\xi^A$ are global coordinates on $\mathcal{G}[1]$ relative to the basis $\{e_A\}$. By definition, $|\xi^A|=-|e_A|+1$. It is easy to see  that the condition $Q^2=0$ is exactly equivalent to the Jacobi identity in $\mathcal{G}$.  Endowed with the action of $Q$, the space of smooth functions $C^\infty(\mathcal{G}[1])$ becomes 
a cochain complex computing  the cohomology of the Lie algebra $\mathcal{G}$ with trivial coefficients, the Chevalley--Eilenberg  (CE) complex. 

Extending $C^\infty(\mathcal{G}[1])$ to the algebra $\mathcal{T}(\mathcal{G}[1])$ of smooth tensor fields on $\mathcal{G}[1]$  wherein $Q$ acts through the operator of Lie derivative $L_Q= i_Q d-d i_Q$, we obtain a more general CE complex with coefficients in the tensor powers of adjoint and coadjoint representations of the Lie algebra $\mathcal{G}$. For instance, considering the action of $Q$ in the space of smooth vector fields on $\mathcal{G}[1]$ yields  the standard CE complex with coefficients in the adjoint representation. The cohomology of this complex controls both the deformation of the homological vector field (\ref{EQ}) and the underlying Lie algebra $\mathcal{G}$. For our present purposes, 
the most interesting is the algebra of exterior differentials forms $\Lambda^\bullet (\mathcal{G}[1])$. This corresponds to the CE complex with coefficients
in the symmetrized tensor powers $S^\bullet \mathcal{G}^\ast$ of the coadjoint module of $\mathcal{G}$, see Sec. \ref{sec:PSHSGRA}. 
\end{example}

\begin{example}\label{E23}
Given a $Q$-manifold $(M,Q)$, consider the shifted tangent bundle $T[n]M$. The operator of Lie derivative $L_Q$ allows one to extend the homological vector $Q$ from $M$ to the total space of $T[n]M$ considered as a $\mathbb{Z}$-graded manifold. Let us denote the latter  by $\mathcal{M}$. If $x^i$ are local coordinates on $M$ and $v^i$ are linear coordinates in the tangent spaces $T_xM$ relative to 
the  natural frame $\{\partial/\partial x^i\}$, then the extended vector field on $\mathcal{M}$ is given by 
\begin{equation}
    \mathcal{Q}=Q^i\frac{\partial }{\partial x^i}+(-1)^nv^j\frac{\partial Q^i}{\partial x^j}\frac{\partial }{\partial v^i}\,.
\end{equation}
By definition,  $|v^i|=|x^i|+n$. We will refer to the $Q$-manifold  $(\mathcal{M}, \mathcal{Q})$ as the {\it first (tangent) prolongation of} $(M,Q)$.  One can obviously iterate this construction  producing bigger and bigger $Q$-manifolds.  Moreover, the shifted tangent bundle may well be replaced by an arbitrary tensor bundle of $M$. 
The resulting  $Q$-manifolds are particular examples of {\it $Q$-vector bundles} \cite{Lyakhovich_2010}, \cite{Grigoriev_2019}.  
\end{example}

In order to define a formal dynamical system we need a pair of $Q$-manifolds: the {\it source} and the {\it target}.  As a source manifold $\mathcal{M}$ we always take the total space of the shifted tangent bundle $T[1]M$ of a space-time manifold $M$ equipped with the canonical homological vector field (\ref{rm-d}).  Let $\mathcal{N}$ denote a target manifold with homological vector field $Q$.  We will treat $\mathcal{N}$ in the sense of {\it formal differential geometry} \cite{Kontsevich:2006jb}, identifying `smooth functions' on $\mathcal{N}$ with formal power series in globally defined coordinates $w^A$:
$$
w^Aw^{B}=(-1)^{|w^{{A}}||w^B|}w^Bw^A\,.
$$
In particular, the homological vector field $Q$ is given by the series 
\begin{equation}\label{Q}
Q=\sum_{n=0}^\infty w^{A_n}\cdots w^{A_1}Q_{A_1\cdots A_n}^A\frac{\partial}{\partial w^A} 
\end{equation}
for some structure constants $Q^A_{A_1\cdots A_n}$. Besides, two more assumptions about the structure of the target $Q$-manifold $\mathcal{N}$ will be made.
\begin{enumerate}
    \item There are no coordinates of negative grade among $\{w^A\}$, so that the manifold $\mathcal{N}$ is actually $\mathbb{N}$-graded.
    \item The expansion (\ref{Q}) starts with quadratic terms in coordinates, that is, $Q^A=Q^A_{A_1}=0$. Such homological vector fields $Q$ are called {\it minimal}.
\end{enumerate}
Notice that the structure constants $Q^A_{A_1A_2}$ of a minimal homological vector field (\ref{Q}) satisfy the Jacobi identity of a graded Lie algebra as in Example \ref{E22}.

Given the pair of $Q$-manifolds above, we identify the classical fields with the smooth maps  ${\mw}: T[1]M\rightarrow \mathcal{N}$ of degree zero. In terms of local coordinates, each map ${\mw}$ is given by a set of relations $w^A={\mw}^A(x,\theta)$, where ${\mw}^A(x,\theta)$ are smooth functions of $x$'s and $\theta$'s with $|{\mw}^A(x,\theta)|=|w^A|$.  
The {\it true} field configurations are, by definition, those relating the homological vector fields, i.e., $\mw_\ast(\mathrm{d})=Q$. Upon identification (\ref{id}), the last condition takes the form of a system of differential equations, namely, 
\begin{equation}\label{EoM}
    d\mw^A=\sum_{n=2} ^\infty Q^A_{A_1\cdots A_n}\mw^{A_n}\wedge \cdots \wedge \mw^{A_1}\,.
\end{equation}
The l.h.s. is given here by the differentials of the forms $\mw^A\in \Lambda^\bullet (M)$, while the r.h.s. involves exterior products of the same forms. Equations  (\ref{EoM}) are thus adopted as field equations for a collection of form fields on $M$. In what follows we will systematically omit the wedge product sign and write Eq. (\ref{EoM}) simply as 
\begin{equation}\label{dWQ}
    E^A\equiv d\mw^A-Q^A(\mw)=0\,,
\end{equation}
$Q^A(\mw)$ being  exterior polynomials in the $\mw$'s. Applying now the de Rham differential to (\ref{dWQ}), one can readily see that the field equations contain no hidden integrability  conditions whenever $Q^2=0$. Besides general covariance, system (\ref{dWQ}) enjoys the gauge symmetry transformations
\begin{equation}\label{GS}
    \delta_{\varepsilon}\mw^A=d\varepsilon^A+ \varepsilon^B\partial_BQ^A\,,\qquad \delta_{\varepsilon}E^A=(-1)^{|w^A|+|w^B|}\varepsilon^C\big(\partial_B\partial_CQ^A\big) E^B\,,
    \end{equation}
where the infinitesimal gauge parameters $\varepsilon^A$ are differential forms of appropriate degrees.  Actually, the general coordinate transformations of the form fields  $\mw^A$ are specifications of  (\ref{GS}). If $\xi$ is a vector field  generating a one-parameter group of diffeomorphisms of $M$, then by Cartan's formula
\begin{equation}\label{diff}
\delta_\xi \mw^A=L_\xi \mw^A=di_\xi \mw^A+i_\xi d\mw^A\approx d(i_\xi \mw^A)+i_\xi Q^A(\mw)\,.
\end{equation}
Hereinafter, the sign $\approx$ means ``equal when the equations of motion hold", i.e., on-shell.  Therefore we can set $\varepsilon^A=i_\xi \mw^A$ to  reproduce the action of an infinitesimal diffeomorphism $\xi$ on the solution space to the field equations (\ref{dWQ}).

\begin{remark}\label{R24}
The gauge symmetry (\ref{GS}) is known to be strong enough to gauge away all local degrees of freedom for any {\it finite} collection of fields $\{\mw^A\}$ provided that $\dim M>1$. 
The last fact can be seen as follows. 
Notice that each form field $\mw^A$ of  degree $>0$ comes `with its own' gauge parameter $\varepsilon^A$ such that $\deg \mw^A=\deg \varepsilon^A+1$. 
In a topologically trivial situation, the corresponding gauge transformation $\delta_{\varepsilon }\mw^A=d\varepsilon^A+\ldots$ allows one 
to remove all physical modes of the field $\mw^A$ subject to the first-order differential equation $d\mw^A=\ldots$ This suggests that all physical degrees of freedom are accommodated in the  zero-form fields, let us enumerate them $\mw^a$.  Without a detailed analysis it is clear that the solution space of the equations $d\mw^a=\ldots$ is parameterized by constants $c^a$  whose number is equal to the  number of zero-forms $\mw^a$'s.\footnote{ In general, the higher-degree forms $\mw^A$ may also add up to the physical sector a finite  number of global degrees of freedom associated with the de Rham cohomology of the space-time manifold $M$.} Therefore if we are interested in non-topological field theories, i.e., models with propagating  degrees of freedom, then we have to consider infinite multiplets of zero-form fields.
\end{remark}

\begin{remark}
It is known that any system of (gauge) partial differential equations can be brought into the form (\ref{EoM}) at the expense of introducing an infinite number of auxiliary fields \cite{Barnich:2004cr,Barnich:2010sw} and references therein. In the context of HSGRA this is known as an {\it unfolded representation}  \cite{Vasiliev:1988sa,Bekaert:2005vh1}. 
Of course, care must be exercised in treating (\ref{EoM}) as a system of partial differential equations whenever an infinite number of fields and interaction vertices are involved  into the game. That is why we refer to (\ref{EoM}) as a {\it formal} dynamical system.  In this paper, we leave aside all subtle  analytical issues related to the field equations (\ref{EoM}) focusing upon their formal consistency. 
\end{remark}

\begin{remark} Notice that the system remains consistent if we omit all the vertices in the r.h.s. of (\ref{EoM}) except the quadratic ones. The resulting system is defined  
solely in terms of the underlying Lie algebra $\mathcal{G}$. Although the truncated equations of motion are still non-linear, one may think of them as describing free field dynamics; the `genuine' interaction vertices start form the cubic order. From this perspective switching on a consistent interaction amounts to deforming a quadratic homological vector field. Hence, all  non-trivial interactions admitted by a free gauge system are fully controlled by the cohomology of the graded Lie algebra $\mathcal{L}$ as discussed in Example \ref{E22}. This leads to a refined version of the {\it N\"oether procedure } (see e.g. \cite{Barnich_1993}) and may be regarded as a main technical advantage of the formal dynamical system approach.  We will illustrate this point in the next section.
\end{remark}

\section{Higher spin gravity in four dimensions}
\label{sec:HSGRA}

The gauge theory of interacting higher spin fields delivers the major class of formal dynamical systems with propagating degrees of freedom. Being general covariant and involving a massless  field  of spin two, they are usually referred to as  Higher Spin Gravities (HSGRA). 
As explained in the previous section, the structure of the corresponding equations of motion is largely controlled by the underlying graded Lie algebra $\mathcal{G}$.  

Below we specify the algebra $\mathcal{G}$ for the case of $4d$ HSGRA. 
\begin{enumerate}
    \item [(1)] $\mathcal{G}$ is concentrated in degrees zero and one, so that $\mathcal{G}=\mathcal{G}_0 \oplus \mathcal{G}_1$ as a vector space and 
    $$
    [\mathcal{G}_0,\mathcal{G}_0]\subset \mathcal{G}_0\,,\qquad [\mathcal{G}_0,\mathcal{G}_1]\subset \mathcal{G}_1\,,\qquad [\mathcal{G}_1,\mathcal{G}_1]=0\,.
    $$
    This allows us to regard  the odd commutative ideal $\mathcal{G}_1$ as a module over the even  subalgebra $\mathcal{G}_0$. 
    \item[(2)] The $\mathcal{G}_0$-module $\mathcal{G}_1$ is given by the adjoint representation of $\mathcal{G}_0$. Therefore $\mathcal{G}_0\simeq
    \mathcal{G}_1$ as vector spaces.  
    \item [(3)] $\mathcal{G}_0=gl_n(\mathfrak{A})$, that is, $\mathcal{G}_0$ is the Lie algebra  of $n\times n$-matrices with entries in an associative algebra $\mathfrak{A}$. 
    
    \item[(4)] $\mathfrak A=\mathcal{A}\otimes \mathcal{A}$, where the complex  associative algebra $\mathcal{A}$ is given by the smash product  $\mathcal{A}=A_1\rtimes \mathbb{Z}_2$ of the first Weyl algebra $A_1$ and the cyclic group $\mathbb{Z}_2$ generated by an involutive automorphism of $A_1$. More precisely, $\mathcal{A}$ is the unital $\mathbb{C}$-algebra on the three generators $y_1$, $y_2$, and $\kappa$ obeying the relations
    \begin{equation}\label{A1Z}
   [ y_1, y_2]=2i\,,\qquad \qquad\{ \kappa, y_\alpha\}=0\,,\qquad \qquad \kappa^2=1\,.
    \end{equation}
   Hereinafter, $\alpha=1,2$ and the brackets (braces) stand for the commutator (anti-com\-mutator).   The first relation in (\ref{A1Z}) is just Heisenberg's commutation relation for the `canonical variables' $y_1$ and $y_2$. Adding the generator $\kappa$ makes the outer automorphism $y_\alpha\rightarrow -y_\alpha$ of $A_1$ into an inner automorphism of $\mathcal{A}$.
\end{enumerate} 
In what follows we will denote the generators of the left and right tensor factors in $\mathfrak{A}=\mathcal{A}\otimes \mathcal{A}$ by $(y_\alpha,\kappa)$ and $(\bar y_{\dot\alpha},\bar{\kappa})$, so that the general element of $\mathfrak{A}$ can be written as 
\begin{equation}\label{ge}
    a=f(y,\bar y)+g(y,\bar y)\kappa + h(y,\bar y)\bar\kappa+ v(y,\bar y)\kappa\bar\kappa\,,
\end{equation}
$f$, $g$, $h$, and $v$ being some ordered complex polynomials in $y$'s and $\bar y$'s. The complex conjugation of the ground field $\mathbb{C}$
extends to the semi-linear anti-involution of $\mathfrak{A}$ that takes $y^\alpha$ and $\kappa$ to $ (y^\alpha)^\ast=\bar y{}^{\dot \alpha}$ and $\kappa^\ast=\bar\kappa$. 

\begin{remark}\label{R31}
The Lie algebra $\mathcal{G}_0$ contains $sp_4(\mathbb{R})\simeq so(3,2)$ as a real subalgebra. The latter is spanned by the unit $n\times n$-matrix $1\!\! 1$ multiplied by real quadratic  polynomials in $y$'s and $\bar y$'s; in so doing, the complex conjugate polynomials $\{y_\alpha, y_\beta\}$ and $\{\bar y_{\dot\alpha},\bar y_{\dot\beta}\}$ generate the complexified Lorentz subalgebra $so(3,1)$, while $y_\alpha\bar y_{\dot \beta}$ correspond to the $AdS_4$ transvections. The algebra $so(3,2)$, being the Lie algebra  of isometries
of $4d$ anti-de~Sitter space, suggests that the empty $AdS_4$  may appear as a natural vacuum solution of $4d$ HSGRA. 
\end{remark}

\begin{remark}\label{R32}

The full associative algebra generated by $y$'s and $\bar y$'s is clearly isomorphic to the second Weyl algebra $A_2=A_1\otimes A_1$. Let $\mathfrak{hs}\subset A_2$ denote the subalgebra of all elements commuting with $\kappa\bar\kappa$. It is spanned by the even polynomials $f(y,\bar y)=f(-y,-\bar y)$  and is called the {\it higher spin algebra} \cite{Fradkin:1986ka,Vasiliev:1988xc}, hence the notation. Since the element $\kappa\bar\kappa$ generates a $\mathbb{Z}_2$ subgroup in the Klein four-group $\mathbb{Z}_2\times \mathbb{Z}_2=\{1,\kappa, \bar\kappa, \kappa\bar\kappa\}$, one can also characterize $\mathfrak{hs}$  as a $\mathbb{Z}_2$-invariant subalgebra of $A_2$. 
Extending $\mathfrak{hs}$ with  $\kappa$ and $\bar\kappa$, we get the smash product algebra $\mathfrak{hs}\rtimes (\mathbb{Z}_2\times \mathbb{Z}_2)$ called usually the {\it  extended higher spin algebra}.
\end{remark}

\begin{remark}
For technical reasons explained in Appendix \ref{app:Hoch}, we  will consider the Lie algebra of `big matrices'. Formally, it is defined as the inductive limit $gl(\mathfrak{A})=\lim\limits_{\rightarrow }gl_n(\mathfrak{A})$ associated with the natural embedding $gl_n\subset gl_{n+1}$ (an $n\times n$-matrix is augmented by zeros). The result is the Lie algebra $\mathcal{G}_0=gl(\mathfrak{A})$ of infinite matrices with only finitely many entries different from zero. The algebra can also be augmented without harm by the matrices proportional to the unit matrix $1\!\!1$. 
\end{remark}

\vspace{3mm}

Given the graded Lie algebra $\mathcal{G}=\mathcal{G}_0\oplus \mathcal{G}_1$ above, we can  define  $4d$ HSGRA as a formal dynamical system with a one-form field $\momega$ and a zero-form field $\mC$, both taking values in $gl(\mathfrak{A})$. Eqs. (\ref{EoM}) assume now the form 
\besubeqs\label{dwdc}
\begin{align}
    d\momega &=\momega\star \momega +\mathcal{V}_3(\momega, \momega, \mC) +\mathcal{V}_4(\momega, \momega, \mC, \mC)+\cdots \,,\label{dw}\\[3mm]
    d\mC&=\momega\star  \mC-\mC\star \momega +\mathcal{V}_3(\momega, \mC, \mC)+\mathcal{V}_4(\momega, \mC, \mC, \mC)+\cdots\label{dC}\,.
\end{align}
\esubeqs
Here $\star$ combines the wedge product of differential forms with the matrix product in $gl(\mathfrak{A})$.
To look for the dynamical equations for interacting massless fields of all spins in this form was first proposed in \cite{Vasiliev:1988sa}.
Geometrically, the target space of fields $\mC$ and $\momega$  is given by an infinite-dimensional $Q$-manifold coordinatized by variables of degree zero and one.   In the smooth setting such  $Q$-manifolds are known to be equivalent to Lie algebroids \cite{Vaintrob}. This allows  one to consider the field equations \eqref{dwdc} as originating from a (formal) Lie algebroid over the target space of fields $\mC$. The natural vacuum solution $\mC=0$ corresponds then to a singular point of the Lie algebroid with isotropy Lie algebra $\mathcal{G}_0=gl(\mathfrak{A})$.

Omitting the interaction vertices $\mathcal{V}_k$, we are left with the system describing a free HSGRA. Indeed, in that case  one can view Eq. (\ref{dw}) as the zero-curvature condition for the connection one-form $\momega$ associated with the gauge algebra $gl(\mathfrak{A})$.  In topologically trivial situation, one can solve the equation  in a purely  gauge form as $\momega=dg\star g^{-1}$, where $g$ is a zero-form with values in $GL(\mathfrak{A})$. 
Then Eq.(\ref{dC}) identifies $\mC$ as a covariantly constant section with values in the adjoint representation of the gauge group $GL(\mathfrak{A})$.  Again, one can write the general solution for $\mC$ as $\mC=g\star C_0\star g^{-1}$, where $C_0$ is an arbitrary element of the algebra $gl(\mathfrak{A})$. We thus see that, modulo gauge invariance, the solutions to the free equations  form a {\it linear} space isomorphic to $gl(\mathfrak{A})$. When endowed with an invariant Hermitian inner product, the space can be identified with the Hilbert space of one-particle states of $4d$ HSGRA.
More precisely, the usual interpretation in terms of particles arises from decomposition of the adjoint representation of $GL(\mathfrak{A})$ into the direct sum of irreducible unitary representations of the anti-de-Sitter group $SO(3,2)\subset GL(\mathfrak{A})$. 

In a slightly different language this is the content of the Flato--Fronsdal theorem \cite{Flato:1978qz} stating that the tensor product of free massless $3d$ scalar and fermion with themselves decomposes into a direct sum of massless fields with all spins. As was shown by Dirac \cite{Dirac:1963ta} many years before that, the free massless $3d$ scalar and fermion, as representations of $so(3,2)$, can be realized as even and odd vectors in the Fock space wherein $A_2$ acts naturally.\footnote{Let us choose creation and annihilation operators, $[a_\ga, a^{\dag}{}^{\gb}]=\delta\fdu{\ga}{\gb}$, instead of $y_\alpha$ and $\bar{y}_{\dot\alpha}$. Then, $sp_4(\mathbb{R})$ is realized by the bilinears in $a$ and $a^\dag$. The scalar/fermion states $|v\rangle$ correspond to the span of $f(a^\dag)|0\rangle$ for even/odd  $f(a^\dag)$ that act on the vacuum $|0\rangle$.} Together with the Flato--Fronsdal theorem, this equips the higher spin multiplet with the action of $A_2$ (understood as a Lie algebra). The appearance of $\mathbb{Z}_2$ is a bit harder to explain. The group  $\mathbb{Z}_2$ realizes an isomorphism that allows one to treat elements from the tensor product $|v\rangle |w\rangle$ (states) as elements $|v\rangle \langle w|$ from the higher spin algebra (operators). As a result one can embed the states into the higher spin algebra. This automorphism is realized by $\kappa$, $\bar\kappa$. The smash product algebra $\mathfrak{A}$, which extends the higher spin algebra $\mathfrak{hs}$ with $\kappa$, $\bar\kappa$, is useful: it is the deformation of $\mathcal{A}$ that allows one to reconstruct all vertices $\mathcal{V}_k$ in \eqref{dwdc}, see below and \cite{Sharapov:2019vyd} for explicit formulas.

\begin{remark}\label{R34}
It should be noted that the spectrum of fields generated by the graded Lie algebra $\mathcal{G}$ above is superfluous, containing more fields than actually needed to describe $4d$ HSGRA. The true physical degrees of freedom are accommodated in the graded subalgebra $\mathcal{G}'\subset \mathcal{G}$, where  $\mathcal{G}'_0=gl(\mathfrak{hs})\Pi$, $\Pi=(1+\kappa\bar\kappa)/2$ and $\mathcal{G}'_1=gl(\mathfrak{hs})K$, $K=(\kappa+\bar\kappa)/2$, $K^2=\Pi$, see Remark \ref{R32}.  The corresponding field configurations are given by $\momega =\momega (y,\bar y)\Pi$ and $\mC=\mC(y,\bar y)K$ such that $\momega(-y,-\bar y)=\momega(y,\bar y)$ and $\mC(-y,-\bar y)=\mC(y,\bar y)$. The use of the `extended' algebra $\mathcal{G}$ offers considerable technical advantages over $\mathcal{G}'$. 
In particular, it makes possible applying  the K\"unneth formula to the tensor product $\mathfrak{A}=\mathcal{A}\otimes \mathcal{A}$. The redundant  fields do not interfere the dynamics of the physical ones and can  easily be excluded at the end of all calculations. This final projection onto the subspace of physical fields will always be implied in the sequel. 
\end{remark}

\begin{remark}\label{R35}
One can also deduce the spectrum of $4d$ HSGRA over $AdS_4$ background  by the conventional field-theoretic analysis. For this end, put ${\momega}=\stackrel{_{\circ}}{\momega}\!\!1\!\!1$, where the one-form field 
\begin{align}\label{AdS}
\stackrel{_{\circ}}{\momega}=-\frac{i}{4}\{y_\alpha, y_\beta\}w^{\alpha\beta}-\frac{i}{4}\{\bar y_{\dot\alpha}, \bar y_{\dot\beta}\}w^{\dot\alpha\dot\beta}-\frac{i}{2}y_\alpha y_{\dot\beta}h^{\alpha\dot\beta}
\end{align} 
takes values in the subalgebra $so(3,2)\subset gl(\mathfrak{A})$, as discussed in Remark \ref{R31}.  The forms $w^{\alpha\beta}$ and $w^{\dot\alpha\dot\beta}$ are then naturally identified with the components of a unique spin-connection  compatible with the vierbein  $h^{\alpha\dot\beta}$ of $AdS_4$, see Example \ref{PG} below for $\lambda=1$. Upon this identification all the structure relations of $AdS_4$ geometry are compactly encoded by the Maurer--Cartan equation $d\!\!\stackrel{_{\circ}}{\momega}=\stackrel{_{\circ}}{\momega}\!\!\star \!\!\stackrel{_{\circ}}{\momega}$. On substituting (\ref{AdS}) into (\ref{dC}) and restricting to the physical sector, one gets an infinite number of relativistic wave equations on the component spin-tensor fields accommodated in
\begin{equation}
    \mC=\sum_{n,m=0}^\infty \mC^{\alpha_1\cdots\alpha_n\dot\beta_1\cdots\dot\beta_m}(x)y_{\alpha_1}\cdots y_{\alpha_n}\bar y_{\dot\beta_1}\cdots\bar y_{\dot\beta_m}K\,.
\end{equation}A closer inspection of these equations shows that they are equivalent to the Bargmann--Wigner equations for the (matrix-valued) massless fields of 
all integer spins, see e.g. \cite{Vasiliev:1988sa,Didenko:2014dwa}. 
\end{remark}

Although $\mC=0$ looks like a natural vacuum solution of $4d$ HSGRA, the most general Lorentz- and $gl$-invariant family of vacuums of Eqs. \eqref{dwdc} is of the form 
\begin{equation}\label{vac}
    \mC=(\mu +\rho\kappa+\bar\rho\bar\kappa + \lambda\kappa\bar\kappa)1\!\!1\,,
\end{equation}
where  $\mu,\rho,\lambda$ are complex parameters. A useful particular solution corresponds to the center of $\mathfrak{A}$, namely, $C=\mu 1\!\!1$.  On substituting this,  Eq. (\ref{dw}) takes the form $d\momega=\momega\circ\momega$, where by $\circ$ we denoted the bilinear operator determining the r.h.s. of the equation, $\mu$ being a formal deformation parameter that is implicit. Eq. \eqref{dC} is trivially satisfied. Since $d^2=0$, the $\circ$-commutator must obey the Jacobi identity, i.e., $[[\momega,\momega],\momega]=0$.
A simple way to satisfy the last condition is to require the $\circ$-product to be associative. Then the r.h.s. of Eq. (\ref{dw}) gives rise to deformation of the associative product in $\mathfrak{A}=\mathcal{A}\otimes \mathcal{A}$ that can depend on several parameters, in principle. The existence, non-triviality and the number of parameters of such a deformation (and hence, interaction) depend on the properties of the algebra itself. As explained in Appendix \ref{B2}, either $\mathcal{A}$ factor in $\mathfrak{A}$ admits a non-trivial deformation, which results in a two-parameter deformation.  This is obtained by alteration of  Heisenberg's commutation relation in (\ref{A1Z}). Now it reads 
\begin{equation}\label{dosc}
    [y_1,y_2]=2i(1+\nu\kappa)\,,
\end{equation}
$\nu$ being a complex parameter. 
The resulting one-parameter family of algebras $\mathcal{A}(\nu)$  is known under the name of {\it deformed oscillator algebra}. Implicitly, it was first introduced by E.~Wigner in  his  1950 paper \cite{Wigner} on foundations of quantum mechanics. See \cite{Yang:1951pyq,Mukunda:1980fv,Vasiliev:1989re} for subsequent discussions. The general element  of 
$\mathfrak{A}(\nu,\bar\nu)=\mathcal{A}(\nu)\otimes \mathcal{A}(\bar\nu)$ is still given by an ordered polynomial in $y$'s, $\bar y$'s, $\kappa$, and $\bar\kappa$.\footnote{In other words, the algebra $\mathfrak{A}(\nu,\bar\nu)$ satisfies the Poincar\'e--Birkhoff--Witt condition. } There are explicit, albeit somewhat  cumbersome, formulas for the product of such polynomials \cite{Pope:1990kc,Bieliavsky:2008mv,Joung:2014qya,Korybut:2014jza,Basile:2016goq,korybut2020star}. Choosing, for example, the symmetric (or Weyl) ordering for $y$'s and $\bar y$'s, while keeping the generators $\kappa$ and $\bar\kappa$ in the rightmost position, one can write the following expansion for the $\circ$-product of two polynomials $a(y,\bar y)$ and $b(y,\bar y)$: 
\begin{equation}\label{dprod}
    a\circ b=a\ast b+\sum_{n+m>0}\phi_{nm}(a,b)(\nu\kappa)^n(\bar\nu\bar\kappa)^m\,.
\end{equation}
Here $\ast$ stands for the usual Weyl--Moyal product in $A_2$ and the collection of bilinear operators $\{\phi_{nm}\}$ defines a two-parameter deformation of the full algebra $\mathfrak{A}$
 `in the directions of $\kappa$ and $\bar\kappa$'. A nice integral representation  for the first-order deformations $\phi_{10}$ and $\phi_{01}$ can be found in Appendix \ref{B2}.

The surprising thing is that knowledge of the interaction vertices  for  a particular vacuum solution $C=\mu 1\!\!1$ permits reconstruction of the r.h.s. of Eqs. (\ref{dw}, \ref{dC}) for arbitrary  $\mC$! As was shown in \cite{Sharapov:2019vyd}, it is possible to express all the $\mathcal{V}$'s  through compositions of the bilinear maps $\phi_{nm}$ entering the expansion (\ref{dprod}). For example,
\begin{align}\label{firstV}
\begin{array}{rcl}
    \mathcal{V}_3(\momega,\momega, \mC)^{i}{}_{j}&=& \displaystyle\sum_{n+m=1}\nu^n\bar\nu^m\phi_{nm}(\momega^i{}_{i_1},\momega^{i_1}{}_{i_2})\star \mC^{i_2}{}_{j} \,,\\[5mm]
   \displaystyle  \mathcal{V}_4(\momega,\momega, \mC, \mC)^i{}_j&=&\displaystyle\sum_{n+m=2}\nu^n\bar\nu^m\phi_{nm}(\momega^i{}_{i_1},\momega^{i_1}{}_{i_2})\star \mC^{i_2}{}_{i_3}\star \mC^{i_3}{}_j \\[5mm]
   \displaystyle& +&\displaystyle\sum_{n+m=k+l=1}\nu^{n+k}\bar\nu^{m+l}\phi_{nm}(\phi_{kl}(\momega^i{}_{i_1},\momega^{i_1}{}_{i_2}), \mC^{i_2}{}_{i_3})\star \mC^{i_3}{}_j\,.
    \end{array}
\end{align}
Here we wrote down explicitly the $gl$-indices $i$'s and $j$'s. As is seen, they are all contracted in chain. Of course, all the vertices are defined modulo field redefinitions. The modulus of the complex parameter $\nu$ can obviously be absorbed by rescaling $\mC$ (or $\mu$ for $C=\mu 1\!\!1$, which is why the value of $\mu$ played no role). Therefore,  setting $\nu=e^{i\theta}$ we are left with the only free 
parameter $\theta$, which allows one to interpolate between the HSGRAs of type $A$ ($\theta=0$) and type $B$ ($\theta=\pi/2$) \cite{Sezgin:2003pt}. In the context of AdS/CFT correspondence, these two theories\footnote{to an extent to which formal dynamical systems represent the actual field theories behind them} should be dual \cite{Klebanov:2002ja,Sezgin:2003pt,Leigh:2003gk}, respectively, to the free boson and fermion vector models on the boundary of $AdS_4$. For general $\theta$, this family of HSGRAs is expected to be dual to Chern--Simons matter theories \cite{Giombi:2011kc}. 

Historically, explicit expressions for the first two vertices $\mathcal{V}_3$ and $\mathcal{V}_4$ were found in the works \cite{Vasiliev:1988sa}, \cite{Vasiliev:1989yr}. In the subsequent paper \cite{Vasiliev:1990cm}, a systematic method was developed for generating all the formal interaction vertices in $4d$ HSGRA. A direct relation of the interaction problem with that of deformations of extended higher spin algebras was established in our papers \cite{Sharapov:2019vyd,Sharapov:2018kjz}; this solves the problem of formal HSGRA for any given higher spin algebra and allows one to construct classical integrable systems out of any family of associative algebras. There is a number of formal HSGRA models available in the literature \cite{Vasiliev:1990cm,Vasiliev:2003ev,Neiman:2015wma,Bonezzi:2016ttk,Arias:2017bvi,Bekaert:2013zya,Bekaert:2017bpy,Grigoriev:2018wrx,Sharapov:2019vyd,Sharapov:2019pdu}, which include original models, variations and different realizations.

\section{Presymplectic AKSZ models}
\label{sec:pre-AKSZ}
Like the previous two sections this one is mostly expository. In order to formulate a class of field-theoretic models in the title  we need to equip the target space $\mathcal{N}$ 
with one more geometric structure discussed below.

First of all, we note that the $\mathbb{N}$-grading on the target space $\mathcal{N}$ can  be conveniently described by means of the Euler vector field  
\begin{equation}\label{vecN}
    N=\sum_A |w^A|w^A\frac{\partial}{\partial w^A}\,.
\end{equation}
By our assumption about target spaces all $|w^A|\geq 0$. We say that $T\in \mathcal{T}(\mathcal{N})$ is a homogeneous tensor field of grade $n$ if $L_NT=nT$, where $L_N$ denotes the Lie derivative along $N$. In particular, the commutation  relations $[N,N]=0$ and $[N,Q]=Q$ mean that $|N|=0$ and $|Q|=1$.  The highest grade of the coordinates $\{w^A\}$ is called the {\it degree} of $\mathcal{N}$ and is denoted by $\deg \mathcal N$. 
For instance, the shifted tangent bundle $T[1]M$ provides  an example of an $\mathbb{N}$-graded $Q$-manifold of degree one.

A {\it presymplectic structure of grade $n$} is, by definition, a closed two-form  $\Omega$ on $\mathcal{N}$ such that
\begin{equation}L_N\Omega=n\Omega\,.
\end{equation}
The pair $(\mathcal{N}, \Omega)$ is called a {\it presymplectic manifold}. If the matrix $(\Omega_{AB})$ of the two-form $\Omega$ happens to be non-degenerate in some (and hence any) coordinate system, then one speaks of a symplectic structure and a symplectic manifold. In that case the inverse tensor $\Pi=\Omega^{-1}$ defines a Poisson structure on $\mathcal{N}$ of grade $(-n)$. 
Denote by $\ker \Omega$ the space of all vector fields $V$ on $\mathcal{N}$ such that $i_V\Omega=0$. Since $d\Omega=0$, the vector fields of $\ker \Omega$ span an integrable distribution on $M$. 
We will refer to it  as the {\it kernel distribution} of $\Omega$. 

A vector field $X$ (a function $H$) is said to be {\it Hamiltonian} if there exists a function $H$ (a vector field X) such that 
\begin{equation}\label{XH}
    i_X\Omega=dH\,.
\end{equation}
The function $H$ is called the Hamiltonian of the vector field $X=X_H$. It follows immediately from the definition that  $(i)$  $|X|+|\Omega|=|H|$; $(ii)$ the presymplectic 
structure is invariant under the action of Hamiltonian vector fields, i.e., $L_X\Omega=0$; $(iii)$ given a Hamiltonian $H$, Eq. (\ref{XH}) defines $X$ up to adding a vector field from $\ker\Omega$; $(iv)$ each Hamiltonian  is invariant under the kernel distribution, i.e., $L_VH=0$ for all $V\in \ker\Omega$. 

An important fact about the geometry of presymplectic manifolds (graded or not) is that the Hamiltonians of Hamiltonian vector fields form a graded Poisson algebra w.r.t. point-wise multiplication  and the Poisson bracket 
\begin{equation}\label{HF}
    \{H,F\}=(-1)^{|H|}L_{X_H}F=(-1)^{|H|}i_{X_H}i_{X_F}\Omega=-(-1)^{(|H|+|\Omega|)(|F|+|\Omega|)}\{F,H\}\,.
\end{equation}
One can easily verify that the bracket is well-defined and satisfies all the required properties: bi-linearity, graded anti-symmetry, the Leibniz rule, and the Jacobi identity. 
In the case of symplectic manifolds  any function is a Hamiltonian and $\{H,F\}=\Pi(dH,dF)$.

\begin{proposition}[\cite{Roytenberg:2006qz}]\label{P41}
For any $\mathbb{N}$-graded presymplectic manifold $(\mathcal{N},\Omega)$ the following  hold. 
\begin{enumerate}
    \item If $|\Omega|=n>0$, then $ \Omega=d\Theta$, where 
    $
 \Theta=\frac1n i_N\Omega\,.
    $
    \item If $V$ is a vector field of degree $m>-n$ such that $L_V\Omega=0$, then  $i_V\Omega = dH$, where 
    $
 H=-(-1)^{m}\frac n{n+m}i_V\Theta
    $\,.
    \item If $\Omega$ is non-degenerate, then 
    $
        \deg \mathcal{N}\leq |\Omega|\leq 2\deg \mathcal{N}
    $\,.
\end{enumerate}
\end{proposition}
The first two statements follow immediately from Cartan's homotopy  formula $L_N=i_Nd+di_N$. 
To prove the last statement, write the presymplectic form in terms of local coordinates as $\Omega=\frac12dw^A\Omega_{AB}(w) dw^B$, whence\footnote{It follows from the definition that $\Omega_{AB}=\Omega_{BA}(-1)^{|w^A||w^B|+|\Omega|(|w^A|+|w^B|)+1}$.}
$$|\Omega|=|w^A|+|\Omega_{AB}| +|w^B|\,.$$
In case $\deg \mathcal{N}>|\Omega|$ some of the differentials $dw^A$ do not enter  $\Omega$ and the two-form is necessarily degenerate. If now $|\Omega|>2\deg \mathcal{N}$, then 
$|\Omega_{AB}|>0$ and the supermatrix $(\Omega_{AB})$ is again non-invertible being composed of elements that are at least linear in $w$'s.  

The one-form $\Theta$ of item  $(1)$ is called a {\it presymplectic potential} for $\Omega$. It is clear that\footnote{To avoid confusion we note that the grade is defined by the Euler vector field \eqref{vecN}; it counts the total degree of $w$'s irrespective of whether they appear as variables $w^A$ or differentials $dw^A$. In other words, the Euler vector field does not take the exterior differential  $d$ into account. } $|\Omega|=|\Theta|$ and the equation \begin{equation}\label{dt}
    \Omega=d\Theta
\end{equation} defines $\Theta$ up to an exact one-form $d\Phi$, e.g. one can always take $\Theta=\tfrac1n i_N\Omega$.

In the following we are interested in presymplectic $Q$-manifolds $(\mathcal{N}, \Omega, Q)$ that satisfy the additional condition $L_Q\Omega=0$.  According to item $(2)$ of the proposition this means that the homological vector field $Q$ is Hamiltonian, i.e.,
\begin{equation}\label{H}
    i_Q\Omega =d\mathcal{H}\,,
\end{equation}
$\mathcal{H}$ being the Hamiltonian. Since $(L_Q)^2=L_{Q^2}=0$, one may also say that $\Omega$ is a cocycle of the differential $L_Q: \Lambda^2(\mathcal{N})\rightarrow \Lambda^2(\mathcal{N})$ increasing the grade of a two-form by one unit. As one might suspect, only non-trivial $Q$-cocycles will be of interest to us below.

Consider now the $\mathbb{N}$-graded presymplectic $Q$-manifold $(\mathcal{N}, Q, \Omega)$  as the target  space of a formal dynamical system (\ref{dWQ}), whose source
manifold $M$ fulfills the only condition \begin{equation}\label{dimM}\dim M=|\Omega|+1\,.\end{equation} Given these data, the AKSZ Lagrangian for the form fields $\mw^A$ reads\footnote{We emphasize that $\mw^A\equiv\mw^A(x, dx)$ are fields, while $w^A$ are just target space coordinates. Correspondingly, $d\mw^A$ is a space-time form of degree $|w^A|+1$ and should not be confused with the 1-form $dw^A$ of grade $|w^A|$ on the target space. The grade (as opposite to degree) of all fields $d\mw^A$ is $1$. Since $d\mw^A$ and $dw^A$ have different grade/degree, the position of various variables is important in the formulas below. }
\begin{equation}\label{LL}
    \mathcal{L}=\Theta_A(\mw) d\mw^A-\mathcal{H}(\mw)\,,
\end{equation}
where $\Theta_A(\mw)$ and $\mathcal{H}(\mw)$ are exterior polynomials in $\mw$'s defined by Eqs. (\ref{dt}) and (\ref{H}). By definition, $\mathcal{L}$ is a form of top degree on $M$.  With the help of (\ref{H}) one can readily bring the variation of $\mathcal{L}$ into the form 
\begin{equation}\label{VL}
    \delta\mathcal{L}=\delta \mw^A\Omega_{AB} ( d\mw^B-Q^B)-d(\Theta_A \delta \mw^A)\,.
\end{equation}
As usual the first term defines the Euler--Lagrange (EL) equations 
\begin{equation}\label{EL}
    E_A\equiv \Omega_{AB}\big( d\mw^B-Q^B(\mw)\big)=0\,,
\end{equation}
while the second term specifies possible boundary conditions. 

If the two-form $\Omega$ happens to be  symplectic, then its matrix $(\Omega_{AB})$ is invertible and the EL equations (\ref{EL}) are fully equivalent to the field equations (\ref{dWQ}). 
In this case, $(\Omega_{AB})$ plays the role of the so-called integrating multiplier in  the inverse problem of variational calculus. 
Moreover, passing to  global  Darboux coordinates on $(\mathcal{N},\Omega)$ considerably simplifies the `kinetic term' in (\ref{LL}), bringing the Lagrangian into the form 
\begin{equation}\label{AKSZ}
    \mathcal{L}=\frac12\Omega_{AB}\mw^Ad\mw^B-\mathcal{H}(\mw)\,,\qquad \Omega_{AB}\in \mathbb{R}\,.
\end{equation}

\begin{remark} The Lagrangians of the form (\ref{AKSZ}) are called AKSZ (sigma-)models after Alexandrov, Kontsevich, Schwartz, and Zaboronsky \cite{Alexandrov:1995kv}, who proposed them in the mid-1990s. 
A good deal of topological field theories can be formulated or re-formulated within the AKSZ approach for an appropriate quartet  $(M,\mathcal{N}, Q, \Omega)$, see \cite{Grigoriev:1999qz, Roytenberg:2006qz, Cattaneo2001OnTA, Ikeda:2012pv, FIORENZA_2012} for reviews and developments. There is also a deep relationship between the AKSZ construction of topological field theories and the Batalin--Vilkovisky (BV) formalism of gauge theories \cite[Ch.~17]{henneaux1994quantization}. This manifests itself in a simple and elegant form of the BV master action associated with the gauge invariant Lagrangian (\ref{AKSZ}).   
In order to construct the corresponding BV action  one simply allows the fields $\mw: T[1]M\rightarrow \mathcal{N}$  to be the maps of {\it arbitrary} $\mathbb{Z}$-degree  and this yields automatically the right spectrum of auxiliary (anti-)fields of the BV formalism.   When evaluated on such promoted  form fields $\mw$ and integrated over $M$, the Lagrangian (\ref{AKSZ}) gives the desired BV master action. 
\end{remark}

\begin{remark}
Contrary to the symplectic case, the general presymplectic AKSZ models have received  much less attention in the literature. Here we should mention the three papers \cite{Alkalaev_2014, grigoriev2016presymplectic, Grigoriev:2020xec}, where the presymplectic AKSZ approach to gauge theories was put forward together with the analysis of some models. In the first paper, the authors discuss the presymplectic AKSZ models from the perspective  of {\it frame-like formulations} known for many gauge theories, including higher spin fields. The second paper introduces the AKSZ formalism into  the {geometric theory of PDEs} through the notion of an {\it intrinsic Lagrangian}.  It was also demonstrated that a large class of Lagrangian gauge theories admits a presymplectic ASKZ action. The papers \cite{grigoriev2016presymplectic, Grigoriev:2020xec}
focus on the presymplectic AKSZ formulation for Einstein's gravity and \cite{Grigoriev:2020xec} adds a BV interpretation. 
We also review the last model and the scalar field in Examples \ref{PG} and \ref{ex:scalar} below. 
\end{remark} 

Turning back to the general case, we need to examine the equality
\begin{equation}\label{EE}
    E_A=\Omega_{AB}E^B
\end{equation}
relating  Eqs. (\ref{dWQ}) and  (\ref{EL}).   Rather than discuss this in full generality, let us only highlight some key points.
First of all, any solution to the equations $E^B = 0$ obviously satisfies  $E_A = 0$ whatever the matrix $(\Omega_{AB})$. Furthermore, if the matrix is degenerate it seems reasonable to declare the EL equations to be `weaker' than the original ones.  This, however, is not always the case. The crux of the matter is hidden integrability conditions that must be allowed for. In order to make them explicit, suppose that all null-vectors of the matrix $\big(\Omega_{AB}(\mw)\big)$ come from the kernel  distribution of the presymplectic structure $\Omega$. Let us further assume that the distribution $\ker \Omega $ is spanned by a set of vector fields $K_a$ on $\mathcal{N}$.  The integrability of $\ker\Omega$ implies the commutation relations
 \begin{equation}\label{XX}
     [K_a,K_b]=f_{ab}^cK_c
 \end{equation}
 for some structure functions $f$'s. Besides, it follows from the identity $L_Qi_{X_a}\Omega=0$ that the kernel distribution
 is $Q$-invariant, i.e., 
 \begin{equation}\label{QX}
     [Q,K_a]=U_a^bK_b
 \end{equation}
 for some functions $U$'s. Now Eq. (\ref{EL}) says that the forms $E^A$ determining the l.h.s. of Eq. (\ref{dWQ}) constitute a null-vector  of the matrix $\big(\Omega_{AB}(\mw)\big)$. Under the assumptions above, this amounts to the equality 
 \begin{equation}\label{Ext}
    d\mw^A-Q^A(\mw)=\lambda^a K^A_a(\mw)\,.
\end{equation}
Here  $\lambda^a$ are new form fields of appropriate degrees and identification (\ref{id}) is implied.   Clearly, the last equations are completely  equivalent to the EL equations (\ref{EL}). Although the new fields $\lambda^a$ enter the equations in a pure algebraic way, their dynamics are not entirely arbitrary.  Indeed, applying the de Rham differential to both sides of (\ref{Ext}) and using Rels. (\ref{XX}) and (\ref{QX}), we find
\begin{equation}\label{dlX}
(d\lambda^a+U^a_b\lambda^b+f^a_{bc}\lambda^b\lambda^c)K_a=0\,.
\end{equation}
If {\it all} the vector fields $K_a$, being  linearly independent, are contained in the r.h.s. of  Eq. (\ref{Ext}), then the last condition amounts to 
\begin{equation}\label{dl}
d\lambda^a+U^a_b\lambda^b+f^a_{bc}\lambda^b\lambda^c =0\,.
\end{equation}
By construction, Eqs. (\ref{Ext}) and (\ref{dl}) are compatible to each other and define a new formal dynamical system extending  (\ref{dWQ}).  The question now is whether the extended system is dynamically equivalent to the original one. If the total number of $\lambda$'s is finite, then the number of physical 
degrees of freedom they can bring in is finite as well, see Remark \ref{R24}.  In case $\dim M>1$, these are global modes associated to the boundary conditions and/or topology of $M$.  Hence, for genuine field theories the formal dynamical systems in question are essentially equivalent to each other (i.e., equivalent up to a finite ambiguity). Another typical situation is when $M\simeq \mathbb{R}^d$, $d>1$, and all the form fields $\lambda^a$ are of strictly positive degree.  Being pure gauge and having no topological modes, the fields $\lambda^a$ can safely be set to zero. In this gauge, the equivalence of the formal dynamical systems at hand is again obvious. 

The analysis of the general case is complicated by two points. First, it may happen that {\it every} generating set  $\{K_a\}$ for $\ker \Omega$ consists of linearly dependent vectors. This implies the existence of left null-vectors $Z_\alpha$ for the rectangular matrix $\{K_a^A\}$, so that  $Z_\alpha^aK_a^A=0$. If the system of null-vectors $\{Z_\alpha\}$ is complete, then we can still divide the l.h.s. of Eq.~(\ref{dlX}) by the $K_a$'s at the expense of introducing new form fields $\xi^\alpha$ and adding the term $\xi^\alpha Z^a_\alpha$ to the r.h.s. of Eq.~(\ref{dl}).  
The verification of formal integrability gives further equations for $\xi^\alpha$ and so on.\footnote{The problem we have to deal with here is quite similar to that of reducible gauge symmetries, see e.g. \cite[Ch.~10]{henneaux1994quantization}. From the algebraic standpoint, each extension defines and is defined by a certain free resolution of the $C^\infty(\mathcal{N})$-module  $\ker\Omega$.} The second point concerns the case where {\it not all} the null-vectors $K_a$ are actually present in Eq. (\ref{Ext}). Here, besides the differential equations, one can find some algebraic constraints on  $\lambda^a$ associated with those $K_a$'s that dropped out of the r.h.s. of Eq. (\ref{Ext}) on account of degree. (Both the points are exemplified below.) 
Under reasonable assumptions  one may repeat this construction once and  again.  The result is a formal dynamical system equivalent to the EL equations (\ref{EL}). All the above comments on  equivalence to the original equations $E^A=0$ are still valid for the fully extended system. We are going to detail this construction elsewhere. 

The last comment concerns gauge symmetries. One should realize that  not every gauge transformation (\ref{GS}) of the  equations  $E^A=0$
can be promoted to a symmetry of the corresponding AKSZ  model (\ref{LL}), unless $\Omega$ is non-degenerate.    Some necessary and sufficient condition for this to happen are 
discussed in \cite{Alkalaev_2014}. In particular, the general coordinate transformations (\ref{diff}) may no longer  be part of (\ref{GS}). On the other hand, the null-vectors of the presymplectic structure $\Omega$ give rise to extra gauge symmetries 
of the form $\delta_\varepsilon\mw^A=\varepsilon^aK_a^A$ provided that $|K^A_a|\leq |\mw^A|$.

\begin{example}[Pure gravity]\label{PG}
In order to construct a first-order formulation of $4d$ gravity with negative cosmological constant, we
can start with the Lie algebra $\mathcal{G}=so(3,2)$. As explained in Example \ref{E22}, it gives rise to the homological vector field 
\begin{equation}\label{QGR}
    Q_{GR}=\omega^a{}_c \omega^{cb} \frac{\partial}{\partial \omega^{ab}}+\omega^a{}_b e^b\frac{\partial}{\partial e^a}+\lambda e^ae^b\frac{\partial}{\partial \omega^{ab}}
\end{equation}
on the graded manifold $\mathcal{G}[1]$ with global coordinates $e^a$ and $\omega^{ab}$. Here the degree-one coordinates $\omega^{ab}=-\omega^{ba}$ correspond
to the generators of the Lorentz subalgebra $so(3,1)\subset so(3,2)$, while $e^a$  are associated with the $AdS_4$ transvections. As usual we raise and lower  the Lorentz indices $a,b,...=0,1,2,3$ with the help of Minkowski metric $\eta_{ab}$. Finally, the parameter $\lambda<0$ is proportional to the cosmological constant.\footnote{As $\lambda\rightarrow 0$ the algebra  $so(3,2)$ contracts to the Poincar\'e algebra $iso(3,1)$. Equally well we could start with the de Sitter algebra $so(4,1)$ taking $\lambda>0$.}  
A relevant $Q$-invariant presymplectic form $\Omega_{GR}$ on $\mathcal{G}[1]$ reads 
\begin{equation}\label{Om}
    \Omega_{GR}=\epsilon_{abcd} e^ade^bd\omega^{cd}\,.
\end{equation}
Notice that it does not depend on $\lambda$. The $2$-form $\Omega_{GR}$, being obviously closed, turns out to be  degenerate.
In degree one and two, the kernel distribution is spanned by the family of null-vectors
\begin{equation}\label{KK}
    K^1_C=C_{ab,cd}e^ae^b\frac{\partial}{\partial \omega_{cd}}\,, \qquad K^2_H=H_{abc,d}e^ae^be^c\frac{\partial}{\partial e_d}\,,
\end{equation}
where the constant parameters $C_{ab,cd}$ and $H_{abc,d}$ form {\it traceless} Lorentz tensors.

In accordance with the general philosophy, the  target space coordinates $e^a$ and $\omega^{ab}$ are promoted to the one-form fields $\me^a=\me^a_\mu dx^\mu$ and ${\momega}^{ab}={\momega}^{ab}_\mu dx^\mu$ associated, respectively, with the vierbein and the spin-connection on $M$.  Applying now the general formulas of Proposition \ref{P41}, we can write the following  AKSZ Lagrangian:
\begin{equation}\label{LGR}
    \mathcal{L}_{GR}= \frac{1}{2}\epsilon_{abcd}\me^c\me^d \Big(d\momega^{ab} -\momega^a{}_n \momega^{nb} -\frac\lambda2 \me^a \me^b\Big)\,.
\end{equation}
Varying it w.r.t. the spin connection and vierbein, we learn that 
\besubeqs\label{EinsteinEq}
\begin{align}
    \epsilon_{abcd} \me^cD\me^d&=0\,,\label{de}\\ \epsilon_{abcd}\me^b(R^{cd}-\lambda \me^c\me^d)&=0\label{R}\,.
\end{align}
\esubeqs
Here $D\me^a=d\me^a-\momega^{a}{}_b\me^b$ is the Lorentz-covariant differential of the vierbein  and $R^{ab}=d\momega^{ab}-\momega^a{}_c\momega^{cb}$ 
is the curvature tensor of $D$. As the presymplectic form (\ref{Om}) has no null-vectors of degree $1$ `in the direction of $e$', the first equation (\ref{de}) is actually equivalent to 
\begin{equation}\label{De0}
    D\me^a=0\,.
\end{equation}
It says that the spin connection  $\momega^{ab}$ is torsion free and can be uniquely expressed  via $\me^a$ and $d\me^a$.     As to the second equation (\ref{R}), being Lorentz and general covariant and depending on second derivatives of the vierbein through $\momega(\me, d\me)$, it may  be nothing  else but the vacuum  Einstein's equation with cosmological constant. Thus, we are lead to conclude that the first-order formulation  (\ref{LGR}) is 
fully equivalent to that relied on the Einstein--Hilbert action.  
But wait a minute, not so fast! What about the kernel of the presymplectic form? 

If the presymplectic form (\ref{Om}) were non-degenerate, we would get just the structure relations of anti-de Sitter geometry, 
\begin{equation}\label{DeR}
    D\me^a=0\,,\qquad R^{ab}=\lambda \me^a\me^b\,,
\end{equation}
encoded in the homological vector field (\ref{QGR}).  Clearly, this leaves no room for local degrees of freedom. The explicit expressions for the null-vectors (\ref{KK}) suggest that Eq. (\ref{R}) is actually equivalent to the following one:
\begin{equation}\label{RC}
    R_{ab}-\lambda \me_a \me_b=C_{ab,cd}\me^c\me^d\,,
\end{equation}
where $C_{ab,cd}$ is now a collection of zero-form fields. Considered as a Lorentz tensor, $C_{ab,cd}$  is anti-symmetric in the first and second pairs of indices and has zero trace.  
Checking the formal integrability of the full system \eqref{De0}, \eqref{RC}, as explained above,  we get both algebraic and differential constraints on $C$'s. The former are given by $\me^a \me^b \me^cC_{ab,cd}=0$ and force the tensor $C_{ab,cd}$ to have the symmetry of  `window' Young diagram. Then Eq. (\ref{RC}) identifies $C_{ab,cd}$ as the Weyl tensor. As for the differential constraints, they have the standard form  $dC_{ab,cd}=\ldots $, where the r.h.s. involves new zero-form fields $C_{ab,cd,e}$ with the symmetry of a two-row Young diagram. These new fields owe their existence to the over-completeness of the null-vector system (\ref{KK}): as $C_{ab,cd,e}$ cannot have three anti-symmetric indices,  the shift $C_{ab,cd}\rightarrow C_{ab,cd}+ C_{ab,cd,n}e^n$ does not affect  $K^1_C$. This gives the null-vectors $Z$'s of the second generation and so on. The above extension procedure  never stops generating an infinite number of zero-form fields $C_{ab,cd,\cdots}$. This results in a formal dynamical system of the Einstein gravity (with infinitely many zero-forms, as promised for a field theory), whose explicit form is not known, but can, in principle, be obtained with the help of \cite{Barnich:2004cr,Barnich:2010sw}.  
\end{example}

\begin{example}\label{ex:scalar}(Gravity + scalar field). The above AKSZ formulation of pure gravity can easily be upgraded to include interaction with  a  scalar field. To illustrate the idea of how one can extend the construction based on an algebra with some of its modules we consider the simplest finite-dimensional module. Specifically, we  extend the underlying Lie algebra $so(3,2)$ to the inhomogeneous anti-de Sitter algebra $iso(3,2)$ and prescribe the abelian ideal of the latter degree one.\footnote{This can also be regarded as a trivial extension of $so(3,2)$ by its fundamental representation.} This results in the graded Lie algebra $\mathcal{G}=\mathcal{G}_0\oplus \mathcal{G}_1$, where $\mathcal{G}_0=so(3,2)$ and $\mathcal{G}_1\simeq \mathbb{R}^{3,2}$ (as vector spaces). The associated homological vector field on $\mathcal{G}[1]$ takes  the form 
\begin{equation}
    Q=Q_{GR}-Q_{S}\,,\qquad Q_{S}=\pi_a e^a\frac{\partial }{\partial\varphi}+(\pi_b\omega^{ba}+\lambda \varphi e^a)\frac{\partial}{\partial \pi^a}\,.
\end{equation}
Here $(\varphi,\pi^a)$ are new coordinates of grade zero associated with the ideal $\mathcal{G}_1$. In order to be $Q$-invariant the presymplectic form  (\ref{Om}) is extended to
\begin{equation}
    \Omega=\Omega_{GR}+d\Theta_S\,,\qquad \Theta_S=\epsilon_{abcd}e^ae^be^c\pi^dd\varphi\,.
\end{equation}
The kernel of the form $\Omega$ is generated by the vector fields (\ref{KK}) together with 
\begin{equation}
   \tilde K^1_C=C_{ab}e^a\frac{\partial}{\partial \pi_b}\,,\qquad \tilde K^2_H=H_{ab}e^ae^b\frac{\partial}{\partial\varphi}\,.
\end{equation}
Here the constant parameters $H_{ab}$ and $C_{ab}$ form, respectively, anti-symmetric and traceless Lorentz tensors. The corresponding  AKSZ Lagrangian now reads
\begin{equation}\label{GRS}
    \mathcal{L}=\mathcal{L}_{GR}+ \epsilon_{abcd}\me^a\me^b\me^c \Big(\mpi^d d\mvarphi+ \big(\frac18 \mpi^a\mpi_a+\frac{\lambda}2\mvarphi^2\big)\me^d \Big)\,.
\end{equation}
Eliminating the auxiliary fields $\mpi^a$ with the help of their equations of motion, one obtains the standard second-order formulation for the scalar $\mvarphi$ coupled to Einstein's gravity with cosmological constant. The mass of $\mvarphi$, however, is below the unitarity bound unless $\lambda\neq 0$.  By definition, the space $\mathcal{G}_1$ corresponds to the vector representation of $so(3,2)$. Instead one can take any finite-dimensional module of $so(3,2)$ associated with  the totally-symmetric traceless tensors of definite rank  and get other discrete values of masses below unitarity. In order to let the mass be arbitrary one has to start with an infinite-dimensional $so(3,2)$-module, see e.g. \cite{Shaynkman:2000ts} and \cite{Alkalaev_2014} for the presymplectic treatment.
\end{example}

\section{Covariant phase space of a presymplectic AKSZ model}
\label{sec:CPhS}

A glance at the AKSZ Lagrangian (\ref{LL}) is enough to observe its similarity with the least action principle in  Hamiltonian mechanics and this is more than just an analogy. Actually, the target-space presymplectic form $\Omega$ induces a presymplectic structure on the covariant phase space of fields, which then  
endows the algebra of physical observables with a Poisson bracket.  
We will not dwell here on the covariant Hamiltonian formalism in field theory referring the reader to the papers \cite{1987thyg.book..676C, Zuckerman:1989cx, Khavkine2013PresymplecticCA, Sharapov:2016qne} for general discussions and examples. 
The basic idea is to treat the boundary term in  (\ref{VL}) as defining a `functional one-form' on the space of fields: 
\begin{equation}\label{hTh}
\hat \Theta=\int_{\Sigma} \Theta_{A}(\mw)\delta \mw^A\,.
\end{equation}
Here $\Sigma\subset M$ is an arbitrary Cauchy surface of initial data for the equations of motion.   Treating now the variation symbol
$\delta$ as an exterior differential on the functional space of fields and applying it to (\ref{hTh}), we get the two-form 
\begin{equation}\label{int}
\hat \Omega=\delta\hat \Theta=\frac12\int_{\Sigma} \delta \mw^A\Omega_{AB}(\mw)\delta \mw^B\,,
\end{equation}
which is just a functional counterpart of the presymplectic form $\Omega$ on  $\mathcal{N}$. By construction, the form $\hat\Omega$ defines a presymplectic structure on the functional space of all field configurations and, through restriction, on every subspace therein. Let  $\mathrm{Sol}(E^A)$ and $\mathrm{Sol}(E_A)$ denote the subspaces of all solutions to Eqs. (\ref{dWQ}) and (\ref{EL}).  In view of Rel. (\ref{EE}) we have the natural embedding $\mathrm{Sol}(E^A)\subset\mathrm{Sol}(E_A)$. Either space is then identified with the {\it covariant phase space} of  the corresponding field theory. 
 It is not hard to see that the restriction of the presymplectic form (\ref{int}) onto $\mathrm{Sol}(E_A)$ (and hence, on $\mathrm{Sol}(E^A)$) does not depend on the choice of a Cauchy surface $\Sigma$; this justifies  the adjective ``covariant'' in the name.  
 
In general, the induced presymplectic structures on the solution spaces above are degenerate even if the original presymplectic form  (\ref{int}) is not. This is due to the gauge symmetries of the equations of motion. Let us check, for example, that the infinitesimal gauge transformations (\ref{GS}) belong to the kernel of the form $\hat\Omega$ restricted to $\mathrm{Sol}(E^A)$. We have 
\begin{equation}
\begin{array}{rcl}
   i_{\delta_\varepsilon \mw}\hat\Omega&=& \displaystyle \int_\Sigma\delta_\varepsilon  \mw^A\Omega_{AB}(\mw)\delta \mw^B=\int_\Sigma \big(d\varepsilon^A+ \varepsilon^C\partial_C Q^A(\mw)\big)\Omega_{AB}(\mw)\delta\mw^B\\[5mm]
   &= &\displaystyle \int_\Sigma (-1)^{|w^A|}\varepsilon^Ad\big(\Omega_{AB}\delta\mw^B\big)+\varepsilon^C\partial_C Q^A\Omega_{AB}\delta\mw^B \\[5mm]
   &\approx& \displaystyle -\int_\Sigma \varepsilon^A(L_Q\Omega)_{AB}\delta\mw^B =0\,.
    \end{array}
\end{equation}
Our convention here is that
$$
\delta d=-d\delta ,\quad \delta\mw^A \mw^B=(-1)^{(|w^A|+1)|w^B|}\mw^B\delta\mw^A,\quad \delta\mw^A d\mw^B=(-1)^{(|w^A|+1)(|w^B|+1)}d\mw^B\delta\mw^A.
$$
It is fit to and motivated by the concept of variational bi-complex, see e.g. \cite{Olver, BigDick, Sharapov2016VariationalTG}.

Following the general recipe, the space of Hamiltonian function(al)s 
 $\hat{H}=\int_\Sigma H(\mw)$, $\mw\in \mathrm{Sol}(E^A)$, is then endowed with the  covariant Poisson bracket (\ref{HF}), which can be viewed as a precursor for the 
canonical quantization of the theory. 

Another important remark is that the correspondence between the presymplectic structure $\Omega$ on the target space $\mathcal{N}$ and its functional counterpart $\hat\Omega$ restricted to  either solution space is far from being one-to-one.  Indeed, shifting the integrand in (\ref{int}) by an on-shell exact form  $\chi_{AB}\delta\mw^A\delta\mw^B\approx d\Psi(\mw)$ does not affect the induced presymplectic structure (Stokes theorem). 
In the case of  $\mathrm{Sol}(E^A)$ this point can further be refined.  The equations of motion $\mw_\ast(\mathrm{d})=Q$ tell us that the action of the de Rham differential is on-shell equivalent to the action of the homological vector field. As a result  $Q$-exact presymplectic forms  $L_Q\Psi$ on the target space $\mathcal{N}$ pass to the $d$-exact forms on the covariant phase space $\mathrm{Sol}(E^A)$. This motivates the following definition: two presymplectic forms $\Omega$ and $\Omega'$ on $\mathcal{N}$ are said to be {\it equivalent} if 
$\Omega-\Omega'=L_Q\Psi$
for some two-form $\Psi$; {correspondingly, the presymplectic structures of the form $\Omega=L_Q\Psi$ are considered trivial}. Equivalent presymplectic structures on $\mathcal{N}$ give rise to the same presymplectic structure on the covariant phase space $\mathrm{Sol}(E^A)$. 
In other words,  the space of non-trivial presymplectic structures for the formal dynamical system (\ref{dWQ}) is identified with the cohomology group $H^{\dim M-1}(L_Q,\Lambda^2(\mathcal{N}))$. It might be well to point out that a non-trivial presymplectic structure on $\mathrm{Sol}(E_A)$ may 
become trivial upon  restriction to the subspace $\mathrm{Sol}(E^A)$. The last point is best exemplified by Einstein's gravity with cosmological constant. 

\begin{example}Proceeding with Example \ref{PG}, we note that the presymplectic structure (\ref{Om}) is $Q$-exact whenever $\lambda\neq 0$: 
\begin{equation}\label{OLP}
    \Omega_{GR}=L_{Q_{GR}}\Psi\,,\qquad \Psi =\frac1{4\lambda}\epsilon_{abcd}d\omega^{ab}d\omega^{cd}\,.
\end{equation}
This is no surprise as the Lie algebra $\mathcal{G}$ underlying the homological vector field (\ref{QGR}) is simple.   
It is easy to realize that  $Q$-invariant presymplectic structures $\Omega$ of grade $3$ correspond to one-cocycles of $\mathcal{G}$ with coefficients in $S^2\mathcal{G}^\ast$, the symmetrized tensor square of the coadjoint representation. But by Whitehead's first lemma the first cohomology of any simple  Lie algebra vanishes.  
The $Q$-exactness of the  form (\ref{OLP}) by no means implies that the induced presymplectic structure  on the solution space of Einstein's equations \eqref{EinsteinEq} is trivial. It is true, however, that the further restriction of $\hat\Omega$ to the subspace of locally (anti-)de Sitter geometries described by Eq. (\ref{DeR}) does lead to the trivial presymplectic structure. This is in contrast with the case $\lambda=0$ when $\mathcal{G}$
degenerates to the Poincar\'e algebra $iso(3,1)$. For the zero cosmological constant the presymplectic form (\ref{Om}) is a non-trivial $Q$-cocycle. (The Poincar\'e algebra being not simple, the first Whitehead's lemma is not applicable anymore.) The moduli spaces of flat geometries are finite-dimensional and depend on the topology of the space-time manifold $M$, see  e.g. \cite{Witten1988Nu} for a physicist-oriented discussion.  Using  $\hat{\Omega}$, one can equip them with presumably non-trivial symplectic  structures.  
\end{example}

\section{Presymplectic AKSZ model for \texorpdfstring{$\boldsymbol{4d}$}{4d} HSGRA }\label{sec:PSHSGRA}
In this section, we put $4d$ HSGRA into the framework of presymplectic AKSZ models and thereby provide its `weakly' Lagrangian description.
All we need is a presymplectic structure $\Omega$ compatible with the homological vector field $Q$ defined by the right-hand sides of the field equations \eqref{dwdc}. We also expect the corresponding Euler--Lagrange equations (\ref{EL}) to be essentially equivalent to the original equations of $4d$ HSGRA and this rules out some trivial choices like $\Omega=0$. Condition (\ref{dimM}) implies that $|\Omega|=3$. Once an appropriate presymplectic $2$-form $\Omega$ is found, the corresponding presymplectic potential $\Theta=\frac13i_N\Omega$ defines immediately the `kinetic'
term of an AKSZ Lagrangian (\ref{LL}) in question as well as the Hamiltonian $\mathcal{H}=\frac34i_Q\Theta$.

The form of the equations of motion \eqref{dwdc} suggests to look for a compatible presymplectic structure as an expansion in powers of $\mC$'s. In order to make contact with the general notation of Sec. \ref{sec:pre-AKSZ} it is convenient to endow the target space $\mathcal{N}=\mathcal{G}[1]$ of fields $\momega$ and $\mC$ with the global coordinate system  $w^A=(\omega^a,C^a)$ w.r.t. some basis $e_A=(e_a^{_{{}^{{}_{-1}}}}, e_a^{_{{}^{0}}})$ in the graded Lie algebra $\mathcal{G}=\mathcal{G}_{-1}\oplus \mathcal{G}_0$ underlying the free theory. To control formal  powers series  $C^a$ we also introduce 
the vector field  
\begin{equation}
N_C=C^a\frac{\partial}{\partial C^a}\,,
\end{equation}
which prescribes degree one to $C$'s and degree zero to $\omega$'s. Now the r.h.s. of Eqs. \eqref{dwdc} come from a homological vector field on $\mathcal{N}$ of the form 
\begin{equation}\label{Qn}
    Q=Q_0+Q_1+\cdots\,, \qquad\qquad [N_C, Q_n]=nQ_n\,,
\end{equation}
where 
\begin{equation}
 Q_n= C^{a_1}\cdots C^{a_n} Q^a_{a_1\cdots a_n bc}\omega^{b}\omega^c\frac{\partial}{\partial \omega^a}+C^{a_0}\cdots C^{a_n} Q^a_{a_0a_1\cdots a_n b}\omega^{b}\frac{\partial}{\partial C^a}\,.
\end{equation}
The leading term $Q_0$, being a homological vector field by itself, determines the graded Lie algebra $\mathcal{G}$ as explained in Example \ref{E22}. 
A similar expansion for the thought-for presymplectic structure can be written as 
\begin{equation}\label{O-exp}
    \Omega=\Omega_m+\Omega_{m+1}+\cdots \,,\qquad \qquad L_{N_C}\Omega_n=n\Omega_n\,,
\end{equation}
\begin{equation}\label{On}
\begin{array}{rcl}
    \Omega_n&=&C^{a_1}\cdots C^{a_n}\Omega^{^{(0)}}_{a_1\cdots a_n abc}\,\omega^a\,d\omega^b \,d\omega^c\\[3mm]
    &+&C^{a_1}\cdots C^{a_{n-1}}\Omega^{^{(1)}}_{a_1\cdots a_n abc}\,\omega^a\,\omega^b \,d\omega^c\,dC^{a_n}\\[3mm]
     &+&C^{a_1}\cdots C^{a_{n-2}}\Omega^{^{(2)}}_{a_1\cdots a_n abc}\,\omega^a\,\omega^b \,\omega^c\,dC^{a_{n-1}}\,dC^{a_{n-2}}\,,
    \end{array}
\end{equation}
where we suppose that $\Omega_m\neq 0$. 
The defining conditions of a presymplectic structure 
\begin{equation}\label{dLQ}
    d\Omega =0\,,\qquad \qquad L_{Q}\Omega=0\,,
\end{equation}
impose an infinite set of linear relations on the structure constants  $\Omega^{^{(0)}}$, $\Omega^{^{(1)}}$, and $\Omega^{^{(2)}}$.
In particular, on substituting expansions (\ref{Qn}) and (\ref{O-exp}) into the condition of $Q$-invariance (\ref{dLQ}), we obtain the chain of equations 
\begin{align}
&&&L_{Q_0} \Omega_m=0\,,\label{dO}\\
   && &L_{Q_0}\Omega_{n}=B_n(\Omega_m,\ldots,\Omega_{n-1})\,,&&B_n\equiv \sum_{k=1}^{m+n} L_{Q_k}\Omega_{n-k}\,, && n=m+1,m+2,  \ldots\label{dOO}
    \end{align}
Since $(L_{Q_0})^2=0$, we are lead to the standard problem of homological perturbation theory. The first equation (\ref{dO})
tells us that  the leading term of the expansion  (\ref{O-exp}) is a $Q_0$-cocycle.  As explained in Sec. \ref{sec:CPhS}, we are interested in non-trivial $Q_0$-cocycles, which then induce non-zero presymplectic structures on the solution space to the field eqiations. Given such a non-trivial cocycle $\Omega_m$,
the remaining equations (\ref{dOO}) are solved one after the other  provided  that no cohomological obstacles arise.  Arguing by induction, one can see that the r.h.s. of the $n$-th equation, $B_n$, is $Q_0$-closed whenever all the previous equations for $\Omega_0,\ldots,\Omega_{n-1}$ are satisfied. Therefore the $2$-form  $B_n$ defines a class of $Q_0$-cohomology, whose vanishing provides the necessary and sufficient condition for solvability of the $n$-th equation (\ref{dOO}).  Notice that Eq. (\ref{dOO}) defines $\Omega_{n}$, if it exists,
up to adding  to it an arbitrary  $Q_0$-cocycle $\Psi_{n}$ (perhaps trivial). If the equivalence class $\Omega_{n}+\Psi_{n}$ contains a $d$-closed representative, then one can take it to extend the sought-for solution one step further.   

On the other hand, the cohomology of the coboundary operator $L_{Q_0}$ coincides with that of the graded Lie algebra $\mathcal{G}=gl(\mathfrak{A})$ as discussed in Example \ref{E22}. In particular,  $\Omega_m$ is determined by an element of $H^\bullet_3(\mathcal{G},S^2\mathcal{G}^\ast)$ and $B_n$ defines a cohomology class of $H^\bullet_{4}(\mathcal{G}, S^2\mathcal{G}^\ast)$, where the subscripts refer to grades. The existence of a non-trivial presymplectic structure thus implies $H^\bullet_3(\mathcal{G},S^2\mathcal{G}^\ast)\neq 0$, while the property that $H^\bullet_4(\mathcal{G}, S^2\mathcal{G}^\ast)=0$ ensures extendibility of the leading term $\Omega_m$ to all higher orders in $C$'s. 
A detailed analysis of the relevant cohomology  is presented in Appendix \ref{app:D}. 

A short summary is that there is no non-trivial solutions to (\ref{dO}) for $m=0$; and hence, expansion (\ref{O-exp}) necessarily starts with terms involving at least one $C$ or $dC$. The solutions with $m>2$ are of little physical interest as they all vanish on the natural higher spin vacuum $\mC=0$, which covers all maximally (higher spin) symmetric backgrounds. The remaining possibilities give a $2$-parameter family of presymplectic structures and Lagrangians. This is in line with Chern--Simons matter theories, which we discuss below. 

\paragraph{Equations recap.} While the general theorems of Appendix \ref{app:D} guarantee that the presymplectic structure to be discussed below is unobstructed, we would like to work out a few terms explicitly as they are the most important ones. To this end let us write down Eq. \eqref{dwdc}  up to the first order in $\mC$ as
\besubeqs\label{uptocubic}
\begin{align}
    d\momega&= \momega\star \momega+ \mathcal{V}_\nu(\momega,\momega,\mC)+\mathcal{O}(\mC^2)\,,\\
    d\mC&= \momega\star \mC-\mC\star \momega+\mathcal{O}(\mC^2)\,.
\end{align}
\esubeqs
The simplest vertex has the factorized form \cite{Sharapov:2017yde,Sharapov:2019vyd}, see \cite{Vasiliev:1988sa} for the explicit expression,
\begin{align}\label{vertexfac}
    \mathcal{V}_\nu(\momega,\momega,\mC)&=\Phi_\nu(\momega,\momega)\star \mC\,,
\end{align}
where $\Phi_\nu$ represents the first deformation, cf. \eqref{dprod}, of the extended higher spin algebra $\mathfrak{A}$:\footnote{Thanks to the product structure  $\mathfrak{A}=\mathcal{A}\otimes \mathcal{A}$ the deformation is built from a single two-cocycle $\phi(-,-)$ of the Weyl algebra $A_1$; by definition, $a\star \phi(b,c)-\phi(a\star b,c)+\phi(a,b\star c)-\phi(a,b)\star \tilde{c}=0$, where $\tilde c(y)=c(-y)$. Then on decomposable elements $\mathbf{a}(y,\bry)=a(y)\otimes \bar a(\bry)\in \mathfrak{A}$,  we have $\phi_{10}(\mathbf{a},\mathbf{b})=\phi(a,b)\otimes \bar a \star \bar b$ and $\phi_{01}(\mathbf{a},\mathbf{b})= a \star b\otimes \phi(\bar a,\bar b)$. }
\begin{align}\label{factorizedform}
    \Phi_\nu\equiv \Phi_\nu(\momega,\momega) =\nu\phi_{10}(\momega,\momega)\kappa+\bar\nu\phi_{01}(\momega,\momega)\bar\kappa\,.
\end{align}
For manipulations below,  it is convenient to rewrite (\ref{uptocubic}) in a more compact form as 
\begin{align}\label{lHSGRA}
    \mR&=\Phi_\nu\star \mC \mod \mC^2\,,& D\mC&=0 \mod \mC\,.
\end{align}
Here
\begin{align}
    D&=d-[\momega,-]_\star\,,& D^2&=[\mR,-]_\star\,,& \mR&=d\momega -\momega\star \momega\,,
\end{align}
and we do not write the $gl$-indices of fields explicitly; it is assumed that all the matrix indices are contracted in chain as in (\ref{firstV}). The main property of the space-time $2$-form $\Phi_\nu(\momega,\momega)$, which provides the formal integrability of the differential equations (\ref{lHSGRA}) modulo $\mC^2$, is 
\begin{equation}\label{dF}
    D\Phi_\nu(\momega,\momega)\approx 0 \mod \mC\,.
\end{equation}
The formulas above will suffice for all formal manipulations. In order to make contact with the field theory approach we have to adjust the cubic vertex 
(\ref{vertexfac}).\footnote{ \label{nondiag} As it stands the vertex (\ref{vertexfac}) does not lead to the desired field equations even at the free level \cite{Vasiliev:1988sa}. What happens is that the free equations mix fields of different spins. Nevertheless, one can diagonalize the equations by performing a linear change of variables; in so doing, the non-commutativity of covariant derivatives, $[\nabla_{a},\nabla_b]\sim \lambda$, plays a crucial role. This example reveals a general fact that a formally integrable $Q$, i.e., $Q^2=0$, has to be adjusted as to make contact with field theory.} We will only need the fact that there exists a better representative $\Vcan_\nu$, one may call it `canonical', such that
\begin{align}\label{canonicalA}
    \Vcan_\nu(\stackrel{_{\circ}}{\momega},\stackrel{_{\circ}}{\momega},C)&= \nu H^{\ga\ga} \pl_\ga \pl_\gb C(y,\bry=0) + \bar\nu H^{\gad\gad} \pl_\gad \pl_\gbd C(y=0,\bry)\,.
\end{align}
Here $\stackrel{_{\circ}}{\momega}$ is the $AdS_4$ connection \eqref{AdS}, $H^{\ga\gb}\equiv h\fud{\ga}{\gdd}\wedge h^{\gb\gdd}$, {\it idem} for $H^{\gad\gbd}$, and we projected the vertex onto the physical sector $\mC=C(y,\bry)K$, see Remark \ref{R34}. The consistency of $\Vcan_\nu$ implies \begin{align}\label{consisV}
    D\Vcan_\nu(\momega, \momega, \mC)&=0 \mod \mC^2\,.
\end{align}
More generally, one can perform a field redefinition in equations \eqref{uptocubic} that maps the vertex  \eqref{vertexfac} to another representative, e.g. $\Vcan_\nu$, which always obeys \eqref{consisV}.

\paragraph{Presymplectic structure and AKSZ to LO.} With the conventions and  notation above the leading order term of the thought-for family  (\ref {O-exp}) of presymplectic structures on $\mathcal{N}=\mathcal{G}[1]$ is given by 
\begin{equation}\label{O1}
   \Omega_1=\Omega_1^{_{(0)}}+ \Omega_1^{_{(1)}}=d\Theta_1\,,\qquad \qquad  \Theta_1= \langle  C\star \Phi_\mu(\omega,\omega)\star d\omega\rangle\,.
\end{equation}
Here $\mu$ is an arbitrary complex parameter and the brackets $\langle -\rangle$ stand for the matrix trace {\it and} the trace in the $\star$-product algebra $\mathfrak{A}$. It is important that $\mu$ differs from $\nu$ in the vertex \eqref{vertexfac}. For the case of HSGRA in $d>4$ the doubling of $\phi$'s in \eqref{vertexfac} does not take place and there is a clear distinction between the cocycle participating in $\Theta$ and the cocycle that deforms the equations, which we will make a few remarks about at the end.

As is clear from the explicit form of $\Theta_1$, one can instead choose 
\begin{align}\label{thetaA}
    \Theta_1= \langle  \mathcal{V}_\mu(\omega,\omega, C)\star d\omega\rangle
\end{align}
for any representative $\mathcal{V}_\mu$ in \eqref{uptocubic} as long as $\mu\neq \nu$. The last inequality ensures non-triviality of the presymplectic structure to the leading order. Explicitly, 
\besubeqs
\begin{align}
    \Omega_1^{_{(0)}}&= \langle  \mathcal{V}_\mu(d\omega,\omega, C)\star d\omega\rangle-\langle  \mathcal{V}_\mu(\omega,d\omega, C)\star d\omega\rangle\,,\\
    \Omega_1^{_{(1)}}&=\langle  \mathcal{V}_\mu(\omega,\omega, dC)\star d\omega\rangle\,.
\end{align}
\esubeqs
Notice that  the component $\Omega_1^{_{(0)}}$ vanishes on the maximally symmetric background $\mC=0$. With the help of \eqref{canonicalA} we find for free fields in $AdS_4$:\footnote{To save letters we denote all symmetric (or to be symmetrized) indices by the same letter. }
\begin{align}\label{freePS}\begin{aligned}
    \Omega&=  \mu H^{\ga\ga} \langle\pl_\ga \pl_\ga dC(y,\bry=0)\star d\omega\rangle + \bar\mu H^{\gad\gad} \langle\pl_\gad \pl_\gad C(y=0,\bry)\star d\omega\rangle=\\[3mm]
    &= \sum_{s=1} \frac{\mu}{(2s-2)!} H^{\ga\ga} dC_{\ga(2s)} d\omega^{\ga(2s-2)} +\frac{\bar\mu}{(2s-2)!} H^{\gad\gad} dC_{\gad(2s)} d\omega^{\gad(2s-2)}\,.
    \end{aligned}
\end{align}
This is an admissible presymplectic structure \cite{Sharapov:2016qne}. For free fields, the parameter $\nu$ in \eqref{canonicalA} does not play any role and can be eliminated by rotating (anti-)holomorphic Weyl tensors ($C_{\gad(2s)}$) $C_{\ga(2s)}$ by the phase of $\nu$. It is also clear that for $\mu=\nu$ the presymplectic structure \eqref{freePS}, being proportional to the r.h.s. of \eqref{canonicalA}, is trivial.  

It is especially interesting to see how the presymplectic AKSZ action look like to the leading order. The equations to this order
\begin{align}\label{veryfree}
    d\momega&= \momega\star \momega\,, &
    d\mC&= \momega\star \mC-\mC\star \momega\,,
\end{align}
describe free fields $\mC$ propagating over a maximally symmetric higher spin background $\momega$. When linearized over $AdS_4$ the $\mC$-equation reduces to a collection of the Bargmann--Wigner equations $\nabla\fdu{\ga}{\gbd}C_{\gad(2s-1)\gbd}=0$, $\nabla\fud{\gb}{\gad}C_{\ga(2s-1)\gb}=0$ for fields of all spins. It is clear that equations \eqref{veryfree} are non-Lagrangian. For $d=3$, the first one  could be obtained from the Chern--Simons action; the second one, however, requires a two-form Lagrange multiplier, say $\boldsymbol{B}$, to write the Lagrangian $\mathcal{L}=\langle\boldsymbol{B}\star(d\mC-\momega\star \mC+\mC\star \momega)\rangle$. In four dimensions, even the first equation does not come from an action. Therefore, the main function of $\Theta_1$ is to provide a presymplectic treatment of \eqref{veryfree}. The corresponding presymplectic AKSZ Lagrangian reads
\begin{align}\label{freeaction}\boxed{\rule[-8pt]{0pt}{24pt}
    \mathcal{L}_1=\big\langle  \mathcal{V}_\mu(\momega,\momega, \mC)\star  \mR\rangle=\big\langle  \mathcal{V}_\mu(\momega,\momega, \mC)\star (d\momega-\momega \star \momega)\rangle}
\end{align}
It acquires a particularly nice form for the canonical vertex $\Vcan_\mu$. Surprisingly, in that case it describes the right propagating degrees of freedom on $AdS_4$ \cite{actions}. (In general, one might expect to obtain only a Lagrangian whose EL equations are weaker than the ones we need). 

The presymplectic structure \eqref{freePS} or, equivalently, the action \eqref{freeaction} can be checked against the well-known Fradkin--Vasiliev action \cite{Vasiliev:1986bq} and its precursor \cite{Vasiliev:1980as}. First, we need to decompose the curvature $\mR=d\momega-\momega\star \momega$ as follows:
\begin{align}
    \mR(y,\bry)&=\sum_{m,n}\frac{i}{m!n!}R_{\ga(m),\gad(n)}\, y^\ga\cdots y^\ga\, \bry^\gad\cdots \bry^\gad=  \mR_- +\mR_0 +\mR_+\,,
\end{align}
where $\mR_-$ ($\mR_+$) takes the $m<n$ ($m>n$) part of the sum and $\mR_0$ corresponds to the terms with $m=n$. Note that the sum is over even $m+n$ since we consider bosonic fields only. The two-form $\mR_0$ contains the (higher spin) torsions.
As with the conventional gravity, one can solve the zero-torsion equation
\begin{equation}
    \mR_0=0
\end{equation}
for some auxiliary fields in a purely algebraic way. With account of this equation, we can write the following integral for the Pontryagin topological invariant:
\begin{align}
    S_{\text{top}}&= \int \Tr{ \mR \star \mR}= \int \Tr{ \mR_+ \star \mR_+}+\int \Tr{ \mR_- \star \mR_-}\,.
\end{align}
Here $\Tr{-}$ combines the matrix trace and the trace on the Weyl algebra $A_2$. The Fradkin--Vasiliev action now reads
\begin{align}\label{FVaction}
    S_{\text{FV}}&= \int \Tr{ \mR \star \sigma(\mR)}= \int \Tr{ \mR_+ \star \mR_+}-\int \Tr{ \mR_- \star \mR_-}\,,
\end{align}
where $\sigma=\mathrm{sgn}(N_y-N_\bry)$ and $N_y=y^\nu \pl_\nu$, $N_\bry=\bar y^{\dot\nu} \pl_{\dot\nu}$ are the number operators for $y$ and $\bry$. The purpose of $\sigma$-map is to project out the torsion and to flip the sign of $\mR_-$ terms. For example, the usual Pontryagin's invariant and the MacDowell--Mansouri action \cite{MacDowell:1977jt} have the form\footnote{We put $i$ in front of MacDowell--Mansouri action assuming the usual reality conditions, $(R_{\ga\ga})^*=R_{\gad\gad}$, but will ignore all such factors below. Note also that the $\mR\mR$-action does not contain the Maxwell/Yang--Mills action. Nevertheless, it can be added by hand as $C_{\ga\ga}C^{\ga\ga}-C_{\gad\gad}C^{\gad\gad}$. The Pontryagin term does include the $s=1$ component. }
\begin{align}
    S_{\text{top}}&=\int R_{\ga\ga} R^{\ga\ga} + R_{\gad\gad} R^{\gad\gad}\,,  & S_{\text{MM}}&=i\int R_{\ga\ga} R^{\ga\ga} - R_{\gad\gad} R^{\gad\gad} \,.
\end{align}
As is well-known in the case of Yang--Mills theory and gravity one can add up the action and the topological invariant to get 
\begin{align}\label{fullaction}
    S&= \tau_+  \int \Tr{ \mR_+ \star \mR_+} + \tau_-\int \Tr{ \mR_- \star \mR_-}\,,
\end{align}
where $\tau_\pm\in \mathbb{C}$ are combinations of the coupling constant $g^2$ and the theta-angle $\theta$, $\tau_\pm=(a/g^2 \pm i\theta /b)$ (here, $a$, $b$ are some numerical constants). This introduces one more coupling $\theta$, which is invisible in the equations of motion. This trick makes the action reminiscent that of $3d$ (higher spin) gravity, which is a non-degenerate sum of two Chern--Simons actions. With the help of $\tau$ the $SL(2,\mathbb{Z})$-action on the space of $3d$ conformal field theories \cite{Witten:2003ya} extends to the  higher spin fields  in bulk \cite{Leigh:2003ez,Giombi:2013yva}.\footnote{It would be interesting to check if the normalization via $a$, $b$ that can be read off from the spin-two part of the action is consistent for higher spin fields as well. } 
Any non-degenerate  combination ($\tau_-\neq \tau_+$) gives an action which is consistent with the required gauge symmetries up to the cubic order since \eqref{FVaction} has this property \cite{Vasiliev:1986bq}. {The cubic part of the action can be evaluated on-shell to}
\begin{align}\label{cubicaction}
    -\tfrac12S_3&= \tau_+  \int \Tr{ \Vcan_{\mu=1,\bar \mu=0}(h,h,\mC) \star (\momega \star \momega)_+} + \tau_-\int \Tr{ \Vcan_{\mu=0,\bar \mu=1}(h,h,\mC) \star (\momega \star \momega)_-}\,,
\end{align}
where we indicated that $\Vcan$ consists of two terms and only one of them contributes to the first (second) part of the action. (Note that \eqref{canonicalA} involves the $AdS_4$ vierbein $h^{\ga\gad}$ rather than  the full $SO(3,2)$-connection $\stackrel{_{\circ}}{\momega}$.)

It is worth noting that each action of the family \eqref{fullaction} fixes the relative coefficient between cubic vertices $V_{s_1,s_2,s_3}$ for various spins $s_{1,2,3}$ that contribute to the action, which is usually not the case for cubic interactions (any linear combination of consistent cubic vertices $V_{s_1,s_2,s_3}$ is consistent again). This rigidity/uniqueness of the action is due to taking the higher spin algebra into account. Moreover, the relative normalization of free kinetic terms is also fixed in the free limit. The latter point is subtle: one could rescale all $\omega_{\ga(n),\gad(m)}$, $n+m=2s-2$, by some spin-dependent constants $a_s$ and change this normalization, but then one would have to change the star-product accordingly. Therefore, the relative normalization makes sense as long as we insist on using the Moyal--Weyl star-product for $\mR$ and for the gauge symmetries $\delta \momega=d\mxi -[\momega,\mxi]_\star$.

After this detour into the Fradkin--Vasiliev action and Pontryagin invariant, we can extract the presymplectic structure and compare it with \eqref{freePS}. The free part of the action \eqref{fullaction} differs from {the usual first order (AKSZ-type) actions} by a total derivative. Varying the action \eqref{fullaction} and extracting the boundary  term, we find the following presymplectic potential: 
\begin{align}
     \hat\Theta&= 2\tau_+  \int_\Sigma \Tr{ \mR_+ \star \delta\momega} + 2\tau_-\int_\Sigma \Tr{ \mR_- \star \delta\momega}\,.
\end{align}
On restricting to the solution space, we can replace the curvature $\mR$ with its on-shell value  
\eqref{canonicalA} and get an equivalent presymplectic potential $\Theta$ on the target space  that yields  
\begin{align}\label{freePSB}
    \Omega=d\Theta&=  2\tau_+ H^{\ga\ga} \langle\pl_\ga \pl_\ga dC(y,\bry=0)\star d\omega\rangle + 2\tau_- H^{\gad\gad} \langle\pl_\gad \pl_\gad C(y=0,\bry)\star d\omega\rangle\,.
\end{align}
This is exactly \eqref{freePS} with $\mu$'s replaced by $2\tau$'s. Furthermore, \eqref{freePSB}, when decomposed into fields with definite spin, is equivalent to the presymplectic form
\begin{align}\label{almostFronsdal}
    \Omega&= i(h\fud{\gb}{\gad}d\omega^{\ga(s-1),\gad(s-1)} d\omega_{\ga(s-1)\gb,\gad(s-2)}- \text{c.c.})\,,
\end{align}
which comes from the free first-order action of Ref. \cite{Vasiliev:1980as}. It coincides in form with the presymplectic structure of the linearized gravity, cf. \eqref{Om}, and is one of the admissible presymplectic structures found in  \cite{Sharapov:2016qne}. 

As another small test of interactions, the presymplectic AKSZ Lagrangian \eqref{freeaction} does reproduce the correct cubic on-shell vertex \eqref{cubicaction} upon appropriate identification between $\tau$ and $\mu$. However, since beyond the free level it is only a weak action (unless otherwise shown) we do not suggest to literally compare the actions in the literature, e.g.  \cite{Sleight:2016dba,Skvortsov:2018uru,Bekaert:2015tva} with \eqref{veryfree}. In general, in physics terms, weak actions can suffer from defects of two kinds: (1) some interactions can disappear because the presymplectic structure is degenerate; (2) some extra interactions can be found due to the action being invariant under a subset of symmetries only. We expect that the presymplectic AKSZ action proposed in the paper does not suffer from either defect, the only real issue being locality, see also discussion in Section \ref{sec:discussion}.

Presymplectic AKSZ Lagrangian \eqref{veryfree} can also clarify\footnote{E. S. is grateful to N. Boulanger for asking this question many years ago.} the very form of the original action \eqref{FVaction} or its refinement \eqref{fullaction}, i.e., the fact that it is very close to a topological one. Free parameters $\tau_\pm$ of \eqref{fullaction} get expressed in terms of a single coupling constant by requiring the action to be Hermitian and parity invariant. On relaxing these conditions (or at least the parity in view of the importance of the $\theta$-term) we get a two-parameter family of actions, which is closely related to the fact that there are two independent deformations of the equations of motion and of the extended higher spin algebra, cf. \eqref{factorizedform}. Therefore, the two-term structure of action \eqref{fullaction} and the appearance of `almost' the trace is explained by \eqref{veryfree} without any reference to $AdS_4$ or to cubic approximation.  

Let us summarize, presymplectic AKSZ action \eqref{freeaction} does reproduce the correct free dynamic over $AdS_4$; it does reproduce the correct presymplectic structure, which leads to the canonical quantization of free higher spin fields; it does reproduce the presymplectic structure coming from cubic action \eqref{fullaction} together with the normalization that is fixed by the higher spin symmetry. The last point should not be too surprising in view of the fact that both  \eqref{freeaction} and \eqref{fullaction} are fixed by the higher spin symmetry. However, \eqref{fullaction} is fixed up to the cubic terms over $AdS_4$, while \eqref{freeaction} is fixed up to ${\mC}^2$-terms, which is much more powerful. Nevertheless, it provides an additional check of the structure of interactions that goes beyond the classification \cite{Joung:2012hz,Francia:2016weg,Metsaev:2018xip}  of cubic vertices $V_{s_1,s_2,s_3}$ within the Noether procedure. 

Let us make a few comments on the presymplectic structure of higher spin fields and its relevance for quantization. In the simplest, non-gauge, case of the scalar field $\mC(x)$, $\Omega$ directly corresponds to the canonical commutation relations:
\begin{align}
    \hat \Omega&= \int_\Sigma \epsilon\fud{a}{bcd} dx^b dx^c dx^d\, \delta \mC \partial_{a} \delta \mC && \Longleftrightarrow && [\mC(x) ,\dot \mC(x')]= i\delta^3(x-x')\,,
\end{align}
with $\Sigma\subset \mathbb{R}^{3,1}$ being a Cauchy surface  $x^0=\mathrm{const}$. In the case of gauge fields, the situation is more complicated. For example, for $s=1$ we have
\begin{align}
    \hat \Omega&=\int_\Sigma \epsilon\fud{a}{bcd} dx^b dx^c dx^d\, \delta \mF_{an}\, \delta \mA^n\,,
\end{align}
which allows one to work out the commutators of physical observables.  In particular, for the components of magnetic field $\mB_i$ and electric field $\mE_j$ we get
\begin{align}\label{spinone}
   [\mB_i(x), \mB_j(x')]= 0\,,& & [\mB_i(x), \mE_j(x')]= i \epsilon_{ijk}\partial^k \delta^3(x-x')\,,&&  [\mE_i(x), \mE_j(x')]=0\,.
\end{align}
Similar commutation relations can be found for the curvatures of higher spin fields. Indeed, let $\phi_{\ga(s),\gad(s)}$ be the traceless part of the Fronsdal field, which can be associated with the totally symmetric part of the higher spin vierbein $e^{\ga(s-1),\gad(s-1)}$. The expressions for the Weyl tensors are very simple in the spinorial language:
\begin{align}
    C_{\ga(2s)}&= \nabla\fdu{\ga}{\gad}\cdots \nabla\fdu{\ga}{\gad} \phi_{\ga(s),\gad(s)}\,, &
    C_{\gad(2s)}&= \nabla\fud{\ga}{\gad}\cdots\nabla\fud{\ga}{\gad} \phi_{\ga(s),\gad(s)}\,.
\end{align}
Upon $3+1$ split we get out of $C_{\ga(2s)}$, $C_{\gad(2s)}$ a pair of  rank-$2s$ spin-tensors, which can be mapped to higher spin electric $\mE_{j_1\cdots j_s}$ and magnetic $\mB_{i_1\cdots i_s}$ fields; both of the fields are symmetric and traceless $SO(3)$-tensors. The only non-trivial commutator implied by $\hat\Omega$ in Minkowski space-time has the form
\begin{align}
   [\mB_{i_1\cdots i_s}(x), \mE_{j_1\cdots j_s}(x')]= i \epsilon_{i_1j_1k_1}\partial^{k_1}\pl_{i_2}\pl_{j_2}...\pl_{i_s}\pl_{j_s} \delta^3(x-x')+\ldots \,,
\end{align}
where the dots complete the r.h.s. to a traceless, symmetric and transverse bi-tensor. Note that \eqref{almostFronsdal} and \eqref{freePSB} are equivalent in $AdS_4$, but not in flat space-time.

The correct quantization of free higher spin fields is guaranteed by the fact that $\Omega_1$ coincides with the one obtained from the standard actions, all of which are equivalent to the Fronsdal action. In particular, the frame-like actions are known \cite{Vasiliev:1980as} to reduce to the Fronsdal action. Note that the presymplectic structure should be used to compute the commutator of observables. The complete classification of observables in $4d$ HSGRA was obtained in \cite{Sharapov:2020quq} with the important results found first in \cite{Sezgin:2005pv,Sezgin:2011hq}. For the free theory one set of gauge invariant observables is given by the components of $\mC$, e.g. by the electric and magnetic fields \eqref{spinone} and by the higher spin generalizations thereof. Another set of observables, currents of the global higher spin symmetry, is considered in Sec. \ref{sec:waves}.

\paragraph{Presymplectic structure and AKSZ to NLO.} For the formal manipulations below, we will use the presymplectic potential $\Theta_1$ in the form (\ref{O1}). 
Applying the homological perturbation theory  (\ref{dO}, \ref{dOO}), one can extend (\ref{O1}) to higher orders in $C$'s to make it invariant w.r.t. the full homological vector field (\ref{Qn}) underlying $4d$ HSGRA. In particular, the sub-leading term of expansion (\ref{O-exp}, \ref{On}) has the following  structure: \begin{equation}\Omega_2=\Omega_2^{_{(0)}}+ \Omega_2^{_{(1)}} +\Omega_2^{_{(2)}}\,.\end{equation} By definition, the  coefficients of the two-form $\Omega_2^{_{(0)}}$ vanish modulo $C^2$; this allows us to ignore the contribution of $\Omega_2^{_{(0)}}$ in the first-order approximation (\ref{lHSGRA}) to the field equations. As for the rest two terms,  we find 
\begin{equation}\label{O2}
\begin{array}{c}
    \Omega_2^{_{(1)}}+ \Omega_2^{_{(2)}}=d\Theta_2\,,\qquad \qquad \Theta_2= \frac12\langle\Lambda_{\mu\nu}\star dC\rangle\,,
    \end{array}
\end{equation}
where
\begin{align}
    \Lambda_{\mu\nu}(\omega,\omega,\omega,C) &=\Phi_\mu(\omega,\omega)\star \Psi_\nu(\omega,C)+\Psi_\mu(\omega,C)\star \Phi_\nu(\omega,\omega)\,,\\[3mm]
    \Psi_\nu(\omega, C)&=\Phi_\nu(\omega,C)-\Phi_\nu(C,\omega)\,.
\end{align}
One can check that 
\begin{equation}\label{dPsi}
    D\Psi_\nu(\momega,\mC)\approx [\mC,\Phi_\nu(\momega,\momega)]_\star \qquad \mod \mC^2\,.
\end{equation}
All this allows us to reconstruct the corresponding AKSZ Lagrangian up to the second order in the zero-form field $\mC$. A straightforward calculation gives 
\begin{equation}\label{L-AKSZ}
\boxed{\rule[-8pt]{0pt}{24pt}
\begin{array}{rcl}
   \mathcal{L}=\big\langle  \mC\star \Phi_\mu(\momega,\momega)\star \mR &-&\frac12\Lambda_{\mu\nu}(\momega,\momega,\momega, \mC) \star D\mC\\[3mm]
   &-&\frac12 \mC\star \Phi_{\mu}(\momega,\momega)\star\Phi_\nu(\momega,\momega)\star\mC\big\rangle  +\mathcal{O}(\mC^3)
    \end{array}}
\end{equation}
Let us check directly that the Lagrangian above leads to the equations of the form (\ref{EL}). We find
\begin{equation}\label{Lagr}
    \begin{array}{rcl}
    \delta \mathcal{L}&=&\langle  \delta \mC\star \Phi_\mu\star \mR + \mC\star \delta \Phi_\mu\star \mR + \mC\star \Phi_\mu\star D\delta\momega  \\[3mm]&-&\frac12\delta \Lambda_{\mu\nu} \star D\mC -\frac12\Lambda_{\mu\nu} \star D\delta \mC \\[3mm]
    &-&\frac12 \delta\mC\star \Phi_{\mu}\star\Phi_\nu\star\mC  -\frac12 \mC\star \Phi_{\mu}\star\Phi_\nu\star\delta \mC\rangle + \mathcal{O}(\mC^2)\\[3mm]
    &\approx&  d\langle \mC\star \Phi_\mu\star \delta\momega \rangle+d\langle\frac12\Lambda_{\mu\nu} \star \delta\mC \rangle \\[3mm]&-&\frac12\langle 
    (D\Lambda_{\mu\nu}-[\mC, \Phi_\mu\star\Phi_\nu]_\star )\star \delta\mC \rangle+ \mathcal{O}(\mC^2)\,.
    \end{array}
\end{equation}
It remains to observe that the last  term  vanishes due to the identity 
\begin{equation}\label{dL}
    D\Lambda_{\mu\nu}\approx [\mC,\Phi_{\mu}\star\Phi_\nu]_\star\quad \mod \mC^2\
    \end{equation}
that follows immediately form (\ref{dF}) and (\ref{dPsi}). 

We see that the EL equations for the Lagrangian (\ref{Lagr}) are satisfied by all solutions to the equations (\ref{lHSGRA}) of $4d$ HSGRA. 
As expected, the total derivatives  in (\ref{Lagr}) exactly reproduce the presymplectic potential $\Theta=\Theta_1+\Theta_2+\ldots$ defined by Rels. (\ref{O1}, \ref{O2}), cf. (\ref{VL}).  We claim that the corresponding on-shell presymplectic structure 
$$
\displaystyle \hat \Omega = \delta\hat \Theta\,,\qquad \hat \Theta =\int_\Sigma \mTheta\,, 
$$
where
\begin{equation}\label{mO}
\begin{array}{c}
    \mTheta=\langle{\mC}\star\Phi_\mu({\momega},{\momega})\star \delta\momega\rangle +\frac12\langle \Lambda_{\mu\nu}({\momega},{\momega},{\momega},\mC)\star\delta\mC\rangle+\mathcal{O}(\mC^2)
    \end{array}
\end{equation} 
and $\momega$, $\mC$ obey equations (\ref{lHSGRA}), becomes trivial modulo $\mC^2$ whenever $\mu=\nu$. Indeed,
\begin{equation}
\begin{array}{rcl}
     \!\!\mTheta&\approx& \!\!\!\langle\Phi_\nu\star \mC\star \delta\momega\rangle  +\langle D\Psi_\nu\star\delta\momega\rangle + \frac12\langle \Lambda_{\nu\nu}\star\delta\mC\rangle+\mathcal{O}(\mC^2)\\[3mm]
     &\simeq& \!\!\!\langle \mR \star \delta\momega\rangle  +\langle \Psi_\nu\star D\delta\momega\rangle + \frac12\langle \Lambda_{\nu\nu}\star\delta\mC\rangle+\mathcal{O}(\mC^2)\\[3mm]
       &\simeq& \!\!\!\langle \mR \star \delta\momega\rangle  -\langle \Psi_\nu\star \delta \mR \rangle + \frac12\langle \Lambda_{\nu\nu}\star\delta\mC\rangle+\mathcal{O}(\mC^2)\\[3mm]
       &=& \!\!\!\frac12d\langle\momega\star \delta\momega\rangle+\delta\big\langle\frac12\momega \star d\momega-\frac13\momega\star \momega\star \momega\big\rangle-\langle \Psi_\nu\star \delta \mR  \rangle  + \frac12\langle \Lambda_{\nu\nu}\star\delta\mC\rangle+\mathcal{O}(\mC^2).
\end{array}
\end{equation}
Denoting  
\begin{equation}\label{bJ}
\begin{array}{c}
    \mJ=\big\langle \small{\frac13}\momega\star\momega\star\momega+\Phi_\nu(\momega,\momega)\star\mC\star\momega\big\rangle+\mathcal{O}(\mC^2)\,,
    \end{array}
\end{equation}
we  proceed as 
\begin{equation}
    \begin{array}{rcl}
   \mTheta &\simeq&\frac 12 \delta \mJ -\langle \Psi_\nu\star\Phi_\nu\star\delta\mC \rangle  + \frac12\langle \Lambda_{\nu\nu}\star\delta\mC\rangle+\mathcal{O}(\mC^2)\\[3mm]
       &=&\frac 12 \delta \mJ+ \frac12\langle (\Phi_\nu\Psi_\nu-\Psi_\nu\Phi_\nu)\star\delta\mC\rangle+\mathcal{O}(\mC^2)\\[3mm]
       &=& \frac 12 \delta \mJ+\frac12\langle [\delta\mC,\Phi_\nu]_\star\star \Psi_\nu+\Psi_\nu\star [\delta\mC,\Phi_\nu]_\star\rangle +\mathcal{O}(\mC^2)\\[3mm]
         &\approx& \frac 12 \delta \mJ+\frac12\langle D\Psi_\nu(\momega,\delta\mC)\star \Psi_\nu(\momega,\mC)+\Psi_\nu(\momega,\mC)\star D\Psi_\nu(\momega,\delta\mC)\rangle +\mathcal{O}(\mC^2)\\[3mm]
         &\approx&  \frac 12 \delta \big( \mJ+\langle [\mC, \Phi_\nu(\momega,\momega)]_\star\star \Psi_\nu(\momega,\mC)\rangle\big) -\frac12 d\langle \Psi_\nu(\momega,\mC)\star  \Psi_\nu(\momega,\delta\mC)\rangle +\mathcal{O}(\mC^2).  
    \end{array}
\end{equation}
Hence, $\mTheta$ degenerates into the sum of on-shell $d$- and $\delta$-exact terms, so that $\hat \Omega =0$ modulo $\mC^2$. The current (\ref{bJ}) is conserved up to the first order in $\mC$, that is,  $d\mJ\approx 0 \; (\mathrm{mod}\;\mC^2)$; its conservation, however, inevitably breaks down at the next order, see \cite{Sharapov:2020quq}. 

The general conclusion is that  for $\mu=\nu$, the presymplectic structure (\ref{mO}) is equivalent to that of the form (\ref{O-exp}) with $m>2$. In other words, even for $\mu=\nu$ the presymplectic structure remains non-trivial, but starts at NNLO. However, this does not seem to be satisfactory from the physical point of view as long as we want to reproduce the correct quantization of free higher spin fields on maximally symmetry backgrounds, $\mC=0$, e.g. on $AdS_4$. Therefore, we keep $\mu\neq \nu$.

Another important comment is that the scalar field of the higher spin multiplet is somewhat separated from $s>0$ fields: its presymplectic structure does not appear at LO and enters at NLO. Of course, the higher spin symmetry relates it to those of higher spin fields, which were found to be the correct ones. In order to extract the correct presymplectic structure for the free scalar field one has to choose $\Theta_1$ of the form \eqref{thetaA} and perform the corresponding redefinitions of $\Theta_2$ \eqref{O2}, which is a problem closely related to the issue of (non-)locality of HSGRA. We hope that many useful statements can be proved first at the formal level, while the non-locality problem awaits its satisfactory resolution. 

Lastly, the presymplectic structure is unobstructed, see Appendix \ref{app:D} for the proof. Therefore, it can be extended to any order and the general expression can be written with the help of the techniques introduced in \cite{Sharapov:2019vyd} for equations. The presymplectic structure and AKSZ action depend on two coupling constants, as required by the AdS/CFT duality with Chern--Simons matter theories (one should set $\mu=\nu={\tilde N}^{-1} e^{i\theta}$ and $\bar\nu=\nu^*$, $\bar\mu=-\mu^*$, where $\theta= \tfrac{\pi}2\lambda$, $\tilde{N}=2N\tfrac{\sin \pi \lambda}{\pi \lambda}$ are expressed in terms of the number of fields $N$ and t'Hooft coupling $\lambda=N/k$ for Chern--Simons level $k$).

\section{Higher spin waves and currents}\label{sec:waves}
As has been mentioned above, the most symmetric vacuum in HSGRA corresponds to $\mC=0$. In this case, the highly non-linear equations of motion \eqref{dwdc} are greatly simplified taking the form 
of zero-curvature condition for the gauge connection $\momega$ associated with the Lie algebra $gl(\mathfrak{A})$. In other words, the solutions to
\begin{equation}\label{vacA}
    d\!\stackrel{_{\circ}}{\momega}=\stackrel{_{\circ}}{\momega}\!\star\!\stackrel{_{\circ}}{\momega}
\end{equation}
describe the most symmetric geometric backgrounds against which the free higher spin fields  can consistently propagate. A particular solution to these equations is provided by the anti-de Sitter space, in which case $\stackrel{_{\circ}}{\momega}\in so(3,2)\subset gl(\mathfrak{A})$.  
We let $\tilde\mC$ and $\tilde\momega$  denote the small fluctuations of fields about the vacuum. Then the linearized equations of motion \eqref{dwdc} read
\begin{equation}\label{hsw}
    \stackrel{_{\circ}}{D}\!\tilde\momega =\Phi_\nu( \stackrel{_{\circ}}{\momega}, \stackrel{_{\circ}}{\momega} )\star \tilde\mC\,,\qquad \qquad  \stackrel{_{\circ}}{D}\!\tilde\mC =0\,.
\end{equation}
Here we introduced the background covariant differential $\stackrel{_{\circ}}{D}=d-[\stackrel{_{\circ}}{\momega},-]_\star$. By definition, $\stackrel{_{\circ}}{D}{\!}^2=0$. 
The equations are invariant under the gauge transformations
\begin{equation}\label{gsym}
    \delta_{\mvarepsilon}\tilde\momega=\stackrel{_{\circ}}{D}\mvarepsilon \,,\qquad\qquad  \delta_{\mvarepsilon}\tilde\mC=0\,,
\end{equation}
$\mvarepsilon$ being an infinitesimal gauge parameter. Besides, system (\ref{hsw}) enjoys the global symmetries of the form
\begin{equation}\label{glsym}
    \delta_\mxi\tilde\momega =[\mxi,\tilde\momega]_\star-\Psi_\nu(\stackrel{_{\circ}}{\momega},\mxi)\star \tilde\mC\,,\qquad\qquad \delta_\mxi \tilde\mC=[\mxi,\tilde\mC]_\star\,,
\end{equation}
where the infinitesimal parameter $\mxi$ obeys the condition $\stackrel{_{\circ}}{D}\mxi=0$. Notice the term involving $\Psi_\nu$, which survives even for the anti-de Sitter background.  In that case the $\kappa$-independent part of $\mxi$ accommodates the whole set of Killing's tensors of $AdS_4$.   Noteworthy also is the fact that the global symmetries form an associative algebra, $\mathrm{Mat}(\mathfrak{A})$, w.r.t. the $\star$-product of $\mxi$'s, and not just the Lie algebra $gl(\mathfrak{A})$ w.r.t. their $\star$-commutators.  

A nice property of (\ref{vacA}, \ref{hsw}) is that this is a fully consistent system on its own. It requires only the Hochschild cocycle $\Phi_\nu$ and does not suffer from non-localities. It describes propagation of higher spin fields on backgrounds more complicated than just the anti-de Sitter space, providing thus  a way to overcome the Aragone--Deser no-go \cite{Aragone:1979hx}. In $d=3$ there is no r.h.s. in \eqref{hsw} since higher spin fields do not have propagating degrees of freedom. For $d>3$, the Hochschild cocycle is required. The simplest solutions beyond empty $AdS_d$ could be topological black holes \cite{Aminneborg:1996iz} and higher spin generalizations thereof \cite{Aros:2019pgj} along the $3d$ lines, see e.g. \cite{Ammon:2012wc}.\footnote{We are grateful to Per Sundell for this suggestion.}

The linearization of (\ref{L-AKSZ}) over the vacuum $\mC=0$ and $\momega\!=\,\stackrel{_{\circ}}{\momega}$ gives the following quadratic Lagrangian:\footnote{Note that to this order it is easy to write the Lagrangian in a more suitable form as 
\begin{align*}
    \mathcal{L}^{\circ}=\langle \Vcan_\mu(\stackrel{_{\circ}}{\momega}, \stackrel{_{\circ}}{\momega},\tilde\mC)\,\star \stackrel{_{\circ}}{D}\tilde\momega-\tfrac12\tilde\Lambda_{\mu\nu}(\stackrel{_{\circ}}{\momega},\stackrel{_{\circ}}{\momega}, \stackrel{_{\circ}}{\momega},\tilde\mC)\,\star \stackrel{_{\circ}}{D}\tilde\mC-\tfrac12 \Vcan_\mu(\stackrel{_{\circ}}{\momega}, \stackrel{_{\circ}}{\momega},\tilde\mC)\star \Vcan_\nu(\stackrel{_{\circ}}{\momega}, \stackrel{_{\circ}}{\momega},\tilde\mC)\rangle\,,
\end{align*}
see footnote \ref{nondiag}. One just needs to perform the field redefinitions that lead from the factorized vertex to $\Vcan$, inducing a certain change of $\Lambda_{\mu\nu}$ into $\tilde\Lambda_{\mu\nu}$. It is the $\Lambda$-term that gives the presymplectic structure for the free scalar field. }
$$
\begin{array}{c}
    \mathcal{L}^{\circ}=\langle \tilde\mC\star \Phi_\mu(\stackrel{_{\circ}}{\momega}, \stackrel{_{\circ}}{\momega}) \,\star \stackrel{_{\circ}}{D}\tilde\momega-\frac12\Lambda_{\mu\nu}(\stackrel{_{\circ}}{\momega},\stackrel{_{\circ}}{\momega}, \stackrel{_{\circ}}{\momega},\tilde\mC)\,\star \stackrel{_{\circ}}{D}\tilde\mC-\frac12 \tilde\mC\star \Phi_{\mu}(\stackrel{_{\circ}}{\momega}, \stackrel{_{\circ}}{\momega})\star\Phi_\nu(\stackrel{_{\circ}}{\momega}, \stackrel{_{\circ}}{\momega})\star\tilde\mC\rangle \,.
    \end{array}
$$
The Lagrangian is obviously invariant under the gauge  transformations (\ref{gsym}), but not under the action of global symmetries (\ref{glsym}) as one can easily check. 
Nonetheless,  we can still use the underlying presymplectic structure to assign conserved currents to the global symmetry transformations. The linearized presymplectic form $\Omega^\circ=d\Theta^\circ$ on the target space  induces that on the configuration space of fields $\tilde \mC$ and $\tilde \momega$. The latter is given by 

\begin{equation}\label{mOO}
\begin{array}{c}
\displaystyle \hat \Omega{}^\circ=\int_\Sigma \mOmega^\circ\,, \qquad\qquad\mOmega^{{\circ}} = \delta\mTheta^{{\circ}}\,, \\[5mm]
     \mTheta^{\circ}=\langle \tilde{\mC}\star\Phi_\mu(\stackrel{_{\circ}}{\momega},\stackrel{_{\circ}}{\momega})\star \delta\tilde\momega\rangle +\frac12\langle \Lambda_{\mu\nu}(\stackrel{_{\circ}}{\momega},\stackrel{_{\circ}}{\momega},\stackrel{_{\circ}}{\momega},\tilde\mC)\star\delta\tilde\mC\rangle\,.
    \end{array}
\end{equation} 
In the case of $AdS_4$ background, this presymplectic structure and the  Lagrangian $\mathcal{L}^\circ$ were first found in  \cite{Sharapov:2016qne}. 

Let $\mX_\mxi$ denote the variational vector field defined by the r.h.s. of equations (\ref{glsym}). We leave it to the reader to check that the vector field $\mX_\mxi$, being tangent to the solution space, is Hamiltonian relative to $\hat\Omega^{{\circ}}$, i.e.,
\begin{equation}
    i_{\mX_\mxi}\mOmega^{{\circ}}\simeq \delta \mJ_\mxi \,,
\end{equation}
where 
\begin{equation}\label{Jxi}
\begin{array}{rcl}
    \mJ_\mxi &=&\big\langle \tilde{\mC}\star\Phi_\mu(\stackrel{_{\circ}}{\momega},\stackrel{_{\circ}}{\momega})\star ([\mxi,\tilde\momega]-\Psi_\nu(\stackrel{_{\circ}}{\momega},\mxi)\star \tilde\mC )       -  \frac12 \Lambda_{\mu\nu}(\stackrel{_{\circ}}{\momega},\stackrel{_{\circ}}{\momega},\stackrel{_{\circ}}{\momega},\tilde\mC)\star    [\mxi,\tilde\mC]     \big  \rangle \\[3mm]
    &+& \frac12\langle\tilde\mC\star \Lambda_{\mu\nu}(\stackrel{_{\circ}}{\momega},\stackrel{_{\circ}}{\momega},\stackrel{_{\circ}}{\momega},\mxi)\star\tilde\mC\rangle\,.
    \end{array}
\end{equation}
By construction, the Hamiltonian $\mJ_\mxi$ defines a conserved current for the higher-spin wave equations (\ref{hsw}). The last fact is easy to verify  directly:
\begin{equation}
    \begin{array}{rcl}
         d\mJ_\mxi &\approx&\langle\tilde\mC\star\Phi_\mu\star ([\mxi,\Phi_\nu\star\tilde\mC]_\star-[\mxi,\Phi_\nu]\star \tilde\mC)\rangle\\[3mm]
         &-&\frac12 \langle [\tilde{\mC},\Phi_\mu\star\Phi_\nu]_\star\star[\mxi,\tilde{\mC}]_\star\rangle - \tilde{\mC}\star[\mxi,\Phi_\mu\star\Phi_\nu]_\star\star \tilde{\mC}\rangle=0\,.
    \end{array}
\end{equation}
Here we used identities (\ref{dF}), (\ref{dPsi}), and (\ref{dL}) together with the cyclicity of the trace.  As with the presymplectic structure (\ref{mOO}), the currents (\ref{Jxi}) become trivial for $\mu=\nu$.

Notice that the conserved currents $\mJ_\mxi$ are not invariant under the gauge transformations (\ref{gsym}) in the sense that
\begin{equation}
    \delta_\mvarepsilon\mJ_\mxi\approx d\langle \tilde\mC \star \Phi_\mu(\stackrel{_{\circ}}{\momega},\stackrel{_{\circ}}{\momega})\star[\mxi,\mvarepsilon]_\star\rangle\neq 0 \,.
\end{equation}
The violation of gauge invariance occurs due to the explicit dependence of $\tilde \momega$ and cannot be removed by adding $d$-exact terms to the currents $\mJ_\mxi$. Of course, this non-invariance does not affect the integrated conserved charges.  It is significant that the currents (\ref{Jxi}) have non-zero projections onto the physical sector if one takes $\mxi\in gl(\mathfrak{hs})$, see Remark \ref{R34}. 

Gauge invariant and non-invariant conserved currents for the free higher spin fields on the $AdS_4$ background have been intensively studied in the literature, see  \cite{Gelfond:2006be, Gelfond:2014pja, universe3040078} and references therein. In  \cite{universe3040078},  an infinite family of gauge non-invariant currents of the form $\tilde\momega\times\tilde \momega$ have been explicitly constructed. Our gauge non-invariant currents (\ref{Jxi}) have a different structure, schematically $\tilde\mC\times \tilde \momega+\tilde\mC\times \tilde\mC$.   The gauge invariant currents were completely classified in Ref. \cite{Gelfond:2006be,Gelfond:2014pja}. Being gauge invariant, they may only depend on the zero-form field  as $\tilde\mC\times\tilde\mC$. It goes without saying that none of the currents above can be extended to or come from the non-linear theory. 
Furthermore, a straightforward cohomological analysis exposed in the Appendices shows that the aforementioned currents $\tilde\momega\times\tilde \momega$ and $\tilde\mC\times\tilde\mC$ cannot be even extended from $AdS_4$ to the general higher spin background (\ref{vacA}). 

The Poisson bracket associated with the presymplectic structure (\ref{mOO}) makes the conserved currents (\ref{Jxi}) into the Lie algebra:
\begin{equation}\label{JJ}
    \{\mJ_\mxi,\mJ_{\mxi'}\}=i_{\mX_\mxi}i_{\mX_{\mxi'}}\mOmega^{\circ}\simeq \delta_\mxi\mJ_{\mxi'}\simeq \mJ_{[\mxi,\mxi']_\star} \,.
\end{equation}
In general, the currents $\mJ_\mxi$ being quadratic in fields, one may expect a non-trivial central charge to appear in the r.h.s. of (\ref{JJ}) upon quantization.    The vanishing of the second cohomology group  $H^2(gl(\mathfrak{A}))$ (see \cite[Sec. B.4]{sharapov2020characteristic}), however, precludes such a possibility, so that  the commutation relations (\ref{JJ}) have to survive quantization as they are.

\section{Final comments and discussion}\label{sec:discussion}
We have constructed a presymplectic AKSZ sigma-model for $4d$ HSGRA at the formal level, that is, we showed that it does exist and depends on the right number of coupling constants. We also worked out the first two orders explicitly and proved that the higher orders are unobstructed. To a great surprise the free action turns out to be a genuine action for higher spin fields rather than just a weak action. The action leads to the right quantum commutators for various observables: the gauge-invariant field strengths and the currents of the global higher spin symmetry (leftover of the gauge symmetry at the free level). The AKSZ action reproduces some of the cubic vertices, while the rest should come from higher orders in the $\mC$-expansion. 

Among the future developments of interest we can mention: $(i)$ to find a compact form for the presymplectic AKSZ action of HSGRA in arbitrary dimension, which should be a variation of the techniques from \cite{Sharapov:2018hnl,Sharapov:2018ioy,Sharapov:2018kjz,Sharapov:2019vyd}; $(ii)$ guided by the examples of lower spin theories, to develop a path-integral quantization of HSGRA by means of presymplectic AKSZ models; $(iii)$ to see if concrete quantum checks of HSGRA can be done already at the formal level, i.e., without paying attention to the non-locality of the models.

\paragraph{Quantization I.} The main purpose of the approach advocated in this paper is to shed some light on the quantization of HSGRA. The basic ingredient of any quantization method is a Poisson structure on the space of physical observables of a classical theory. This is, for example, a starting point of the deformation quantization technique and the canonical quantization. In other approaches, the Poisson brackets manifest themselves and can be recovered through the semi-classical limit of equal-time commutation relations. Each Lagrangian field theory enjoys a canonical presymplectic structure that induces a non-degenerate Poisson bracket on gauge-invariant functionals of fields. 

It is particularly remarkable that one can define and classify all suitable presymplectic structures
independently of Lagrangians; the only input one needs for that are classical equations of motion. In this paper, we solved this classification problem for $4d$ HSGRA.  Applying cohomological analysis, we found that all physically relevant presymplectic structures form a two-parameter family  that can be read off or encoded in an AKSZ-type sigma-model (\ref{L-AKSZ}). Moreover, we were able to find the explicit expressions (\ref{mO}) for (some representatives of) these presymplectic structures up to the second-order in $\mC$. As an immediate application of these presymplectic structures, we computed the Poisson brackets of the conserved currents (\ref{Jxi}) in free theory. We also argued that 
the classical commutation relations of the currents (\ref{JJ}) should survive quantization. 

The deformation quantization of finite-dimensional presymplectic manifolds has been considered in the works \cite{2002JMP43283V, 2020TMP204.1079G}.  Appropriate adaptation of this method to the context of field theory (see e.g. \cite{88c5409ac2d1498fa6234cbfe818edba} and references therein) will hopefully work for the model at hand.  As an alternative, one can try to apply the path-integral quantization to the presymplectic AKSZ model of Sec. \ref{sec:PSHSGRA}. It is well to bear in mind that the weak Lagrangian (\ref{L-AKSZ}) may not be a Lagrangian in the ordinary sense and its relevance to the path-integral quantization of HSGRA calls for further investigation. 

\paragraph{Quantization II.} More abstractly one could check what are the possible counterterms in a HSGRA, which does not require actual quantization. The beauty of the counterterm argument is that one might be able to prove renormalizability or finiteness even without having to quantize anything. However, the naive argument -- the more symmetries, the less counterterms -- does not seem to apply without further fuss: there exists infinitely many invariants of the higher spin symmetry that can serve as potential counterterms \cite{Sharapov:2020quq}. Nevertheless, additional physical restrictions might potentially rule most of them out. Therefore, the direct quantization of HSGRA is an important open problem too.

\paragraph{Vacuum corrections.} The on-shell AKSZ action is proportional to the corresponding Hamiltonian; the latter is an example of on-shell observables which were classified in \cite{Sharapov:2020quq}. The on-shell action starts with $\momega^4\mC^2$-type terms and appears to vanish on any maximally symmetric background ($\mC=0$). 

However, the value of the classical action on $AdS_4$ vacuum should not be zero. Moreover, it is known to contain a rather peculiar numerical factor $\tfrac{1}{16} \big(\log (4)-\frac{3 \zeta (3)}{\pi ^2}\big)$ for the $\Delta=1$ boundary conditions \cite{Klebanov:2002ja,Sezgin:2003pt}, the free energy of the free scalar field on the three-sphere. Some contribution might come from the boundary terms (a higher spin analog of the Gibbons--Hawking term) that the action should be supplemented with. All possible on-shell non-trivial three-forms can be found in \cite{Sharapov:2020quq} and again there is no candidate among $\momega^3$-type observables. A more rigorous approach to the problem  would be to carefully take into account the near boundary behaviour of fields and gauge parameters, which may lead us to reconsider the problem of boundary terms.

It should also be borne in mind that
the on-shell action is actually ill-defined and requires regularization. Formally, it is the indeterminate form $0\cdot \infty$, where $0$ corresponds to the on-shell value of the integrand, while $\infty$ comes from integration over an infinite volume. To evaluate the integral properly, one could choose a family of solutions approaching the $AdS_4$ vacuum; in so doing, each $\mC$ of the family must satisfy an appropriate fall-off condition to ensure convergence. The limit of the integral as $\mC\rightarrow 0$ may then be identified with the value of the presymplectic AKSZ action (\ref{L-AKSZ}) on the $AdS_4$ vacuum. Unfortunately, not much exact solutions to HSGRA  are available in the literature to implement the above regularization procedure.  It seems reasonable to use a family of solutions that preserves as much symmetries as possible, e.g. as in \cite{Sezgin:2005pv,Iazeolla:2017dxc}.

Note that in the ordinary (non-higher-spin) cases the on-shell value is proportional to the volume of (Euclidian) $AdS_d$, which is divergent and needs to be regularized. One can also regard the space-time integral as the volume of the quotient $SO(d+1,1)/SO(d,1)$, which relates it to the `gauge' group. In the higher spin case the group is not yet well-defined, see e.g. \cite{Anninos:2020hfj}. Moreover, higher spin transformations mix the metric $g_{\mu\nu}$ with all higher spin fields and the on-shell value of the action should not be thought of as the volume of $AdS_d$, rather it should be related to the (regularized) volume of the higher spin group.   

\paragraph{How degenerate the presymplectic AKSZ action is?} In general, as is discussed in Sec. \ref{sec:pre-AKSZ}, presymplectic actions cannot reproduce all of the equations of motion. Nevertheless, in some cases, e.g. gravity, the hidden integrability conditions allow one to get equations that are completely equivalent to the one we started with. In order to have such a miracle one needs the right balance between fields and equations (e.g. in gravity the torsion constrain has the same number of components as spin-connection $\omega^{a,b}$ and the Einstein equations can, in principle, be reproduced from the variation of vielbein $e^a$, which is what happens). However, in any HSGRA in $d>2$ this balance is broken: the equations for $\mC$ and $\momega$ are one- and two-forms valued in a higher spin algebra. This argument is not precise enough since we do not have to reproduce all of the equations, only the ones that lead to the differential equations for the dynamical fields (for example, torsion-like constraints can be imposed by hand). 

Nevertheless, the following observation can save the day: variation with respect to $\momega$ can, in principle, reproduce all of the $\mC$-equations, \eqref{dc}. Suppose this is true. Then we can show, see Appendix \ref{app:integrability}, that the $\momega$-equations resulting from the integrability of the $\mC$-equations have to be exactly \eqref{dw} up to a two-form $\boldsymbol{B}$
\begin{align}
    d\momega&= \momega\star \momega+ \mathcal{V}(\momega,\momega,\mC)+\mathcal{O}(\mC^2) + \boldsymbol{B}\,,
\end{align}
where $\boldsymbol{B}$ belongs to the center of the higher spin algebra, i.e. it is $\boldsymbol{B}= \mathrm{1} \cdot B$ and $B$ is a non-trivial two-form belonging to the de Rham cohomology, if any.\footnote{We are grateful to Ergin Sezgin for a useful discussion on the ambiguities of the equations, see also \cite{Boulanger:2015kfa} for the discussion of interactions' ambiguities related to de Rham cohomology.} 

As was mentioned in Sec. 4, weak Lagrangians associated with degenerate presymplectic structures may violate some of the gauge symmetries of the original equations of motions. The surprising thing is that this does not happen at the free level. The Lagrangian for higher-spin waves is obviously invariant under linearized gauge transformations (\ref{gsym}) (while some of the global symmetries are broken). Whether this observation extends to the  non-linear level is not clear at the moment and requires a more detailed study of the kernel of the presymplectic structure. The information about the kernel is also needed for the identification of physical observables that admit consistent quantization using the presymplectic structure. We are going to address all these issues in future publications.
\paragraph{Higher dimensions.} It is instructive to have a look at the presymplectic AKSZ model for the $d$-dimensional formal HSGRA. As argued in \cite{Sharapov:2020quq}, the basis of the Hochschild cohomology of the (extended) higher spin algebra is given by a two-cocycle $\phi$ and a $(d-2)$-cocycle $\psi$. The two-cocycle is the one that drives the deformation of the free equations. The $(d-2)$-cocycle contributes to higher form observables. We see that at $d=4$ both of these, rather different in nature, cocycles happen to be two-cocycles. Algebraically the doubling of two-cocycles is due to the fact that the extended higher spin algebra $\mathfrak{A}=\mathcal{A}\otimes \mathcal{A}$ is the tensor square of $\mathcal{A}=A_1\rtimes \mathbb{Z}_2$.  

With the cocycle $\psi$ we can write the following natural expression for the presymplectic potential on fields:
\begin{align}
    \mTheta_1=\langle \psi(\momega,\ldots ,\momega)\star \mC\star \delta \momega\rangle\,.
\end{align}
It is also plausible that there exists a field redefinition that brings this $\mTheta_1$ into the form
\begin{align}
    \mTheta_1=\langle \mathcal{V}(\momega,\ldots,\momega, \mC)\star \delta \momega\rangle
\end{align}
such that for the anti-de Sitter vacuum $\stackrel{\circ}{\momega}= h^a P_a +\tfrac12 w^{a,b}L_{ab}$  we have
\begin{align}
    \mTheta_1&= \epsilon_{v_1\cdots v_{d-2}u_1 u_2}h^{v_1}\cdots h^{v_{d-2}} \mC^{u_1 a(s-1), u_2 b(s-2)} \delta \tilde\momega_{a(s-1),b(s-1)}\,.
\end{align}
Here, $h^a$ is a vielbein, $w^{a,b}$ is a spin-connection, $P_a$ and $L_{ab}$ are the generators of transvections and Lorentz transformations, and $\tilde\momega$ is the fluctuation about $\stackrel{\circ}{\momega}$.
This gives the canonical presymplectic form 
\begin{align}
    \mOmega^\circ&= \epsilon_{v_1\cdots v_{d-2}u_1 u_2}h^{v_1}\cdots h^{v_{d-2}} \delta\tilde \mC^{u_1 a(s-1), u_2 b(s-1)} \delta\tilde \momega_{a(s-1),b(s-1)}
\end{align}
for free fields $\tilde \momega$, $\tilde \mC$ on $\mC=0$ background.
The non-linear presymplectic AKSZ action should have then the form  
\begin{align}\label{daction}
    S= \int \langle \mathcal{V}(\momega,\ldots,\momega, \mC)\star (d \momega -\momega\star \momega)\rangle +\mathcal{O}(\mC^2)\,.
\end{align}
We are going to detail this construction elsewhere, but it is tempting to compare it with the Einstein--Hilbert action in the frame-like formulation
\begin{align}
    S[e,\omega]&= \int \epsilon_{v_1\cdots v_{d-2} u_1u_2} e^{v_1}\cdots e^{v_{d-2}} (d\omega^{u_1,u_2} - \omega\fud{u_1}{c} \omega^{c,u_2})\,.
\end{align}
The first factor is nothing else but the Chevalley--Eilenberg cocycle of the Poincar\'e algebra. 

In $3d$, one should not be surprised that to leading order the only presymplectic structure is the one of Chern--Simons theory and the presymplectic AKSZ action is just the Chern--Simons action. It would be interesting to explore the $3d$ case further. 

As a step towards the complete AKSZ sigma-model it is worth noting that the non-linear equations of any HSGRA are integrable in the sense of being equivalent to the following `free system' \cite{Sharapov:2019vyd}:
\begin{align}
    d\hat\momega&=\hat\momega \ast \hat\momega\,, &
    d \hat\mC&=\hat\momega\ast \hat\mC-\hat\mC\ast \hat\momega\,.
\end{align}
Here $\ast$ is the product in the deformed (extended) higher spin algebra $A_\hbar$, $$a\ast b= a\star b +\hbar\, \phi(a,b)+\cdots\,,$$ and the fields $\hat\momega$, $\hat\mC$ take values in $A_\hbar$. The deformed algebra features the same structure of Hochschild cocycles, e.g. it has $2$-cocycle $\varphi(a,b)= \pl_\hbar (a\ast b)$, it also has an invariant trace $\langle - \rangle$ and should have an appropriate cocycle $\widehat{\mathcal{V}}(\hat\momega,\ldots,\hat\momega, \hat\mC)$. Now, one can write a complete action as
\begin{align}
    S= \int \langle \widehat{\mathcal{V}}(\hat\momega,\ldots,\hat\momega, \hat\mC)\ast (d \hat\momega -\hat\momega\ast \hat\momega)\rangle \,,
\end{align}
which may be a good starting point for exploring the quantum properties of HSGRA in the future.

\section*{Acknowledgments}
\label{sec:Aknowledgements}
We are grateful to Nicolas Boulanger, Maxim Grigoriev, Carlo Iazeolla, Ergin Sezgin and Per Sundell for useful discussions and correspondence. E. S. is grateful to the Erwin Schr{\"o}dinger Institute in
Vienna for hospitality during the program ``Higher Structures and Field Theory'' while this work was in progress.  The work of A. Sh. was supported by the Ministry of Science and Higher Education of the Russian Federation, Project No. 0721-2020-0033. The work of E. S. was supported by the Russian Science Foundation grant 18-72-10123 in association with the Lebedev Physical Institute. Each author blames the faults that remain on the other.\footnote{We are grateful to Karapet Mkrtchyan for suggesting this peaceful resolution.}

\appendix

\section{Hochschild, cyclic, and Lie algebra cohomology}
\label{app:Hoch}
In this appendix, we collect some basic definitions and constructions related to the cohomology of associative and Lie algebras. 
For a more coherent  exposition of the material we refer the reader to  \cite{Loday}, \cite{Co}, \cite{Feigin1987}.
A word about notation: all unadorned tensor products $\otimes$ and Hom's are taken over $k$, a ground field of characteristic zero. We 
systematically follow the Koszul sign convention. As in the main text, the grade of a homogeneous element $a$ is denoted by $|a|$. Many formulas below 
are considerably simplified if one uses the shifted grade $\bar a=|a|-1$.

The Hochschild cohomology $HH^\bullet(A,M)$ of a graded associative $k$-algebra $A$  with coefficients in a graded $A$-bimodule $M$ is the cohomology 
of the Hochschild cochain complex $C^\bullet(A,M)$ composed by the vector spaces 
$$
C^p=\mathrm{Hom}(A^{\otimes p}, M)\,,\qquad A^{\otimes p}=\underbrace{A\otimes \cdots \otimes A}_p
$$
and the homomorphisms $\partial: C^p\rightarrow C^{p+1}$ defined by 
\begin{equation}\label{HD}
(\partial f)(a_1,\ldots, a_{p+1})=(-1)^{(\bar a_1+1)(\bar f+1)}a_1 f(a_2,\ldots,a_{p+1})  - (-1)^{\bar a_1+\cdots +\bar a_p}f(a_1,\ldots,a_{p})a_{p+1}
\end{equation}
$$
+\sum_{k=1}^{p}(-1)^{\bar a_1+\cdots +\bar a_k} f(a_1,\ldots,a_ka_{k+1},\ldots, a_{p+1})\,.
$$
In the special case $M=A^\ast$ it is convenient to identify the spaces $C^p(A,A^\ast)$ with $\mathrm{Hom}(A^{\otimes(p+1)},k)$. Then the formula for the Hochschild differential takes the form   
\begin{equation}\label{HD1}
\begin{array}{rl}
(\partial g)(a_0, a_1,\ldots, a_{p+1})=&\displaystyle \sum_{k=0}^{p}(-1)^{\bar a_0+\cdots+\bar a_k} g(a_0, a_1,\ldots,a_ka_{k+1},\ldots, a_{p+1})\\[5mm]
+&(-1)^{(\bar a_0+1)(\bar a_1+\cdots +\bar a_{p+1})} g(a_1, \ldots, a_p,a_{p+1}a_0)\,,
\end{array}
\end{equation}
where, by definition,  $g(a_0,\ldots,a_{p-1},a_{p})=(-1)^{|a_{p}|}f(a_0,\ldots,a_{p-1})(a_{p})$.

As was first observed by A. Connes, the complex $ C^\bullet(A,A^\ast)$ contains a subcomplex $C_{\mathrm{cyc}}^\bullet(A)$ of {\it cyclic cochains}, i.e., 
cochains $g\in \mathrm{Hom}(A^{\otimes(p+1)},k)$ satisfying the additional condition 
\begin{equation}\label{cyclic}
g(a_0,a_1,\ldots, a_p )=(-1)^{\bar a_0(\bar a_1+\cdots+\bar a_p)}g(a_1,\ldots, a_{p},a_0)\,.
\end{equation}
The cohomology of the complex $C_{\mathrm{cyc}}^\bullet(A)$ is called the {\it cyclic cohomology} of $A$ and the corresponding cohomology groups are denoted by $HC^\bullet(A)$.  Upon restricting to cyclic cochains, one can bring the differential (\ref{HD1}) into a more familiar form\footnote{Notice that the signs in either form obey the {\it Koszul sign rule} if one shifts the grade of all $a$'s by $-1$, so that the dot product acquires degree $1$. Upon this interpretation the  dot product and cyclic permutation of $a$'s go first and the map $g$ after.}
\begin{equation}\label{CD}
\begin{array}{rl}
(\partial g)(a_0, a_1,\ldots, a_{p+1})=&\displaystyle \sum_{k=0}^{p}(-1)^{\bar a_0+\cdots+\bar a_k} g(a_0, a_1,\ldots,a_ka_{k+1},\ldots, a_{p+1})\\[5mm]
+&(-1)^{\bar a_{p+1}(\bar a_0+\cdots +\bar a_{p}+1)} g(a_{p+1}a_0, a_1, \ldots, a_p)\,.
\end{array}
\end{equation}
Considering, for example, the ground field $k$ as a one-dimensional algebra over itself one readily concludes  that $C^{2n}_{\mathrm{cyc}}(k)\simeq k$ and $C^{2n+1}_{\mathrm{cyc}}(k)=0$. Hence, $HC_{\mathrm{cyc}}^{2n}(k)\simeq k$ and $HC_{\mathrm{cyc}}^{2n+1}(k)=0$.

It follows from the definition that the  cyclic cohomology groups are contravarinat functors of the algebra, so that any algebra homomorphism $h:A\rightarrow  B$ induces a homomorphism $h^\ast: HC^p(B)\rightarrow HC^p(A)$ in cohomology.   

Depending on coefficients some standard operations can be defined on Hochschild and cyclic cohomology. 
Below we discuss two such operations: the {cup product} and the action of derivations.  
The former relates the cohomology  of algebras $A$ and $B$ with that of their tensor product $A\otimes B$ and  is defined as follows.  

Let $M$ and $N$ be bimodules over algebras $A$ and $B$, respectively. Then for any $f\in C^{q}(A,M)$ and $g\in C^{p}(B,N)$ we put 
\begin{equation}\label{cp}
\begin{array}{c}
    (f\sqcup g)(a_1\otimes b_1,\ldots,a_{q+p}\otimes b_{q+p})\\[5mm]
    =(-1)^\epsilon f(a_1,\ldots,a_q)a_{q+1}\cdots a_{q+p}\otimes b_1\cdots b_{q}g(b_{q+1},\ldots, b_{q+p})\,,
    \end{array}
\end{equation}
where  $(-1)^\epsilon$ is the Koszul sign resulting from permutations of $a$'s, $b$'s, and $g$. By definition, $f\sqcup g\in C^{q+p}(A\otimes B, M\otimes N)$. 
The cup product (\ref{cp}) is differentiated by the Hochschild coboundary operator (\ref{HD}) thereby inducing a product in cohomology:\footnote{Not to be confused  with  Gerstenhaber's $\cup$-product on $HH^\bullet(A,A)$.}
\begin{equation}\label{cup-hoh}
    \sqcup : HH^q(A,M)\otimes HH^p(B,N)\rightarrow HH^{q+p}(A\otimes B, M\otimes N)\,.
\end{equation}
If all the cohomology groups $HH^p(B,N^\ast)$ turn out to be finite-dimensional, then  (\ref{cup-hoh}) defines a natural isomorphism
\begin{equation}\label{KunHH}
 HH^n(A\otimes B,M\otimes N)\simeq \bigoplus_{p+q=n} HH^p(A,M)\otimes HH^q(B,N)    
\end{equation}
for any $A$-bimodule $M$. This  is just the dual version of the K\"unneth formula  for Hochschild homology, see \cite[Ch. X, Thm. 7.4]{MacLane}. 

Unlike the Hochschild cohomology,  the cup product
\begin{equation}\label{cup-cyc}
    \sqcup : HC^q(A)\otimes HC^p(B)\rightarrow HH^{q+p}(A\otimes B)
\end{equation}
for cyclic cohomology groups cannot be defined at the level of complexes; one has to multiply cyclic cocycles. 
The explicit formula for this cup product is somewhat cumbersome and we do not present it here. The reader can found it in many places, e.g. \cite[Sec. 4.4.10]{Loday}, \cite[II.1]{Connes:85}.  The cyclic analog of the isomorphism (\ref{KunHH}) is given now by the exact sequence
\begin{equation}\label{KS}
 0\rightarrow  HC^\bullet(A)\bigotimes_{HC^\bullet(k)} HC^\bullet(B)\stackrel{\!\!\sqcup}{\rightarrow}  HC^{\bullet}(A\otimes B){\rightarrow}                                              \mathrm{Tor}^{HC^\bullet(k)}(HC^\bullet (A), HC^\bullet(B))\rightarrow 0
\end{equation}
under the assumption  that all groups $HC^p(B)$ are finite-dimensional, see \cite[Thm. 1]{Kassel1986}.  (The commutative algebra $HC^\bullet(k)$ and its left/right action on cyclic cohomology are defined below.) As is seen, the $\sqcup$-product homomorphism from the tensor product of $HC^\bullet(k)$-modules, being  injective, is not generally surjective, yet it becomes an isomorphism whenever either of the $HC^\bullet(k)$-modules is torsion free.

By way of illustration let us take $B$ to be the matrix algebra $\mathrm{Mat}_n(k)$ viewed as a bimodule over itself. Then $A\otimes B=\mathrm{Mat}_n(A)$
and $M\otimes N=\mathrm{Mat}_n(M)$. Since the algebra $\mathrm{Mat}_n(k)$ is {\it separable} \cite[Sec. 1.2.12]{Loday}, 
\begin{equation}
    HH^\bullet(\mathrm{Mat}_n(k), \mathrm{Mat}_n(k) )\simeq HH^0(\mathrm{Mat}_n(k), \mathrm{Mat}_n(k) )\simeq k\,,
\end{equation}
where the group $HH^0 (\mathrm{Mat}_n(k),\mathrm{Mat}_n(k))$, being isomorphic to the centre of $\mathrm{Mat}_n(k)$, is generated by the unit matrix $1\!\! 1$.  By the K\"unneth formula (\ref{KunHH}),
\begin{equation}
     HH^p(A, M )\simeq  HH^p(\mathrm{Mat}_n(A), \mathrm{Mat}_n(M))\,.
\end{equation}
At the level of cochains the isomorphism is induced by the so-called {\it cotrace map}:  $\mathrm{cotr}(f)=f\sqcup 1\!\!1$ for any $f\in C^{p}(A,M)$. As Rel. (\ref{cp}) suggests, 
\begin{equation}\label{cotrace}
    \mathrm{cotr}(f)(a_1\otimes m_1,\ldots, a_p\otimes m_p)=f(a_1,\ldots, a_p)\otimes m_1\cdots m_p
\end{equation}
for $a_i\otimes m_i\in A\otimes \mathrm{Mat}_n(k)$.

In the case of cyclic $p$-cochains (\ref{cyclic}) the map (\ref{cotrace}) takes the form
\begin{equation}\label{cotrc}
    \mathrm{cotr}(g)(a_0\otimes m_0,\ldots, a_p\otimes m_p)=g(a_0,\ldots, a_p)\mathrm{tr}(m_0\cdots m_p)
\end{equation}
and gives rise to the isomorphism 
\begin{equation}\label{cotrch}
    HC^p(A)\simeq HC^{p}(\mathrm{Mat}_n(A))
\end{equation}
of cyclic cohomology groups.

As was mentioned above  $HC^{2n}(k)\simeq k$. Let $\sigma$ denote the basis $2$-cocyle for $HC^2(k)$ obeying the normalization condition
$\sigma(1,1,1)=1$. Note that for each $k$-algebra $A$, there is the natural isomorphism $A\otimes k\simeq A$. Using this isomorphism and the $2$-cocycle $\sigma$, we can define a homomorphism 
\begin{equation}\label{per}
S:HC^{p}(A)\rightarrow HC^{p+2}(A)
\end{equation}
by setting
\begin{equation}\label{u=S}
Sf=\sigma\sqcup f=f\sqcup\sigma\,,\qquad \forall f\in HC^{p}(A)\,.
\end{equation}
The homomorphism $S$ of degree $2$ is called the {\it periodicity map}. For example, applying $S$ to a $1$-cocycle $\phi$, one obtains
$$
(S\phi)(a_0,a_1,a_2,a_3)
=(-1)^{|a_3|+(|a_2|+|a_3|)(|a_0|+|a_1|)}\phi(a_2a_3a_0,a_1)+(-1)^{|a_1|}\phi(a_0a_1a_2,a_3)\,.
$$
It is instructive to verify the cyclic property (\ref{cyclic}) of the resulting $3$-cocycle $S\phi$. 
If we take $A=k$, then (\ref{per}) makes $HC^\bullet(k)$ into an associative  commutative algebra; in fact $HC^\bullet(k)\simeq k[S]$. This allows one to regard each $k$-vector space $HC^\bullet(A)$ as a bimodule over the $k$-algebra $HC^\bullet(k)$. 

Let $\mathrm{Der}(A)$ denote the space of all derivations of the graded algebra $A$. By definition, homogeneous elements of $\mathrm{Der}(A)$
are homomorphism $D: A\rightarrow A$ obeying the graded Leibniz rule $$D(ab)=(Da)b+(-1)^{|a||D|}a(Db)\,.$$ The derivations are known to form a graded  Lie algebra w.r.t. the commutator. Furthermore, each derivation  $D$ gives rise to a cochain transformation $L_D: C^p_{\mathrm{cyc}}(A)\rightarrow C^p_{\mathrm{cyc}}(A)$ defined by 
\begin{equation}\label{LD}
    (L_D g)(a_0,a_1,\ldots, a_p)=\sum_{k=0}^p(-1)^{\bar{D}(\bar a_0+\cdots+\bar{a}_{k-1})}g(a_0,\ldots, Da_k,\ldots, a_p)\,.
\end{equation}
As usual, this induces a homomorphism $L_D^\ast:  HC^p(A)\rightarrow HC^p(A)$ in cohomology. The induced homomorphism is known to be trivial for {\it inner }
derivations. A similar action of derivations can be defined for Hochschild cohomology as well. 

Each derivation $D$ of an algebra $A$ trivially extends to the derivation $\hat D$ of the tensor product $A\otimes B$ by setting $\hat D(a\otimes b)=Da\otimes b$. This extension appears to be compatible with the cup product of cyclic cocycles in the most natural way: 
\begin{equation}
    L_{\hat D}(f\sqcup g)=L_Df\sqcup g \,.
\end{equation}
As  a result, each $D\in \mathrm{Der}(A)$ generates a homomorphism $L_{\hat D}^\ast : HC^p(A\otimes B)\rightarrow HC^p(A\otimes B)$ in  cohomology.

The fact that the cyclic complex $C^\bullet_{\mathrm{cyc}}(A)$ is a subcomplex of the Hochschild complex $C^\bullet (A,A^\ast)$ gives rise to the long exact
sequence in cohomology  
\begin{equation}\label{CEPS}
\xymatrix{\cdots\ar[r]& HH^p (A,A^\ast)\ar[r]^-B&HC^{p-1}(A)\ar[r]^-S&HC^{p+1}(A)\ar[r]^-I&HH^{p+1}(A,A^\ast)\ar[r]&\cdots}\,.
\end{equation}
The sequence involves the periodicity map (\ref{per}) and is known as Connes' Periodicity Exact Sequences. In many interesting cases it reduces the problem of computation of cyclic cohomology to that of  Hochschild cohomology. The map $I$ is induced by the inclusion $C^\bullet_{\mathrm{cyc}}(A)\rightarrow  C^\bullet (A,A^\ast)$, while the definition of $B$ is more complicated, see \cite{Co}.

Among important applications of cyclic cohomology is computation of the cohomology of the Lie algebra  $gl(A)$ of `big matrices'. By definition, the algebra $gl(A)$ consists of infinite matrices with only finitely many entries different from zero. Formally, it is defined through the inductive limit $gl(A)=\lim\limits_{\rightarrow}gl_n(A)$ corresponding to the natural inclusions $gl_n(A)\subset gl_{n+1}(A)$ (an $n\times n$-matrix is augmented by zeros). 
A precise  relationship between the cohomology of the Lie algebra of matrices and cyclic  cohomology is established by the  Tsygan--Loday--Quillen theorem \cite{Tsygan_1983}, \cite{LQ}.  

In order to formulate this theorem precisely we need some more terminology. Recall that the Chevalley--Eilenberg cochain complex of a graded Lie algebra $L=\bigoplus L^n$ consists of the sequence of groups $C^p(L)=\mathrm{Hom}(\Lambda^pL,k)$ endowed with a coboundary operator  $\delta : C^p(L)\rightarrow C^{p+1}(L)$. By definition, 
 \begin{equation}\label{transp}
     c(a_1,\ldots,a_k,a_{k+1},\ldots, a_{p})=(-1)^{\bar a_k\bar a_{k+1}}c(a_1,\ldots, a_{k+1}, a_k,\ldots,a_p)
 \end{equation}
 and
\begin{equation}\label{ChE}
\displaystyle (\delta c)(a_1, \ldots,  a_{p+1})= \sum_{1\leq k<l\leq p+1}(-1)^{\epsilon_{kl}}c([a_k,a_l],  a_1, \ldots,\hat{a}_k,\ldots,\hat{a}_l,\ldots, a_{p+1})\,,
\end{equation}
where 
$$
\epsilon_{kl}=\bar a_k+\bar a_k(\bar a_1+\cdots +\bar a_{k-1})+\bar a_l(\bar a_1+\cdots+\bar a_{k-1}+\bar a_{k+1}+\cdots + \bar a_{l-1})\,.
$$
As usual the hats indicate omitting of the corresponding arguments.  By definition, the Lie algebra cohomology with trivial coefficients is the cohomology of the Chevalley--Eilenberg complex above.  The corresponding cohomology groups are denoted by $H^\bullet (L)$. 

Viewing  the cochains (\ref{transp}) as exterior forms on the graded vector space $L$ we can make $C^\bullet(L)$ into  a differential graded algebra w.r.t. the exterior product of forms and the differential (\ref{ChE}).  As a result, the cohomology space $H^{\bullet}(L)$ acquires the structure of graded commutative  algebra:
\begin{equation}\label{d-prod}
[c_1]\cdot [c_2]=[c_1\cdot c_2 ]=(-1)^{\bar c_1\bar c_2}[c_2\cdot  c_1]\,,
\end{equation}
$c_{i}$ being cocycles representing the cohomology classes $[c_{i}]$. Denoting $H_+^\bullet (L)=\bigoplus_{p>0} H^p(L)$, we define the space of {\it indecomposable} elements of the algebra 
$H^{\bullet}(L)$
as the quotient  $$\mathrm {Indec}\, H^{\bullet}(L)= H^\bullet (L)/H_+^\bullet (L)\cdot H_+^\bullet (L)\,.$$

Given an associative algebra $A$, denote by $L(A)$ the associated Lie algebra with the Lie bracket given by the commutator in $A$.  Restricting a cyclic  $p$-cochain   $f: A^{\otimes (p+1)}\rightarrow k$
to the subspace of anti-symmetric chains $\Lambda^{(p+1)} A\subset A[-1]^{\otimes (p+1)}$ gives then a cochain of the Chevalley--Eilenberg complex associated to the Lie algebra $L(A)$. Moreover, the restriction appears to be a cochain map, so that 
$$
(\partial f)(a_0\wedge a_1\wedge \ldots\wedge a_{p+1})=(\delta f)(a_0\wedge a_1\wedge\ldots\wedge a_{p+1})
$$
for any $f\in C^p(A)$. As a result we have a homomorphism of cohomology groups 
\begin{equation}
    \varepsilon^\ast_p: HC^p(A)\rightarrow H^p (L(A))
\end{equation}
induced by the inclusion $\varepsilon_{p}: \Lambda^{p+1}A\rightarrow A[-1]^{\otimes (p+1)}$. This is known as an {\it antisymmetrization map} \cite[Sec. 1.3.4]{Loday}. 

Composing now the antisymmetrization map with the cotrace (\ref{cotrace}),  one can define a homomorphism from the cyclic cohomology of $A$ to the cohomology of the Lie algebra $gl_n(A)$:
\begin{equation}\label{TLQ}
   \varphi^\ast=\varepsilon^\ast\circ \mathrm{cotr}^\ast: HC^{\bullet-1} (A)\rightarrow H^{\bullet} (gl_n(A))\,.
\end{equation}
At the level of cocycles the homomorphism is given by the formula\footnote{There is no need to antisymmetrise all $p+1$ arguments due to cyclicity of $g$ and $\mathrm{tr}$.}
\begin{equation}
    \varphi(g)(a_0\otimes m_0,\ldots, a_p\otimes m_p)=\sum_{\sigma\in S_p}(-1)^{\epsilon_\sigma}g(a_0,a_{\sigma(1)},\ldots, a_{\sigma(p)})\mathrm{tr}(m_0m_{\sigma(1)}\cdots m_{\sigma(p)})\,.
\end{equation}
Here $(-1)^{\epsilon_\sigma}$ is the Koszul sign caused  by elementary transpositions (\ref{transp}) of arguments.

Now we are in position to state the aforementioned  Tsygan--Loday--Quillen theorem.

\begin{theorem}\label{TA1}
{\it The image of the map (\ref{TLQ}) lies in the indecomposable part of the algebra $H^\bullet (gl_n(A))$ and induces an isomorphism
$$
HC^{p-1}(A)\simeq \mathrm{Indec}\, H^{p}(gl_n(A))
$$
for all $n\geq p$. As an exterior algebra, $H^\bullet(gl(A))$ is freely generated by the graded vector space $HC^{\bullet-1}(A)$. }
\end{theorem}

In other words, the cohomology group $H^p (gl_n(A))$ does not depend on the size of matrices provided it is large enough.

\section{Cohomology of  Weyl algebras and their smash products}\label{B2}

The polynomial Weyl algebra $A_n$ over $\mathbb{C}$ is a unital algebra on $2n$ generators $q^i$ and $p_j$ subject to Heisenberg's commutation relations
\begin{equation}\label{qp}
    [q^i,q^j]=0\,,\qquad [p_i,p_j]=0\,,\qquad [q^i,p_j]=\delta^i_j1\,.
\end{equation}
It is known to be a simple Noetherian domain  with  a $k$-basis  consisting of the ordered monomials in $q$'s and $p$'s, see e.g. \cite{coutinho_1995}.

The Hochschild cohomology groups of Weyl algebras are known  for various coefficients.  For instance, applying the Koszul resolution (see e.g. \cite{AFLS}, \cite{Pinczon}) yields  
\begin{equation}\label{HH0}
    HH^\bullet(A_n,A_n)\simeq HH^0(A_n,A_n)\simeq \mathbb{C}
\end{equation}
and 
\begin{equation}\label{dual}
HH^p(A_{n}, M)\simeq HH^{2n-p}(A_n, M^\ast)\,,\qquad HH^p(A_n, M)=0 \qquad \forall p>2n
\end{equation}
for any bimodule $M$. Among other things, the isomorphisms (\ref{HH0}) mean that  all Weyl algebras are rigid, have only inner derivations, and their center is  generated by the unit element. Combining (\ref{HH0}) and  (\ref{dual}), one also obtains 
\begin{equation}
       HH^\bullet(A_n,A^\ast_n)\simeq HH^{2n}(A_n,A^\ast_n)\simeq \mathbb{C}\,.
\end{equation}
An explicit formula for a non-trivial $2n$-cocycle $\tau_{2n}$ generating the group $HH^{2n}(A_n,A_n^\ast)$ was found in the 2005 paper \cite{FFS} by Feigin, Felder, and Shoikhet.  It was derived as a consequence of Shoikhet's proof \cite{Shoikhet:2000gw} of Tsygan's formality conjecture. It should be noted that fifteen years earlier Vasiliev had found an explicit expression for $\tau_2$ in the context of $4d$ HSGRA \cite{Vasiliev:1988sa}.  
In order to present the cocycle $\tau_{2}$ explicitly  it is convenient to identify the elements of $A_1$ with the polynomials $a(q,p)$ in (commuting) indeterminates  $q$ and $p$ endowed with the Weyl--Moyal star-product 
\begin{equation}
 a\star b =m \exp \alpha (a\otimes b)\,,     
\end{equation}
where 
\begin{equation}
    \alpha=\frac12\left(\frac{\partial}{\partial p}\otimes \frac{\partial}{\partial q}-\frac{\partial}{\partial q}\otimes \frac{\partial}{\partial p}\right)\in \mathrm{End} (A_1\otimes A_1)
\end{equation}
and $m(a\otimes b)=ab$. We also introduce the maps $\alpha_{01}$, $\alpha_{12}$, $\alpha_{02}$:
\begin{equation}
    \alpha_{01}(a_0\otimes a_1\otimes a_2) =\frac12 \left(\frac{\partial a_0}{\partial p}\otimes  \frac{\partial a_1}{\partial q}\otimes   a_2
    -\frac{\partial a_0}{\partial q}\otimes  \frac{\partial a_1}{\partial p}\otimes   a_3\right)\in \mathrm{End}(A_1\otimes A_1\otimes A_1)
\end{equation}
and similarly for $\alpha_{12}$ and $\alpha_{02}$. 
 Finally, we define the homomorphism $\mu: A_1\otimes A_1\otimes A_1\rightarrow \mathbb{C}$ by
\begin{equation}
    \mu(a_0\otimes a_1\otimes a_2)=a_0(0)a_1(0)a_2(0)\,.
\end{equation}
Here $a(0)$ is the constant term of the polynomial $a(q,p)$. Now the expression for the $2$-cocycle reads 
\begin{equation}\label{tau}
    \tau_2(a_0,a_1,a_2)=\mu\circ F(\alpha_{01},\alpha_{12},\alpha_{02})(a_0\otimes a_1\otimes a_2)\,,
\end{equation}
where the operator $F(\alpha_{01},\alpha_{12},\alpha_{02})\in \mathrm{End}(A_1\otimes A_1\otimes A_1)$ is determined by the following entire analytic function of three variables: 
\begin{equation}
  F (x,y,z)=\frac{(z^2-y^2)e^{z+y-x}+(y^2-x^2)e^{y+x-z}+(x^2-z^2)e^{x+z-y}}{(x-z)(z-y)(y-x)}\,.
\end{equation}
Clearly, the action of the operator  $F$ is well defined on polynomials. Unlike the Hochschild $2$-cocycles of Refs. \cite{FFS} and \cite{Vasiliev:1988sa}, the cocycle (\ref{tau}) enjoys cyclic invariance,
\begin{equation}
    \tau_2(a_0,a_1,a_2)=\tau_2(a_2,a_0,a_1)\,,
\end{equation}
thereby generating the cyclic cohomology group $HC^2(A_1)\simeq \mathbb{C}$.

In the context of HSGRA,  Weyl algebras usually appear  in  
smash products with finite groups of their automorphisms. Recall that, given an associative $k$-algebra $A$ and a finite group $G\subset \mathrm{Aut}(A)$,  
the {\it skew group algebra} $A\rtimes G$ (aka {\it smash product algebra}) is defined to be the $k$-vector space $A\otimes k[G]$ endowed with the product 
\begin{equation}\label{smpr}
    (a_1\otimes g_1)(a_2\otimes g_2)=a_1a_2^{g_1}\otimes g_1g_2\,.
\end{equation}
Here $a^g$ denotes the action of $g\in G$ on $a\in A$. 

Notice that the automorphism group of $A_n$ contains a subgroup $Sp_{2n}(\mathbb{C})$ acting by linear transformations on the $2n$-dimensional complex space $V$ spanned by the generators $q$'s and $p$'s.  If $G$ is a finite subgroup of $Sp_{2n}( k)$, then  $g^{|G|}=e$ for any $g\in G$ and the action of $g$ is diagonalizable in $V$.  Denote by $2\mu_g$ the multiplicity of the eigenvalue $1$ of the operator $g: V\rightarrow V$. Notice that  $\mu_g=\mu_{hgh^{-1}}$ and the set of all element $g\in G$ with $2\mu_g=p$ is invariant under conjugation.
The next theorem is due to  Alev, Farinati, Lambre, and Solotar \cite{AFLS} (see also \cite{Pinczon}).

\begin{theorem}\label{ThAFLS}
Let $n_p(G)$ denote the number of conjugacy classes  of elements $g\in G $ with $2\mu_g=p$, then 
$$
\dim HH^{2n-p}(A_n\rtimes G, A_n\rtimes G)=\dim HH^{p}\big(A_n\rtimes G, (A_n\rtimes G)^\ast\big)=n_p(G)\,.
$$
\end{theorem}

As an example, consider the most simple skew group algebra $\mathcal{A}=A_1\rtimes \mathbb{Z}_2$. Here the group $\mathbb{Z}_2=\{e, \varkappa\}$ acts on $A_1$ by the involution 
\begin{equation}\label{k}
    q^\varkappa=-q\,,\qquad p^\varkappa=-p\,.
\end{equation}
Since $2\mu_e=2$ and $2\mu_\varkappa=0$, all non-trivial groups of Hochschild cohomology mentioned in the theorem above  are 
\begin{equation}\label{AA02}
   HH^2(\mathcal{A},\mathcal{A})\simeq HH^0(\mathcal{A},\mathcal{A}^\ast)\simeq \mathbb{C}\simeq HH^2(\mathcal{A},\mathcal{A}^\ast)\simeq HH^0(\mathcal{A},\mathcal{A})\,.
\end{equation}
In particular, we see that the algebra $\mathcal{A}$ admits a unique non-trivial deformation. 

Since $HH^n(\mathcal{A},\mathcal{A}^\ast)=0$ for $n>2$ one  can easily  see from Connes' exact sequence (\ref{CEPS})  that 
\begin{equation}\label{HCA}
HC^{2k-1}(\mathcal{A})=0\,,\qquad 
     HC^0(\mathcal{A})\simeq \mathbb{C}\,,\qquad
HC^{2k}(\mathcal{A})\simeq\mathbb{C}^2\,,\qquad
k=1,2,3,\ldots\,,
\end{equation}
see \cite[Appendix B.4]{sharapov2020characteristic} for details. Furthermore, the groups $HC^\bullet(\mathcal{A})$ form a free $HC^\bullet(\mathbb{C})$-module generated (via $S$) by the pair of elements $\phi_0\in HC^0(\mathcal{A})$ and $\phi_2\in HC^2(\mathcal{A})$.  The first one is the trace $\phi_0=\mathrm{Tr}:\mathcal{A}\rightarrow \mathbb{C}$ defined by the projection onto the one-dimensional subspace $\mathbb{C}(1\otimes \varkappa)\subset \mathcal{A}$. We can normalize it by setting $\mathrm{Tr}(1\otimes \varkappa)=1$. The trace is known to give rise to a non-degenerate inner product on $\mathcal{A}$ defined by $(a,b)=\mathrm{Tr}(ab)$. This inner product allows one to identify the $\mathcal{A}$-bimodules $\mathcal{A}$ and $\mathcal{A}^\ast$. 
An explicit expression for a  $2$-cocycle representing the class $\phi_2$ is given in \cite[Eq. (4.15)]{sharapov2020characteristic}.
We refer to $\phi_0$ and $\phi_2$ as {\it primary classes} of cyclic cohomology.
All the other classes in (\ref{HCA}) are obtained from these two by successive application of the periodicity operator:
$
S^n\phi_0\in HC^{2n}(\mathcal{A})$, $S^n\phi_2\in HC^{2n+2}(\mathcal{A})
$.

As discussed in Sec. \ref{sec:HSGRA}, the skew group algebra $\mathcal{A}=A_1\rtimes \mathbb{Z}_2$ is a building block for the extended  higher spin algebra $\mathfrak{A}$ 
underlying  $4d$ HSGRA. The latter is given  by the tensor square   
\begin{equation}\label{AA}
\mathfrak{A}=\mathcal{A}\otimes\mathcal{A}=A_2\rtimes (\mathbb{Z}_2\times \mathbb{Z}_2)\,.
\end{equation}
One may also regard it as the smash product of the Weyl algebra $A_2$ and the Klein four-group $\mathbb{Z}_2\times\mathbb{Z}_2$ acting on $A_2$ by symplectic reflections. By the K\"unneth formula (\ref{KunHH}) for  Hochschild cohomology
$$
HH^n(\mathfrak{A}, \mathfrak{A})=\bigoplus_{q+p=n}HH^q(\mathcal{A},\mathcal{A})\otimes HH^p(\mathcal{A},\mathcal{A})\,,
$$
whence
\begin{equation}\label{HHAA}
    HH^0(\mathfrak{A},\mathfrak{A})\simeq \mathbb{C}\,,\qquad HH^2(\mathfrak{A},\mathfrak{A})\simeq \mathbb{C}^2\,,\qquad HH^4(\mathfrak{A},\mathfrak{A})\simeq\mathbb{C}\,,
\end{equation}
and the other groups vanish. Again, the standard interpretations of the second and third groups of Hochschild cohomology suggest that the algebra $\mathfrak{A}$ admits a two-parameter family of formal  deformations. A representative cocycle generating  the group $HH^0(\mathfrak{A},\mathfrak{A})$ defines a non-degenerate trace on $\mathfrak{A}$, which implies the isomorphism $\mathfrak{A}\simeq\mathfrak{A}^\ast$. Hence,  
\begin{equation}\label{HHAA*}
     HH^0(\mathfrak{A},\mathfrak{A}^\ast)\simeq \mathbb{C}\,,\qquad HH^2(\mathfrak{A},\mathfrak{A}^\ast)\simeq \mathbb{C}^2\,,\qquad HH^4(\mathfrak{A},\mathfrak{A}^\ast)\simeq\mathbb{C}\,.
\end{equation}
We could also arrive at these isomorphisms by the direct application of Theorem \ref{ThAFLS}.

Now the cyclic cohomology of $\mathfrak{A}$ can  be computed by means of  the  Connes exact sequence (\ref{CEPS}) or by 
the K\"unneth formula 
\begin{equation}\label{HCHC}
    HC^\bullet(\mathcal{A})\bigotimes_{HC^\bullet(\mathbb{C})} HC^\bullet(\mathcal{A})\simeq HC^\bullet(\mathfrak{A})\,.
\end{equation}
In either approach  one finds 
\begin{equation}\label{HHHH}
\begin{array}{lll}
     HC^0(\mathfrak{A})\simeq\mathbb{C}\,,&\qquad HC^2(\mathfrak{A})\simeq \mathbb{C}^3\,, &\qquad  HC^{4+2k}(\mathfrak{A})\simeq \mathbb{C}^4\,, \\[5mm]
       HC^{2k+1}(\mathfrak{A})=0\,,&\qquad  k=0,1,2\ldots\,. &
\end{array}
\end{equation}
For detail, see \cite[Appendix B.5]{sharapov2020characteristic}. As we already know $HC^\bullet(\mathcal{A})$ is a free $HC^\bullet(\mathbb{C})$-module of rank two generated by the primary cohomology classes $\phi_0$  and $\phi_2$ in degrees $0$ and $2$. Denoting by $\bar \phi_0$ and $\bar \phi_2$ the same cohomology classes for the second copy of $\mathcal{A}$ in (\ref{HCHC}), we see that 
 $HC^\bullet(\mathfrak{A})$ is a rank-four $HC^\bullet(\mathbb{C})$-module freely generated by the cup products
\begin{equation}\label{prime}
   \Phi_0=\phi_0\sqcup\bar\phi_0\,, \qquad \Phi_2=\phi_0\sqcup\bar\phi_2\,, \qquad \bar{\Phi}_2=\phi_2\sqcup\bar\phi_0\,, \qquad \Phi_4=\phi_2\sqcup\bar\phi_2\,.
\end{equation}
These are the primary cocycles of $HC^\bullet(\mathfrak{A})$.

\section{Cohomology of Grassmann algebras}\label{CGA}

Let $\Lambda_n$ denote the Grassmann algebra over $\mathbb{C}$ on $n$ generators $\theta^1,\ldots,\theta^n$ subject to the relations
$$
\theta^i\theta^j=-\theta^j\theta^i\,.
$$
As the algebra $\Lambda_n$ is supercommutative one concludes immediately that $$HH^0(\Lambda_n, \Lambda_n)=Z(\Lambda_n)= \Lambda_n\quad \mbox{and}\quad HH^1(\Lambda_n, \Lambda_n)=\mathrm{Der}(\Lambda_n)\,.$$ Geometrically, one can think of $\Lambda_n$ as the algebra of smooth functions on a  supermanifold $\mathcal{G}$ with odd coordinates $\theta$'s.  Then the Lie superalgebra  
$\mathrm{Der}(\Lambda_n)$ of derivations of $\Lambda_n$ can be identified with the 
algebra of smooth vector fields on $\mathcal{G}$ w.r.t. the supercommutator.   The last fact is a particular manifestation of the   Hochschild--Kostant--Rosenberg theorem  for smooth graded-commutative algebras \cite[Sec. 5.4.5]{Loday}. In the case under consideration it states the isomorphism 
\begin{equation}\label{HKR}
    HH^p(\Lambda_n,\Lambda_n)\simeq \Lambda^p \big(\mathrm{Der}(\Lambda_n)\big)\,,
\end{equation}
the r.h.s. being the space of polyvector fields on $\mathcal{G}$.  In terms of the odd coordinates $\theta^i$, each $p$-vector $\phi$ is given by 
\begin{equation}\label{PV}
   \phi=\phi^{i_1\cdots i_p}(\theta)\partial_{i_1}\wedge \cdots \wedge \partial_{i_p}\,, \qquad \partial_i\equiv \frac{\partial}{\partial\theta^i}\,,
\end{equation}
where the coefficients $\phi^{i_1\cdots i_p}\in \Lambda_n$ are totally symmetric in permutation of indices.

The cyclic cohomology of $\Lambda_n$ is also well known. As was shown by Kassel \cite[Prop. 2]{Kassel1986}
\begin{equation}\label{HCV}
HC^p(\Lambda_n)\simeq HC^p(\mathbb{C})\oplus V^p\,,
\end{equation}
where $HC^{2k}(\mathbb{C})\simeq\mathbb{C}$, $HC^{2k+1}(\mathbb{C})=0$, and the complex dimensions of the vector spaces $V^p$ are encoded by the Poincar\'e series
\begin{equation}
    \mathrm{ch}(V)=\sum_{p=0}^\infty t^p\dim V^p=\frac{2^n-(1-t)^n}{(1+t)(1-t)^n}\,.
\end{equation}
In particular, $\dim HC^0(\Lambda_n)=\dim \Lambda_n=2^n$. Actually, $\Lambda_n$ is a bialgebra with the standard coproduct $$\Delta \theta^i=1\otimes \theta^i+\theta^i\otimes 1\,.$$
This enables us to equip the complex vector space $HC^\bullet(\Lambda_n)$ with the structure of a graded-commutative algebra.  Indeed, composing the $k$-algebra homomorphism  $\Delta: \Lambda_n\rightarrow \Lambda_n\otimes \Lambda_n$ with the cup product (\ref{cup-cyc}), we get a new product\footnote{Recall that cyclic cohomology is a contravariant functor of algebra.}
\begin{equation}\label{dc}
    \Delta^\ast\circ \sqcup: HC^\bullet(\Lambda_n)\otimes HC^\bullet(\Lambda_n)\rightarrow HC^\bullet (\Lambda_n)
\end{equation}
that makes $HC^\bullet(\Lambda_n)$ into a graded associative algebra.  The algebra  $HC^\bullet(\Lambda_n)$ contains $HC^\bullet(\mathbb{C})$ as a subalgebra and the action of  the periodicity map $S$ on $HC^\bullet(\Lambda_n)$ is induced by the inclusion $HC^2(\mathbb{C})\subset HC^\bullet(\Lambda_n)$. The map $S$ generates the first summand in (\ref{HCV}) and acts trivially on  the $V^p$'s. 

As was first shown in \cite{COQUEREAUX1995333} cyclic cocycles representing the spaces $ V^p$ admit a nice interpretation in terms of the polyvector fields (\ref{PV}). See \cite{grensing2004berezin} for subsequent discussions. Specifically, to each $p$-vector $\phi$ one first associates a Hochschild $p$-cocycle by the rule 
\begin{equation}\label{phi}
    \phi(a_0,a_1,\ldots, a_p)=\int a_0\phi^{i_1\cdots i_p}\partial_{i_1}a_1\cdots \partial_{i_p}a_p\in C^p(\Lambda_n,\Lambda_n^\ast)\,.
\end{equation}
Here $a_k\in \Lambda_n$ and the integral sign stands for the Berezin integral on the Grassmann algebra. One can easily check that the property $\partial \phi=0$ is automatically satisfied for any $\phi$.  The cyclicity condition
\begin{equation}
    \phi(a_0,a_1,\ldots, a_p)=(-1)^{(|a_0|+1)(|a_1|+\cdots+|a_p|+p)} \phi(a_1,\ldots, a_n, a_0)\,,
\end{equation}
however, requires the $p$-vector $\phi$ to be divergence-free, i.e.,
\begin{equation}\label{div}
    \partial_{i_1}\phi^{i_1\cdots i_p}(\theta)=0\,.
\end{equation}
In order to analyse the last condition, it is convenient to identify the polyvector fields with the elements of the Berezin algebra $\mathcal{B}_n$ on $n$ anti-commuting generators $\theta^i$ and the same number of commuting generators $y_i$:
$$
\theta^i\theta^j=-\theta^j\theta^i\,,\qquad y_iy_j=y_jy_i\,,\qquad \theta^iy_j=y_j\theta^i\,.
$$
Clearly, there is the one-to-one correspondence 
\begin{equation}
    \phi^{i_1\cdots i_p}(\theta)\partial_{i_1}\wedge \cdots \wedge \partial_{i_p}\quad \Longleftrightarrow \quad \phi(\theta,y)=
    \phi^{i_1\cdots i_p}(\theta)y_{i_1}\cdots y_{i_p}\,.
\end{equation}
Upon this identification the operator of divergence (\ref{div}) passes to the odd Laplace operator
\begin{equation}
    \Delta =\frac{\partial^2}{\partial\theta^i\partial y_i}\,.
\end{equation}
Since $ \Delta^2=0$, we may say that the divergence-free polyvectors correspond to the cocycles (perhaps trivial) of the odd Laplacian:
$$
\Delta \phi=0\,.
$$
The non-trivial cocycle can easily be computed with the help of the homotopy operator $h=\theta^iy_i\cdot$. Clearly,
\begin{equation}
    \Delta h+h\Delta= n+y_i\frac{\partial}{\partial y_i}-\theta^i\frac{\partial}{\partial \theta^i}\,.
\end{equation}
In view of this relation there is the only (up to a multiplicative constant) non-trivial $\Delta$-cocyle 
\begin{equation}\label{C}
C=\theta^1\cdots\theta^n\,.
\end{equation}
All other cocycles are of the form $\phi=\Delta \psi$. 

Since $\Delta$ is a second-order differential operator, the product of two $\Delta$-cocyles is not a cocycle in general. Instead, the $\Delta$-cocycles form a graded Lie algebra w.r.t. the bracket 
\begin{equation}\label{C13}
(\phi_1,\phi_2)=\Delta(\phi_1\phi_2)-(\Delta\phi_1)\phi_2-(-1)^{|\phi_1|}\phi_1\Delta\phi_2\,.    
\end{equation}
This is just the Schouten bracket on polyvector fields.  
It follows from the definition that the bracket is  differentiated by $\Delta$; and hence, it maps $\Delta$-cocycles to $\Delta$-cocycles.  Taken together the odd Laplacian and the bracket endow $\mathcal{B}_n$ with the structure of Batalin--Vilkovisky algebra. 
 
 One can use the $0$-vector (\ref{C}) to write non-trivial $2m$-cocycles representing the first direct summand in (\ref{HCV})  
 in terms of the Berezin integral, namely, 
\begin{equation}\label{c2m}
    c_{2m}(a_0,\ldots,a_{2m})=\int C a_0a_1\cdots a_{2m}=a_0a_1\cdots a_{2m}|_{\theta=0}\,. 
\end{equation}
Notice that the first member of this family, $c_0$,  also comes from  (\ref{phi}).

Let us now specify the above constructions to the case of $\Lambda_2$. The corresponding Berezin algebra $\mathcal{B}_2$ is generated by $\theta$, $\bar \theta$,
$y$, and $\bar y$. 
As a complex vector space, $\mathcal{B}_2$ is spanned by the monomials 
$$
y^n\bar y^m\,, \qquad \theta y^n\bar y^m,\qquad \bar \theta y^n\bar y^m,\qquad \theta\bar\theta y^n\bar y^m\,.
$$
Applying $\Delta$ to them, we conclude that the space of $\Delta$-cocycles is generated by
\begin{equation}\label{AB}
    A_{nm}=n\bar\theta y^{n-1}\bar y^m-m\theta y^n\bar y^{m-1}\,,\qquad B_{nm}=y^n\bar y^m\,,\qquad C=\theta\bar\theta
\end{equation}
for $m,n=0,1,2,\ldots$. 
They form the following Lie  superalgebra:
$$
(C,C)=0\,,\qquad (C,A_{nm})=0\,,\qquad (C,B_{nm})=A_{nm}\,,\qquad (B_{nm},B_{kl})=0\,,
$$
$$(A_{nm}, B_{kl})=(ln-mk)B_{n+k-1,m+l-1}\,, \qquad (A_{nm}, A_{kl})=(ln-mk)A_{n+k-1,m+l-1}\,.
$$

Thus, we are lead to conclude that the cyclic cohomology of $\Lambda_2$ is generated by the $\Delta$-exact polynomials (\ref{AB}) together with the cyclic cocycles  (\ref{c2m}) associated with the non-trivial $\Delta$-cocycle $C=\theta\bar\theta$.

\section{Presymplectic structures in  \texorpdfstring{$\boldsymbol{4d}$}{4d} HSGRA}\label{app:D}

We begin with an algebraic reformulation of the problem. At the free level the homological vector field underlying $4d$ HSGRA comes from the graded Lie algebra $\mathcal{G}$ described  in items (1) - (4) of Sec. \ref{sec:HSGRA}.
The construction of the algebra $\mathcal{G}$ also admits the following geometric interpretation.  Starting from the Lie algebra 
$\mathcal{G}_0=gl(\mathfrak{A})$ of `big matrices' associated with the extended higher spin algebra $\mathfrak{A}$, we can define the canonical homological vector field 
\begin{equation}\label{D1}
Q=\frac12\omega^A\omega^Bf_{AB}^C\frac{\partial}{\partial \omega^C}
\end{equation}
on the $\mathbb{N}$-graded manifold $N=\mathcal{G}_0[1]$, as explained in Example \ref{E22}.  Here $\omega^A$ are global coordinates on $N$ associated with a basis $\{e_A\}\subset \mathcal{G}_0$ wherein the commutation relations take the form $[e_A,e_B]=f_{AB}^Ce_C$. Next, following the recipe of Example \ref{E23}, we construct the first prolongation  of  (\ref{D1}) to the total space of the shifted tangent bundle $\mathcal{N}=T[-1]N$. This is given by the homological  vector field
\begin{equation}\label{D2}
    \mathcal{Q}=\frac12\omega^A\omega^Bf_{AB}^C\frac{\partial}{\partial \omega^C}+\omega^AC^Bf^K_{AB}\frac{\partial}{\partial C^K}\,,
\end{equation}
 $C^A$ being linear coordinates in the  tangent spaces. By definition,  $|C^A|=0$. Since $\mathcal{Q}$ is quadratic, it defines and is defined by some graded Lie algebra $\mathcal{G}=\mathcal{G}_{-1}\oplus \mathcal{G}_0$, so that $\mathcal{N}=\mathcal{G}[1]$. One can regard $\mathcal{G}$ as the trivial extension of the Lie algebra $\mathcal{G}_0$ with its adjoint module  put in degree $-1$ (adjoint extension). It is the homological vector field (\ref{D2})
 that determines the free equations of $4d$ HSGRA. One can iterate this construction  to produce higher prolongations of the homological vector field (\ref{D1}). Of particular interest to us is the second prolongation of $Q$ to the homological vector field $\mQ$ on the total space $\mN$ of the tangent bundle\footnote{This can also be seen as a  {\it double tangent bundle} of $N$. The construction of the homological vector field $\mQ$ by $\mathcal{Q}$ is a particular example of a {\it tangent prolongation  Lie algebroid} \cite[Ch. 9]{mackenzie_2005}. } $T[1]\mathcal{N}=T[1]T[-1]N$. From the algebraic viewpoint
 this yields the double adjoint extension $\mG=\mathcal{G}_{-1}\oplus \mathcal{G}_{0}\oplus \mathcal{G}_0\oplus\mathcal{G}_1$ of the Lie algebra $\mathcal{G}_0$. Geometrically, we  can identify the algebra of smooth functions on $\mN=\mathcal{\mG}[1]$ with the algebra of exterior differential
 forms $\Lambda(\mathcal{N})$, see Example \ref{E21}. Upon such identification the $\mathcal{Q}$-invariant differential forms on $\mathcal{N}$ correspond to $\mQ$-invariant functions on $\mN$ and vice versa. To emphasise this correspondence we  denote the linear coordinates  in the tangent spaces of $T[1]\mathcal{N}$ by $\delta \omega^A$ and $\delta C^A$. Then
 \begin{equation}\label{Deq}
    C^\infty (\mN)\ni f(\omega, C,
 \delta \omega, \delta C)\quad \Longleftrightarrow\quad f(\omega, C,
 d \omega, d C)\in \Lambda^\bullet(\mathcal{N})\,.
 \end{equation}
 It should be pointed out that $|\delta \omega^A|=2$ and $|\delta C^A|=1$, while $|d \omega^A|=1$ and $|dC^A|=0$. Therefore, equivalence (\ref{Deq}) does not mean the equality of $\mathbb{N}$-degree, if one regards the symbol $\delta$ as 
 the `exterior differential' of the coordinates $\omega^A$ and $C^A$.   

As an intermediate  summary of our discussion we state the following isomorphisms of cohomology groups: 
\begin{equation}
    H^\bullet(L_{\mathcal{Q}},\Lambda(\mathcal{N}))\simeq  H^\bullet(\mQ,C^\infty(\mN))\simeq H^\bullet(\mG)\,.
\end{equation}
This reduces the classification of non-trivial presymplectic structures on the $NQ$-manifold $(\mathcal{N}, \mathcal{Q})$ to the computation of 
certain Lie algebra cohomology groups with trivial coefficients. Notice that the form degree on the left induces an additional grading in the Lie algebra cohomology  groups $H^\bullet(\mG)$, which will be introduced in a moment under the name of {\it weight}.  Looking for $\mathcal{Q}$-invariant presymplectic structures on $\mathcal{N}$, we are thus interested in certain elements of $H^\bullet(\mG)$ of weight two. 

The next step is to reinterpret the Lie algebra $\mG$ -- the double adjoint extension of $\mathcal{G}_0=gl(\mathfrak{A})$ -- in terms of the underlying associative algebra $\mathfrak{A}$. A simple observation is that $\mG\simeq gl(\mathfrak{A}\otimes \Lambda_2)$, where $\Lambda_2$ is the Grassmann algebra on two odd generators $\theta$ and $\bar\theta$. For our purposes, it is convenient to prescribe them the following $\mathbb{Z}$-degrees:
\begin{equation}
|\theta|=-1\,, \qquad |\bar\theta|=1\,.
\end{equation}
Since 
\begin{equation}
    \theta^2=0\,,\qquad \bar\theta{}^2=0\,,\qquad \theta\bar\theta+\bar\theta\theta=0\,,
\end{equation}
the general element of $gl(\mathfrak{A}\otimes \Lambda_2)$  has the form
\begin{equation}\label{D6}
    f=f_{0}+\theta f_{-1}+\bar\theta f_{1}+\bar\theta\theta \bar f_0\,,\qquad f_i\in gl(\mathfrak{A})\,.
\end{equation}
It is easy to see that the commutation relations in $gl(\mathfrak{A}\otimes \Lambda_2)$ coincide exactly with those in $\mG=\mathcal{G}_{-1} \oplus \mathcal{G}_0\oplus \mathcal{G}_0\oplus\mathcal{G}_1$ if we set $f_0,\bar f_0\in \mathcal{G}_0$, $f_{-1}\in \mathcal{G}_{-1}$, and $f_1\in \mathcal{G}_1$.
On passing to field theory, the element (\ref{D6}) is promoted to the `superfield'
\begin{equation}\label{D7}
    \mf(\theta,\bar\theta)=\momega +\theta \mC+\bar\theta \delta\momega+\bar\theta\theta \delta\mC\,,
\end{equation}
which accommodates the zero- and one-form fields $\mC$ and $\momega$ together with their variational differentials $\delta\mC$  and $\delta\momega$. 

By Theorem \ref{TA1}, $H^\bullet (gl(\mathfrak{A}\otimes \Lambda_2))$ is a graded associative algebra freely generated by the elements of the subspace 
\begin{equation}
    \mathrm{Indec}\, H^\bullet(gl(\mathfrak{A}\otimes \Lambda_2))\simeq HC^{\bullet-1}(\mathfrak{A}\otimes \Lambda_2)\,.
\end{equation}
We also know that the  cyclic cohomology groups $HC^\bullet(\mathfrak{A})$  constitute a free 
$HC^\bullet(\mathbb{C})$-module  generated by the four primary classes (\ref{prime}).
This implies the K\"unneth isomorphism
\begin{equation}\label{D9}
    HC^\bullet(\mathfrak{A}\otimes \Lambda_2)\simeq HC^\bullet(\mathfrak{A})\bigotimes_{HC^\bullet(\mathbb{C})} HC^\bullet(\Lambda_2)
\end{equation}
defined  by the cup product (\ref{cup-cyc}).
The problem thus reduces to identifying those cohomology classes on the right that correspond to free presymplectic structures and their obstructions to deformation. We proceed with a closer examination of the right tensor  factor in (\ref{D9}).

Keeping the notation of Appendix \ref{CGA},  we endow the Berezin algebra $\mathcal{B}_2$ with an auxiliary $\mathbb{Z}$-grading  by setting
\begin{equation}
    w(\theta)=w(y)=0\,,\qquad w(\bar \theta) =-1\,,\qquad w(\bar y)=1\,.
\end{equation}
We will refer to this grading as the {\it weight}, lest one confuse it with many other degrees.   
Besides, we introduce the differential  
\begin{equation}\label{D10}
    \delta a=(a, \bar y)=\frac{\partial a}{\partial \bar\theta }\,,\qquad \forall  a\in \mathcal{B}_2\,.
\end{equation}
Clearly,  $w(\delta)=1$, $\delta^2=0$, and $[\Delta,\delta]=0$. Taken together with the Schouten bracket (\ref{C13}) this differential makes the space $\mathcal{B}_2$ into a differential graded Lie algebra. When restricted to  $\Lambda_2\subset \mathcal{B}_2$, $\delta$ becomes a  derivation of the Grassmann algebra $\Lambda_2$. 
By formula  (\ref{LD}) it induces a homomorphism $L_\delta^\ast: HC^p(\Lambda_2)\rightarrow HC^p(\Lambda_2)$ in cyclic cohomology, which then trivially extends to the tensor product (\ref{D9}).  As should be evident from  (\ref{D7}) the differential (\ref{D10}) just mimics the action of the de Rham differential  on $\Lambda(\mathcal{N})$. The non-trivial presymplectic structures come from those  elements of $\mathcal{B}_2$ that are both $\Delta$- and $\delta$-closed and have weight one.\footnote{It is well to bear in  mind that the Berezin integral (\ref{phi}) implies one more differentiation by $\bar\theta$, so that the resulting cyclic cocycles have weight two and correspond to $2$-forms on $\mathcal{N}$.}   It follows from (\ref{AB}) that  all $\Delta$-cocycles of weight one are given by linear combinations of the divergence-free polyvector fields   
\begin{equation}\label{alpha-beta}
    \alpha_{n+1}=A_{n2}=n\bar \theta y^{n-1}\bar y^2-2\theta y^n\bar y\,,\qquad \beta_{n+1}=B_{n1}=y^n\bar y\,.
\end{equation}
Applying to them $\delta$, we readily find 
\begin{equation}\label{da}
   \delta \alpha_{n+1}=ny^{n-1}\bar y^2\,,\qquad \delta \beta_{n+1}  =0\,.
\end{equation}
Hence, only the first element $\alpha_1=-2\theta\bar y$ of the $\alpha$-series and all elements of the $\beta$-series  (\ref{alpha-beta}) may generate  presymplectic structures.   Notice that 
\begin{equation}
    \beta_{n+1}=\delta(\bar\theta y^n\bar y)=\delta\Big(\frac1{n+1}A_{n+1,1}\Big)\,.
\end{equation}
Translated into the language of presymplectic geometry the last equality means that the  divergence-free polyvectors  
\begin{equation}
    \gamma_{n+1}=\bar\theta y^n\bar y-\frac{1}{n+1}\theta y^{n+1}\,,\qquad w(\gamma_{n+1})=0\,,
\end{equation} generate $Q_0$-invariant potentials for the presymplectic forms $\Omega_{n+1}$ on $\mathcal{N}$ associated with $\beta_{n+1}$, see Rels. (\ref{Qn}, \ref{On}).
 The corresponding cyclic cocycles read 
 \begin{equation}\label{c}
\begin{array}{c}
 \displaystyle   \gamma_{n}(a_0,\ldots,a_{n})=\sum_{k=1}^{n}\int a_0\bar\theta \partial a_1\cdots \bar \partial a_k\cdots \partial a_{n}-\int a_0\theta \partial a_1\cdots\partial a_{n}\\[5mm]
\displaystyle =-(-1)^{|a_0|}\Big(\bar\partial(a_0\partial a_1\cdots\partial a_{n})+\sum_{k=1}^{n}\partial (a_0\partial a_1\cdots\bar\partial a_k\cdots\partial a_n )\Big)_{\theta=\bar\theta=0}
\\[5mm]
\displaystyle =\sum_{k=0}^{n} \partial a_0\cdots\bar\partial a_k\cdots\partial a_{n}\Big|_{\theta=\bar\theta=0}\,,
\end{array}
\end{equation}

\begin{equation}\label{a}
\begin{array}{c}
 \displaystyle    \beta_n(a_0,a_1,\ldots,a_n) =(L_\delta\gamma_n)(a_0,a_1,\ldots,a_n)=\sum_{k=1}^n\int a_0\partial a_1\cdots \bar\partial a_k\cdots \partial a_n\\[5mm]\displaystyle =\sum_{k=1}^n\bar\partial\partial  ( a_0\partial a_1\cdots \bar\partial a_k\cdots \partial a_n)
   =\sum_{k=0}^n\bar\partial (\partial a_0\cdots\bar\partial a_k\cdots\partial a_{n})\,,
    \end{array}
\end{equation}
\begin{equation}\label{D18}
\alpha_1(a_0,a_1)= \int a_0 \theta \bar\partial a_1= (-1)^{|a_0|}\bar\partial a_0\bar\partial a_1|_{\theta=\bar\theta=0}\,.
\end{equation}
This agrees with explicit computations for $n=0,1,2$ presented in \cite{Kast, COQUEREAUX1995333}.   The fact that $\alpha_1$ is a non-trivial $\delta$-cocycle in the space of $\Delta$-closed elements of $\mathcal{B}_2$ means that any presymplectic structure associated with $\alpha_1$ admits no $Q_0$-invariant presymplectic potential in distinction to the case of $\beta_n$'s.

Turning now to the left tensor factor in (\ref{D9}), we recall that the 
$HC^\bullet(\mathbb{C})$-module   $HC^\bullet(\mathfrak{A})$ is freely  generated by the four primary classes (\ref{prime}).  Of these, only $\Phi_2$ and $\bar\Phi_2$ can generate presymplectic structures on $\mathcal{N}$ of degree $3$. In other words, all 
`free' presymplectic structures of $4d$ HSGRA come from the two infinite series of  
cyclic cohomology classes  
\begin{equation}\label{ww}
     \varpi_n=\Phi_2 \sqcup \beta_n\,,\qquad \bar \varpi_n=\bar \Phi_2 \sqcup \beta_n\,.
\end{equation}
The direct computation of the $\sqcup$-products shows that the resulting presymplectic  structures are of the form $\Omega_n^{^{(1)}}$, see Eq. (\ref{On}). 
The corresponding presymplectic potentials originate from the classes 
\begin{equation}
    \vartheta_n=\Phi_2 \sqcup \gamma_n\,,\qquad \bar \vartheta_n=\bar \Phi_2 \sqcup \gamma_n\,. 
\end{equation}
Notice that the remaining class (\ref{D18}) gives no presymplectic structure in degree $3$.

We claim that all free presymplectic structures associated with the series (\ref{ww}) survive upon switching on interaction.  More precisely, by means of homological perturbation theory of Sec. \ref{sec:PSHSGRA},   
they can always be deformed  so as to become compatible with non-linear field equations (\ref{dw}, \ref{dC}), whatever the interaction vertices. The existence of such a deformation is ensured by the absence of obstructing cocycles. The last fact can be seen in two equivalent ways. First,  one can try to deform a free
presymplectic $2$-form by itself. For reasons of degree, all the obstructing cohomology classes, if any,  must  belong to the linear span of  
$
    \Phi_2\sqcup \alpha_n
$ and $\bar\Phi_2\sqcup \alpha_n$. However,
Eq. (\ref{da}) says that the corresponding $2$-forms are not closed for $n>1$; and hence, they cannot appear as obstructions to deformation.
The remaining case $n=1$ is also excluded as the corresponding $2$-forms do not depend on $C$, while the interaction vertices do. An alternative possibility is to deform the free presymplectic structure through the deformation of its presymplectic potential. 
Again, by degree considerations, all potential obstructions are spanned by the classes $
    \Phi_4\sqcup B_{n0}=L^\ast_\delta (\Phi_4\sqcup B_{n1})
$, so that the corresponding $1$-forms turn out to be  exact. The exact $1$-forms represent natural ambiguity in the choice of a presymplectic potential and can thus be disregarded. The details are left to the reader.   All in all, we see that the classes (\ref{ww}) span the space of all presymplectic structures in $4d$ HSGRA.  

Finally, let us note that the map $I: HC^2(\mathfrak{A})\rightarrow HH^2(\mathfrak{A}, \mathfrak{A}^\ast)$ of the long exact sequence  (\ref{CEPS})  is actually an isomorphism, so that each Hochschild $2$-cocycle is cohomologous to a cyclic one.  The existence of a non-degenerate trace on $\mathfrak{A}$ implies  
further isomorphisms $\mathfrak{A}\simeq\mathfrak{A}^\ast$ and $HH^2 (\mathfrak{A},\mathfrak{A}^\ast)\simeq  HH^2(\mathfrak{A},\mathfrak{A})$. 
In view of these isomorphisms it is little wonder that the same pair of the Hochschild $2$-cocycles of $HH^2(\mathfrak{A},\mathfrak{A})$ define the cubic vertices (\ref{vertexfac}) and the presymplectic structure (\ref{O1}) of $4d$ HSGRA gravity.

\section{Integrability of zero-form equations}
\label{app:integrability}
Since equations \eqref{dwdc} for $\momega$ and $\mC$ do not seem to be Lagrangian, it is  interesting to to ask the following question: to which extent do the equations of motion for $\mC$'s control those for $\momega$'s?  The equations in question have the form
\besubeqs\label{EEE}
\begin{align}
    d\momega^\alpha &=\frac12f_{\alpha\beta}^\gamma(\mC)\momega^\alpha\momega^\beta\,,\label{dwA}\\ d\mC^i&=V^i_\alpha(\mC)\momega^\alpha\,.\label{dCA}
\end{align}
\esubeqs
Here the indices $\alpha$ and $i$, labelling the fields, are essentially equivalent and originate from the same algebra $gl(\mathfrak{A})$. At this point, however, it is useful to consider them as independent. 
We also assume (and this is indeed the case for HSGRA) that the right sides of equations (\ref{EEE}) define a Lie algebroid with  anchor $V$.  In other words, the vector fields $V_\alpha=V_\alpha^i\partial/\partial C^i$ form an integrable distribution 
on a manifold coordinatized by $C^i$:
\begin{align}\label{VV}
    [V_\alpha,V_\beta]=f_{\alpha\beta}^\gamma V_\gamma\,.
\end{align}
Notice that  we do not assume the vector fields $V_\alpha$ to be linearly independent in general position. Nevertheless, the Jacobi identity $[[V,V],V]=0$ is presumably satisfied in the strongest form:
\begin{equation}\label{ff}
    f^\lambda_{\alpha\mu}f^\mu_{\beta\gamma}+V_{\alpha}f_{\beta\gamma}^\lambda+cycle (\alpha,\beta,\gamma)=0\,.
\end{equation}

Let us now examine the integrability conditions for the second equation (\ref{dC}). Applying the de Rham differential $d$ to both sides, we find
\begin{align}\label{Int}
    \left(\frac12f_{\alpha\beta}^\gamma\omega^\alpha\omega^\beta -d\omega^\gamma\right)V_\gamma=0\,.
\end{align}
If the $V_\alpha$'s are linearly independent, then the last condition implies the first equation (\ref{dwA}). In that case Eq. (\ref{dCA}) `knows' about Eq. (\ref{dwA}), and we may omit 
the later without any consequences for the system. In the general case, there may be some null-vectors $Z_A^\alpha=Z_A^\alpha(\mC)$ spanning the kernel of the anchor $V$, that is,
\begin{align}\label{nullvec}
    Z_A^\alpha\, V_\alpha^i=0\,.
\end{align}
Then Eq. (\ref{Int}) implies that 
\begin{align}
    d\momega^\lambda =f_{\alpha\beta}^\lambda \momega^\alpha\momega^\beta+ Z^\lambda_A\boldsymbol{B}^A
\end{align}
for some collection of $2$-forms $\boldsymbol{B}^A$. These $2$-forms should be regarded as new independent variables  describing arbitrariness in the dynamics of $\momega$'s whenever equations (\ref{dwA}) are omitted.  
The dynamics of $\boldsymbol{B}$'s are not completely arbitrary. Checking the integrability condition and assuming the null-vectors $Z_A$ to be linearly independent, one can find
\begin{align}\label{dFA}
    d\boldsymbol{B}^A=\momega^\alpha U_{\alpha B}^A(\mC)\boldsymbol{B}^B
\end{align}
for some structure functions $U$'s. In such a way we arrive at the natural extension of the original system (\ref{EEE}) by the $2$-form fields $\boldsymbol{B}^A$ subject to (\ref{dFA}). 

When applied to $4d$ HSGRA, these ideas instruct us to look for null-vectors \eqref{nullvec}. To leading order in $\mC$,  one gets then the equation  $Z\star \mC-\mC\star \tilde Z=0$, where $\tilde{Z}(y,\bar y)=Z(y,-\bar y)$ (we deal with the physical fields only). It is easy to see that the last equation has only constant solutions, $Z=\mathrm{const}\cdot \mathrm{1\!\!1}$, which also satisfy the entire equation \eqref{nullvec}.

\footnotesize
\providecommand{\href}[2]{#2}\begingroup\raggedright\endgroup


\begin{thebibliography}{100}

\bibitem{Blencowe:1988gj}
M.~Blencowe, ``{A Consistent Interacting Massless Higher Spin Field Theory in
  $D$ = (2+1)},''
{\em Class.Quant.Grav.} {\bfseries 6} (1989) 443.

\bibitem{Bergshoeff:1989ns}
E.~Bergshoeff, M.~P. Blencowe, and K.~S. Stelle, ``{Area Preserving
  Diffeomorphisms and Higher Spin Algebra},''
{\em Commun. Math. Phys.} {\bfseries 128} (1990) 213.

\bibitem{Campoleoni:2010zq}
A.~Campoleoni, S.~Fredenhagen, S.~Pfenninger, and S.~Theisen, ``{Asymptotic
  symmetries of three-dimensional gravity coupled to higher-spin fields},''
  {\em JHEP} {\bfseries 1011} (2010) 007,
\href{http://arxiv.org/abs/1008.4744}{{\ttfamily arXiv:1008.4744 [hep-th]}}.

\bibitem{Henneaux:2010xg}
M.~Henneaux and S.-J. Rey, ``{Nonlinear $W_{\infty}$ as Asymptotic Symmetry of
  Three-Dimensional Higher Spin Anti-de Sitter Gravity},'' {\em JHEP}
  {\bfseries 1012} (2010) 007,
\href{http://arxiv.org/abs/1008.4579}{{\ttfamily arXiv:1008.4579 [hep-th]}}.

\bibitem{Pope:1989vj}
C.~N. Pope and P.~K. Townsend, ``{Conformal Higher Spin in (2+1)-dimensions},''
{\em Phys. Lett.} {\bfseries B225} (1989) 245--250.

\bibitem{Fradkin:1989xt}
E.~S. Fradkin and V.~{\relax Ya}. Linetsky, ``{A Superconformal Theory of
  Massless Higher Spin Fields in $D$ = (2+1)},'' {\em Mod. Phys. Lett.}
  {\bfseries A4} (1989) 731.
[Annals Phys.198,293(1990)].

\bibitem{Grigoriev:2019xmp}
M.~Grigoriev, I.~Lovrekovic, and E.~Skvortsov, ``{New Conformal Higher Spin
  Gravities in $3d$},'' {\em JHEP} {\bfseries 01} (2020) 059,
\href{http://arxiv.org/abs/1909.13305}{{\ttfamily arXiv:1909.13305 [hep-th]}}.

\bibitem{Alday:2020qkm}
L.~F. Alday, J.-B. Bae, N.~Benjamin, and C.~Jorge-Diaz, ``{On the Spectrum of
  Pure Higher Spin Gravity},''
  \href{http://dx.doi.org/10.1007/JHEP12(2020)001}{{\em JHEP} {\bfseries 12}
  (2020) 001},
\href{http://arxiv.org/abs/2009.01830}{{\ttfamily arXiv:2009.01830 [hep-th]}}.

\bibitem{Segal:2002gd}
A.~Y. Segal, ``{Conformal higher spin theory},'' {\em Nucl. Phys.} {\bfseries
  B664} (2003) 59--130,
\href{http://arxiv.org/abs/hep-th/0207212}{{\ttfamily arXiv:hep-th/0207212
  [hep-th]}}.

\bibitem{Tseytlin:2002gz}
A.~A. Tseytlin, ``{On limits of superstring in $AdS_5\times S^5$},'' {\em
  Theor. Math. Phys.} {\bfseries 133} (2002) 1376--1389,
  \href{http://arxiv.org/abs/hep-th/0201112}{{\ttfamily arXiv:hep-th/0201112
  [hep-th]}}.
[Teor. Mat. Fiz.133,69(2002)].

\bibitem{Bekaert:2010ky}
X.~Bekaert, E.~Joung, and J.~Mourad, ``{Effective action in a higher-spin
  background},'' {\em JHEP} {\bfseries 02} (2011) 048,
\href{http://arxiv.org/abs/1012.2103}{{\ttfamily arXiv:1012.2103 [hep-th]}}.

\bibitem{Joung:2015eny}
E.~Joung, S.~Nakach, and A.~A. Tseytlin, ``{Scalar scattering via conformal
  higher spin exchange},''
  \href{http://dx.doi.org/10.1007/JHEP02(2016)125}{{\em JHEP} {\bfseries 02}
  (2016) 125},
\href{http://arxiv.org/abs/1512.08896}{{\ttfamily arXiv:1512.08896 [hep-th]}}.

\bibitem{Beccaria:2016syk}
M.~Beccaria, S.~Nakach, and A.~A. Tseytlin, ``{On triviality of S-matrix in
  conformal higher spin theory},''
  \href{http://dx.doi.org/10.1007/JHEP09(2016)034}{{\em JHEP} {\bfseries 09}
  (2016) 034},
\href{http://arxiv.org/abs/1607.06379}{{\ttfamily arXiv:1607.06379 [hep-th]}}.

\bibitem{Adamo:2018srx}
T.~Adamo, S.~Nakach, and A.~A. Tseytlin, ``{Scattering of conformal higher spin
  fields},'' \href{http://dx.doi.org/10.1007/JHEP07(2018)016}{{\em JHEP}
  {\bfseries 07} (2018) 016},
\href{http://arxiv.org/abs/1805.00394}{{\ttfamily arXiv:1805.00394 [hep-th]}}.

\bibitem{Metsaev:1991mt}
R.~R. Metsaev, ``{Poincare invariant dynamics of massless higher spins: Fourth
  order analysis on mass shell},''
{\em Mod. Phys. Lett.} {\bfseries A6} (1991) 359--367.

\bibitem{Metsaev:1991nb}
R.~R. Metsaev, ``{$S$ matrix approach to massless higher spins theory. 2: The
  Case of internal symmetry},''
{\em Mod. Phys. Lett.} {\bfseries A6} (1991) 2411--2421.

\bibitem{Ponomarev:2016lrm}
D.~Ponomarev and E.~D. Skvortsov, ``{Light-Front Higher-Spin Theories in Flat
  Space},'' {\em J. Phys.} {\bfseries A50} no.~9, (2017) 095401,
\href{http://arxiv.org/abs/1609.04655}{{\ttfamily arXiv:1609.04655 [hep-th]}}.

\bibitem{Skvortsov:2018jea}
E.~D. Skvortsov, T.~Tran, and M.~Tsulaia, ``{Quantum Chiral Higher Spin
  Gravity},'' {\em Phys. Rev. Lett.} {\bfseries 121} no.~3, (2018) 031601,
\href{http://arxiv.org/abs/1805.00048}{{\ttfamily arXiv:1805.00048 [hep-th]}}.

\bibitem{Skvortsov:2020wtf}
E.~Skvortsov, T.~Tran, and M.~Tsulaia, ``{More on Quantum Chiral Higher Spin
  Gravity},'' {\em Phys. Rev.} {\bfseries D101} no.~10, (2020) 106001,
\href{http://arxiv.org/abs/2002.08487}{{\ttfamily arXiv:2002.08487 [hep-th]}}.

\bibitem{Skvortsov:2020gpn}
E.~Skvortsov and T.~Tran, ``{One-loop Finiteness of Chiral Higher Spin
  Gravity},''
\href{http://arxiv.org/abs/2004.10797}{{\ttfamily arXiv:2004.10797 [hep-th]}}.

\bibitem{Gopakumar:2011qs}
R.~Gopakumar, R.~K. Gupta, and S.~Lal, ``{The Heat Kernel on $AdS$},'' {\em
  JHEP} {\bfseries 11} (2011) 010,
\href{http://arxiv.org/abs/1103.3627}{{\ttfamily arXiv:1103.3627 [hep-th]}}.

\bibitem{Tseytlin:2013jya}
A.~A. Tseytlin, ``{On partition function and Weyl anomaly of conformal higher
  spin fields},'' {\em Nucl. Phys.} {\bfseries B877} (2013) 598--631,
\href{http://arxiv.org/abs/1309.0785}{{\ttfamily arXiv:1309.0785 [hep-th]}}.

\bibitem{Giombi:2013fka}
S.~Giombi and I.~R. Klebanov, ``{One Loop Tests of Higher Spin AdS/CFT},'' {\em
  JHEP} {\bfseries 12} (2013) 068,
\href{http://arxiv.org/abs/1308.2337}{{\ttfamily arXiv:1308.2337 [hep-th]}}.

\bibitem{Giombi:2014yra}
S.~Giombi, I.~R. Klebanov, and A.~A. Tseytlin, ``{Partition Functions and
  Casimir Energies in Higher Spin $AdS_{d+1}/CFT_d$},'' {\em Phys. Rev.}
  {\bfseries D90} no.~2, (2014) 024048,
\href{http://arxiv.org/abs/1402.5396}{{\ttfamily arXiv:1402.5396 [hep-th]}}.

\bibitem{Beccaria:2014jxa}
M.~Beccaria, X.~Bekaert, and A.~A. Tseytlin, ``{Partition function of free
  conformal higher spin theory},'' {\em JHEP} {\bfseries 08} (2014) 113,
\href{http://arxiv.org/abs/1406.3542}{{\ttfamily arXiv:1406.3542 [hep-th]}}.

\bibitem{Beccaria:2014xda}
M.~Beccaria and A.~A. Tseytlin, ``{Higher spins in AdS$_{5}$ at one loop:
  vacuum energy, boundary conformal anomalies and AdS/CFT},'' {\em JHEP}
  {\bfseries 11} (2014) 114,
\href{http://arxiv.org/abs/1410.3273}{{\ttfamily arXiv:1410.3273 [hep-th]}}.

\bibitem{Beccaria:2015vaa}
M.~Beccaria and A.~A. Tseytlin, ``{On higher spin partition functions},'' {\em
  J. Phys.} {\bfseries A48} no.~27, (2015) 275401,
\href{http://arxiv.org/abs/1503.08143}{{\ttfamily arXiv:1503.08143 [hep-th]}}.

\bibitem{Gunaydin:2016amv}
M.~G{\"u}naydin, E.~D. Skvortsov, and T.~Tran, ``{Exceptional $F(4)$
  higher-spin theory in AdS$_{6}$ at one-loop and other tests of duality},''
  {\em JHEP} {\bfseries 11} (2016) 168,
\href{http://arxiv.org/abs/1608.07582}{{\ttfamily arXiv:1608.07582 [hep-th]}}.

\bibitem{Bae:2016rgm}
J.-B. Bae, E.~Joung, and S.~Lal, ``{One-loop test of free SU(N) adjoint model
  holography},'' {\em JHEP} {\bfseries 04} (2016) 061,
\href{http://arxiv.org/abs/1603.05387}{{\ttfamily arXiv:1603.05387 [hep-th]}}.

\bibitem{Skvortsov:2017ldz}
E.~D. Skvortsov and T.~Tran, ``{AdS/CFT in Fractional Dimension and Higher Spin
  Gravity at One Loop},'' \href{http://dx.doi.org/10.3390/universe3030061}{{\em
  Universe} {\bfseries 3} no.~3, (2017) 61},
\href{http://arxiv.org/abs/1707.00758}{{\ttfamily arXiv:1707.00758 [hep-th]}}.

\bibitem{Ponomarev:2019ltz}
D.~Ponomarev, E.~Sezgin, and E.~Skvortsov, ``{On one loop corrections in higher
  spin gravity},''
\href{http://arxiv.org/abs/1904.01042}{{\ttfamily arXiv:1904.01042 [hep-th]}}.

\bibitem{deMelloKoch:2018ivk}
R.~de~Mello~Koch, A.~Jevicki, K.~Suzuki, and J.~Yoon, ``{AdS Maps and Diagrams
  of Bi-local Holography},'' {\em JHEP} {\bfseries 03} (2019) 133,
\href{http://arxiv.org/abs/1810.02332}{{\ttfamily arXiv:1810.02332 [hep-th]}}.

\bibitem{Aharony:2020omh}
O.~Aharony, S.~M. Chester, and E.~Y. Urbach, ``{A Derivation of AdS/CFT for
  Vector Models},''
\href{http://arxiv.org/abs/2011.06328}{{\ttfamily arXiv:2011.06328 [hep-th]}}.

\bibitem{Skvortsov:2018uru}
E.~Skvortsov, ``{Light-Front Bootstrap for Chern-Simons Matter Theories},''
  {\em JHEP} {\bfseries 06} (2019) 058,
\href{http://arxiv.org/abs/1811.12333}{{\ttfamily arXiv:1811.12333 [hep-th]}}.

\bibitem{Giombi:2011kc}
S.~Giombi, S.~Minwalla, S.~Prakash, S.~P. Trivedi, S.~R. Wadia, and X.~Yin,
  ``{Chern-Simons Theory with Vector Fermion Matter},'' {\em Eur. Phys. J.}
  {\bfseries C72} (2012) 2112,
\href{http://arxiv.org/abs/1110.4386}{{\ttfamily arXiv:1110.4386 [hep-th]}}.

\bibitem{Maldacena:2012sf}
J.~Maldacena and A.~Zhiboedov, ``{Constraining conformal field theories with a
  slightly broken higher spin symmetry},''
\href{http://arxiv.org/abs/1204.3882}{{\ttfamily arXiv:1204.3882 [hep-th]}}.

\bibitem{Aharony:2012nh}
O.~Aharony, G.~Gur-Ari, and R.~Yacoby, ``{Correlation Functions of Large N
  Chern-Simons-Matter Theories and Bosonization in Three Dimensions},'' {\em
  JHEP} {\bfseries 12} (2012) 028,
\href{http://arxiv.org/abs/1207.4593}{{\ttfamily arXiv:1207.4593 [hep-th]}}.

\bibitem{Aharony:2015mjs}
O.~Aharony, ``{Baryons, monopoles and dualities in Chern-Simons-matter
  theories},'' {\em JHEP} {\bfseries 02} (2016) 093,
\href{http://arxiv.org/abs/1512.00161}{{\ttfamily arXiv:1512.00161 [hep-th]}}.

\bibitem{Karch:2016sxi}
A.~Karch and D.~Tong, ``{Particle-Vortex Duality from 3d Bosonization},'' {\em
  Phys. Rev.} {\bfseries X6} no.~3, (2016) 031043,
\href{http://arxiv.org/abs/1606.01893}{{\ttfamily arXiv:1606.01893 [hep-th]}}.

\bibitem{Seiberg:2016gmd}
N.~Seiberg, T.~Senthil, C.~Wang, and E.~Witten, ``{A Duality Web in 2+1
  Dimensions and Condensed Matter Physics},'' {\em Annals Phys.} {\bfseries
  374} (2016) 395--433,
\href{http://arxiv.org/abs/1606.01989}{{\ttfamily arXiv:1606.01989 [hep-th]}}.

\bibitem{Dempster:2012vw}
P.~Dempster and M.~Tsulaia, ``{On the Structure of Quartic Vertices for
  Massless Higher Spin Fields on Minkowski Background},''
  \href{http://dx.doi.org/10.1016/j.nuclphysb.2012.07.031}{{\em Nucl. Phys.}
  {\bfseries B865} (2012) 353--375},
\href{http://arxiv.org/abs/1203.5597}{{\ttfamily arXiv:1203.5597 [hep-th]}}.

\bibitem{Bekaert:2015tva}
X.~Bekaert, J.~Erdmenger, D.~Ponomarev, and C.~Sleight, ``{Quartic AdS
  Interactions in Higher-Spin Gravity from Conformal Field Theory},'' {\em
  JHEP} {\bfseries 11} (2015) 149,
\href{http://arxiv.org/abs/1508.04292}{{\ttfamily arXiv:1508.04292 [hep-th]}}.

\bibitem{Maldacena:2015iua}
J.~Maldacena, D.~Simmons-Duffin, and A.~Zhiboedov, ``{Looking for a bulk
  point},'' {\em JHEP} {\bfseries 01} (2017) 013,
\href{http://arxiv.org/abs/1509.03612}{{\ttfamily arXiv:1509.03612 [hep-th]}}.

\bibitem{Sleight:2017pcz}
C.~Sleight and M.~Taronna, ``{Higher-Spin Gauge Theories and Bulk Locality},''
  {\em Phys. Rev. Lett.} {\bfseries 121} no.~17, (2018) 171604,
\href{http://arxiv.org/abs/1704.07859}{{\ttfamily arXiv:1704.07859 [hep-th]}}.

\bibitem{Ponomarev:2017qab}
D.~Ponomarev, ``{A Note on (Non)-Locality in Holographic Higher Spin
  Theories},'' {\em Universe} {\bfseries 4} no.~1, (2018) 2,
\href{http://arxiv.org/abs/1710.00403}{{\ttfamily arXiv:1710.00403 [hep-th]}}.

\bibitem{Klebanov:2002ja}
I.~R. Klebanov and A.~M. Polyakov, ``{AdS dual of the critical $O(N)$ vector
  model},'' {\em Phys. Lett.} {\bfseries B550} (2002) 213--219,
\href{http://arxiv.org/abs/hep-th/0210114}{{\ttfamily arXiv:hep-th/0210114}}.

\bibitem{Sezgin:2003pt}
E.~Sezgin and P.~Sundell, ``{Holography in 4D (super) higher spin theories and
  a test via cubic scalar couplings},'' {\em JHEP} {\bfseries 0507} (2005) 044,
\href{http://arxiv.org/abs/hep-th/0305040}{{\ttfamily arXiv:hep-th/0305040
  [hep-th]}}.

\bibitem{Vasiliev:1988sa}
M.~A. Vasiliev, ``Consistent equations for interacting massless fields of all
  spins in the first order in curvatures,''
{\em Annals Phys.} {\bfseries 190} (1989) 59--106.

\bibitem{Fronsdal:1978rb}
C.~Fronsdal, ``Massless fields with integer spin,''
{\em Phys. Rev.} {\bfseries D18} (1978) 3624.

\bibitem{Kazinski:2005eb}
P.~O. Kazinski, S.~L. Lyakhovich, and A.~A. Sharapov, ``{Lagrange structure and
  quantization},'' {\em JHEP} {\bfseries 07} (2005) 076,
\href{http://arxiv.org/abs/hep-th/0506093}{{\ttfamily arXiv:hep-th/0506093
  [hep-th]}}.

\bibitem{Boulanger:2011dd}
N.~Boulanger and P.~Sundell, ``{An action principle for Vasiliev's
  four-dimensional higher-spin gravity},'' {\em J. Phys.} {\bfseries A44}
  (2011) 495402,
\href{http://arxiv.org/abs/1102.2219}{{\ttfamily arXiv:1102.2219 [hep-th]}}.

\bibitem{Kaparulin:2011zz}
D.~S. Kaparulin, S.~L. Lyakhovich, and A.~A. Sharapov, ``{On Lagrange structure
  of unfolded field theory},'' {\em Int. J. Mod. Phys.} {\bfseries A26} (2011)
  1347--1362,
\href{http://arxiv.org/abs/1012.2567}{{\ttfamily arXiv:1012.2567 [hep-th]}}.

\bibitem{Kaparulin:2011aa}
D.~S. Kaparulin, S.~L. Lyakhovich, and A.~A. Sharapov, ``{Lagrange Anchor and
  Characteristic Symmetries of Free Massless Fields},'' {\em SIGMA} {\bfseries
  8} (2012) 021,
\href{http://arxiv.org/abs/1112.1860}{{\ttfamily arXiv:1112.1860 [hep-th]}}.

\bibitem{MISUNA2019134956}
N.~Misuna, ``On unfolded off-shell formulation for higher-spin theory,''
  \href{http://dx.doi.org/https://doi.org/10.1016/j.physletb.2019.134956}{{\em
  Physics Letters B} {\bfseries 798} (2019) 134956}.

\bibitem{misuna2020offshell}
N.~G. Misuna, ``{Off-shell higher-spin fields in $AdS_{4}$ and external
  currents},'' \href{http://arxiv.org/abs/2012.06570}{{\ttfamily
  arXiv:2012.06570 [hep-th]}}.

\bibitem{1987thyg.book..676C}
C.~{Crnkovic} and E.~{Witten}, ``{Covariant description of canonical formalism
  in geometrical theories.},'' in {\em Three Hundred Years of Gravitation},
  pp.~676--684.
\newblock 1987.

\bibitem{Zuckerman:1989cx}
G.~J. Zuckerman, ``{Action principles and global geometry},'' {\em Conf. Proc.
  C} {\bfseries 8607214} (1986) 259--284.

\bibitem{2013JMP....54k1502K}
I.~{Khavkine}, ``{Presymplectic current and the inverse problem of the calculus
  of variations},'' \href{http://dx.doi.org/10.1063/1.4828666}{{\em Journal of
  Mathematical Physics} {\bfseries 54} no.~11, (Nov., 2013) 111502--111502},
  \href{http://arxiv.org/abs/1210.0802}{{\ttfamily arXiv:1210.0802 [math-ph]}}.

\bibitem{Boulanger:2015kfa}
N.~Boulanger, E.~Sezgin, and P.~Sundell, ``{4D Higher Spin Gravity with
  Dynamical Two-Form as a Frobenius-Chern-Simons Gauge Theory},''
\href{http://arxiv.org/abs/1505.04957}{{\ttfamily arXiv:1505.04957 [hep-th]}}.

\bibitem{Bonezzi:2016ttk}
R.~Bonezzi, N.~Boulanger, E.~Sezgin, and P.~Sundell,
  ``{Frobenius--Chern--Simons gauge theory},'' {\em J. Phys.} {\bfseries A50}
  no.~5, (2017) 055401,
\href{http://arxiv.org/abs/1607.00726}{{\ttfamily arXiv:1607.00726 [hep-th]}}.

\bibitem{Alkalaev_2014}
K.~Alkalaev and M.~Grigoriev, ``{Frame-like Lagrangians and presymplectic
  AKSZ-type sigma models},''
  \href{http://dx.doi.org/10.1142/s0217751x14501036}{{\em Int. J. Mod. Phys. A}
  {\bfseries 29} no.~18, (Jul, 2014) 1450103},
  \href{http://arxiv.org/abs/1312.5296}{{\ttfamily arXiv:1312.5296 [hep-th]}}.

\bibitem{grigoriev2016presymplectic}
M.~Grigoriev, ``{Presymplectic structures and intrinsic Lagrangians},''
  \href{http://arxiv.org/abs/1606.07532}{{\ttfamily arXiv:1606.07532
  [hep-th]}}.

\bibitem{Alexandrov:1995kv}
M.~Alexandrov, M.~Kontsevich, A.~Schwarz, and O.~Zaboronsky, ``{The Geometry of
  the Master Equation and Topological Quantum Field Theory},'' {\em Int. J.
  Mod. Phys.} {\bfseries A12} (1997) 1405--1429,
\href{http://arxiv.org/abs/hep-th/9502010}{{\ttfamily arXiv:hep-th/9502010
  [hep-th]}}.

\bibitem{Sharapov:2016qne}
A.~A. Sharapov, ``{On presymplectic structures for massless higher-spin
  fields},'' {\em Eur. Phys. J.} {\bfseries C76} no.~6, (2016) 305,
\href{http://arxiv.org/abs/1602.06393}{{\ttfamily arXiv:1602.06393 [hep-th]}}.

\bibitem{Sharapov:2020quq}
A.~Sharapov and E.~Skvortsov, ``{Characteristic Cohomology and Observables in
  Higher Spin Gravity},'' \href{http://dx.doi.org/10.1007/JHEP12(2020)190}{{\em
  JHEP} {\bfseries 12} (2020) 190},
\href{http://arxiv.org/abs/2006.13986}{{\ttfamily arXiv:2006.13986 [hep-th]}}.

\bibitem{Sharapov:2018hnl}
A.~A. Sharapov and E.~D. Skvortsov, ``{A simple construction of associative
  deformations},'' {\em Letters in Mathematical Physics} (Jul, 2018) ,
  \href{http://arxiv.org/abs/1803.10957}{{\ttfamily arXiv:1803.10957
  [math-ph]}}.

\bibitem{Sharapov:2018ioy}
A.~A. Sharapov and E.~D. Skvortsov, ``{On deformations of
  $A_\infty$-algebras},'' {\em J. Phys.} {\bfseries A52} no.~47, (2019) 475203,
\href{http://arxiv.org/abs/1809.03386}{{\ttfamily arXiv:1809.03386 [math-ph]}}.

\bibitem{Sharapov:2018kjz}
A.~Sharapov and E.~Skvortsov, ``{$A_\infty$ algebras from slightly broken
  higher spin symmetries},''
  \href{http://dx.doi.org/10.1007/JHEP09(2019)024}{{\em JHEP} {\bfseries 09}
  (2019) 024},
\href{http://arxiv.org/abs/1809.10027}{{\ttfamily arXiv:1809.10027 [hep-th]}}.

\bibitem{Sharapov:2019vyd}
A.~Sharapov and E.~Skvortsov, ``{Formal Higher Spin Gravities},'' {\em Nucl.
  Phys.} {\bfseries B941} (2019) 838--860,
\href{http://arxiv.org/abs/1901.01426}{{\ttfamily arXiv:1901.01426 [hep-th]}}.

\bibitem{Vasiliev:1986bq}
M.~A. Vasiliev and E.~S. Fradkin, ``Gravitational interaction of massless high
  spin ($s > 2$) fields,''
{\em JETP Lett.} {\bfseries 44} (1986) 622--627.

\bibitem{Sleight:2016dba}
C.~Sleight and M.~Taronna, ``{Higher Spin Interactions from Conformal Field
  Theory: The Complete Cubic Couplings},'' {\em Phys. Rev. Lett.} {\bfseries
  116} no.~18, (2016) 181602,
\href{http://arxiv.org/abs/1603.00022}{{\ttfamily arXiv:1603.00022 [hep-th]}}.

\bibitem{Roytenberg:2006qz}
D.~Roytenberg, ``{AKSZ-BV Formalism and Courant Algebroid-induced Topological
  Field Theories},'' \href{http://dx.doi.org/10.1007/s11005-006-0134-y}{{\em
  Lett. Math. Phys.} {\bfseries 79} (2007) 143--159},
  \href{http://arxiv.org/abs/hep-th/0608150}{{\ttfamily arXiv:hep-th/0608150}}.

\bibitem{Cattaneo_2006}
A.~Cattaneo, D.~Fiorenza, and R.~Longoni, ``{Graded Poisson Algebras},''
  \href{http://dx.doi.org/10.1016/b0-12-512666-2/00434-x}{{\em Encyclopedia of
  Mathematical Physics} (2006) 560–567},
  \href{http://arxiv.org/abs/1811.07395}{{\ttfamily arXiv:1811.07395
  [math.SG]}}.

\bibitem{Voronov2019}
T.~T. Voronov, ``{Graded Geometry, Q-Manifolds, and Microformal Geometry},''
  {\em Fortschritte der Physik} {\bfseries 67} no.~8--9, (May, 2019) 1910023,
  \href{http://arxiv.org/abs/1903.02884}{{\ttfamily arXiv:1903.02884
  [hep-th]}}.

\bibitem{schwarz1993}
A.~Schwarz, ``{Semiclassical approximation in Batalin-Vilkovisky formalism},''
  {\em Comm. Math. Phys.} {\bfseries 158} no.~2, (1993) 373--396,
  \href{http://arxiv.org/abs/hep-th/9210115}{{\ttfamily arXiv:hep-th/9210115}}.

\bibitem{Lyakhovich_2010}
S.~Lyakhovich, E.~Mosman, and A.~Sharapov, ``{Characteristic Classes of
  Q-manifolds: Classification and Applications},''
  \href{http://dx.doi.org/10.1016/j.geomphys.2010.01.008}{{\em Journal of
  Geometry and Physics} {\bfseries 60} no.~5, (May, 2010) 729–759},
  \href{http://arxiv.org/abs/0906.0466}{{\ttfamily arXiv:0906.0466 [math-ph]}}.

\bibitem{Grigoriev_2019}
M.~Grigoriev and A.~Kotov, ``{Gauge PDE and AKSZ-type Sigma Models},''
  \href{http://dx.doi.org/10.1002/prop.201910007}{{\em Fortschritte der Physik}
  {\bfseries 67} no.~8-9, (May, 2019) 1910007},
  \href{http://arxiv.org/abs/1903.02820}{{\ttfamily arXiv:1903.02820
  [hep-th]}}.

\bibitem{Kontsevich:2006jb}
M.~Kontsevich and Y.~Soibelman, ``{Notes on $A_\infty$-Algebras,
  $A_\infty$-Categories and Non-Commutative Geometry},''
  \href{http://dx.doi.org/10.1007/978-3-540-68030-7\_6}{{\em Lect. Notes in
  Physics} {\bfseries 757} (2009) 153--220},
  \href{http://arxiv.org/abs/math/0606241}{{\ttfamily arXiv:math/0606241}}.

\bibitem{Barnich:2004cr}
G.~Barnich, M.~Grigoriev, A.~Semikhatov, and I.~Tipunin, ``{Parent field theory
  and unfolding in BRST first-quantized terms},'' {\em Commun. Math. Phys.}
  {\bfseries 260} (2005) 147--181,
\href{http://arxiv.org/abs/hep-th/0406192}{{\ttfamily arXiv:hep-th/0406192
  [hep-th]}}.

\bibitem{Barnich:2010sw}
G.~Barnich and M.~Grigoriev, ``{First order parent formulation for generic
  gauge field theories},'' {\em JHEP} {\bfseries 1101} (2011) 122,
\href{http://arxiv.org/abs/1009.0190}{{\ttfamily arXiv:1009.0190 [hep-th]}}.

\bibitem{Bekaert:2005vh1}
X.~Bekaert, S.~Cnockaert, C.~Iazeolla, and M.~Vasiliev, ``{Nonlinear higher
  spin theories in various dimensions},'' in {\em {1st Solvay Workshop on
  Higher Spin Gauge Theories}}, pp.~132--197.
\newblock 2004.
\newblock \href{http://arxiv.org/abs/hep-th/0503128}{{\ttfamily
  arXiv:hep-th/0503128}}.

\bibitem{Barnich_1993}
G.~Barnich and M.~Henneaux, ``{Consistent couplings between fields with a gauge
  freedom and deformations of the master equation},''
  \href{http://dx.doi.org/10.1016/0370-2693(93)90544-r}{{\em Phys. Lett.}
  {\bfseries B311} no.~1-4, (Jul, 1993) 123--129}.

\bibitem{Fradkin:1986ka}
E.~S. Fradkin and M.~A. Vasiliev, ``Candidate to the role of higher spin
  symmetry,''
{\em Ann. Phys.} {\bfseries 177} (1987) 63.

\bibitem{Vasiliev:1988xc}
M.~A. Vasiliev, ``Equations of motion of interacting massless fields of all
  spins as a free differential algebra,''
{\em Phys. Lett.} {\bfseries B209} (1988) 491--497.

\bibitem{Vaintrob}
A.~Y. Vaintrob, ``Lie algebroids and homological vector fields,'' {\em Russian
  Mathematical Surveys} {\bfseries 52} no.~2, (1997) 428.

\bibitem{Flato:1978qz}
M.~Flato and C.~Fronsdal, ``{One Massless Particle Equals Two Dirac Singletons:
  Elementary Particles in a Curved Space. 6.},''
{\em Lett.Math.Phys.} {\bfseries 2} (1978) 421--426.

\bibitem{Dirac:1963ta}
P.~A.~M. Dirac, ``{A Remarkable representation of the 3 + 2 de Sitter group},''
{\em J. Math. Phys.} {\bfseries 4} (1963) 901--909.

\bibitem{Didenko:2014dwa}
V.~Didenko and E.~Skvortsov, ``{Elements of Vasiliev theory},''
\href{http://arxiv.org/abs/1401.2975}{{\ttfamily arXiv:1401.2975 [hep-th]}}.

\bibitem{Wigner}
E.~P. Wigner, ``{Do the Equations of Motion Determine the Quantum Mechanical
  Commutation Relations?},'' {\em Phys. Rev.} {\bfseries 77} (Mar, 1950)
  711--712.

\bibitem{Yang:1951pyq}
L.~M. Yang, ``{A Note on the Quantum Rule of the Harmonic Oscillator},''
{\em Phys. Rev.} {\bfseries 84} no.~4, (1951) 788.

\bibitem{Mukunda:1980fv}
N.~Mukunda, E.~C.~G. Sudarshan, J.~K. Sharma, and C.~L. Mehta,
  ``{Representations and properties of parabose oscillator operators. I. Energy
  position and momentum eigenstates},''
  \href{http://dx.doi.org/10.1063/1.524695}{{\em J. Math. Phys.} {\bfseries 21}
  (1980) 2386--2394}.

\bibitem{Vasiliev:1989re}
M.~A. Vasiliev, ``Higher spin algebras and quantization on the sphere and
  hyperboloid,''
{\em Int. J. Mod. Phys.} {\bfseries A6} (1991) 1115--1135.

\bibitem{Pope:1990kc}
C.~N. Pope, L.~J. Romans, and X.~Shen, ``{A New Higher Spin Algebra and the
  Lone Star Product},''
\href{http://dx.doi.org/10.1016/0370-2693(90)91782-7}{{\em Phys. Lett.}
  {\bfseries B242} (1990) 401--406}.

\bibitem{Bieliavsky:2008mv}
P.~Bieliavsky, S.~Detournay, and P.~Spindel, ``{The Deformation quantizations
  of the hyperbolic plane},''
  \href{http://dx.doi.org/10.1007/s00220-008-0697-9}{{\em Commun. Math. Phys.}
  {\bfseries 289} (2009) 529--559},
\href{http://arxiv.org/abs/0806.4741}{{\ttfamily arXiv:0806.4741 [math-ph]}}.

\bibitem{Joung:2014qya}
E.~Joung and K.~Mkrtchyan, ``{Notes on higher-spin algebras: minimal
  representations and structure constants},'' {\em JHEP} {\bfseries 05} (2014)
  103,
\href{http://arxiv.org/abs/1401.7977}{{\ttfamily arXiv:1401.7977 [hep-th]}}.

\bibitem{Korybut:2014jza}
A.~V. Korybut, ``{Covariant structure constants for a deformed oscillator
  algebra},'' {\em Theor. Math. Phys.} {\bfseries 193} no.~1, (2017)
  1409--1419,
\href{http://arxiv.org/abs/1409.8634}{{\ttfamily arXiv:1409.8634 [hep-th]}}.

\bibitem{Basile:2016goq}
T.~Basile, N.~Boulanger, and F.~Buisseret, ``{Structure constants of
  shs$[\lambda]$ : the deformed-oscillator point of view},'' {\em J. Phys.}
  {\bfseries A51} no.~2, (2018) 025201,
\href{http://arxiv.org/abs/1604.04510}{{\ttfamily arXiv:1604.04510 [hep-th]}}.

\bibitem{korybut2020star}
A.~V. Korybut, ``{Star product for deformed oscillator algebra
  $\mathsf{Aq}(2,\nu)$},'' \href{http://arxiv.org/abs/2006.01622}{{\ttfamily
  arXiv:2006.01622 [hep-th]}}.

\bibitem{Leigh:2003gk}
R.~G. Leigh and A.~C. Petkou, ``{Holography of the N=1 higher spin theory on
  AdS(4)},'' {\em JHEP} {\bfseries 0306} (2003) 011,
\href{http://arxiv.org/abs/hep-th/0304217}{{\ttfamily arXiv:hep-th/0304217
  [hep-th]}}.

\bibitem{Vasiliev:1989yr}
M.~A. Vasiliev, ``Dynamics of massless higher spins in the second order in
  curvatures,''
{\em Phys. Lett.} {\bfseries B238} (1990) 305--314.

\bibitem{Vasiliev:1990cm}
M.~A. Vasiliev, ``{Closed equations for interacting gauge fields of all
  spins},''
{\em JETP Lett.} {\bfseries 51} (1990) 503--507.

\bibitem{Vasiliev:2003ev}
M.~A. Vasiliev, ``{Nonlinear equations for symmetric massless higher spin
  fields in (A)dS(d)},'' {\em Phys. Lett.} {\bfseries B567} (2003) 139--151,
\href{http://arxiv.org/abs/hep-th/0304049}{{\ttfamily arXiv:hep-th/0304049
  [hep-th]}}.

\bibitem{Neiman:2015wma}
Y.~Neiman, ``{Higher-spin gravity as a theory on a fixed (anti) de Sitter
  background},'' {\em JHEP} {\bfseries 04} (2015) 144,
\href{http://arxiv.org/abs/1502.06685}{{\ttfamily arXiv:1502.06685 [hep-th]}}.

\bibitem{Arias:2017bvi}
C.~Arias, R.~Bonezzi, and P.~Sundell, ``{Bosonic Higher Spin Gravity in any
  Dimension with Dynamical Two-Form},'' {\em JHEP} {\bfseries 03} (2019) 001,
\href{http://arxiv.org/abs/1712.03135}{{\ttfamily arXiv:1712.03135 [hep-th]}}.

\bibitem{Bekaert:2013zya}
X.~Bekaert and M.~Grigoriev, ``{Higher order singletons, partially massless
  fields and their boundary values in the ambient approach},'' {\em Nucl.
  Phys.} {\bfseries B876} (2013) 667--714,
\href{http://arxiv.org/abs/1305.0162}{{\ttfamily arXiv:1305.0162 [hep-th]}}.

\bibitem{Bekaert:2017bpy}
X.~Bekaert, M.~Grigoriev, and E.~D. Skvortsov, ``{Higher Spin Extension of
  Fefferman-Graham Construction},'' {\em Universe} {\bfseries 4} no.~2, (2018)
  17,
\href{http://arxiv.org/abs/1710.11463}{{\ttfamily arXiv:1710.11463 [hep-th]}}.

\bibitem{Grigoriev:2018wrx}
M.~Grigoriev and E.~D. Skvortsov, ``{Type-B Formal Higher Spin Gravity},'' {\em
  JHEP} {\bfseries 05} (2018) 138,
\href{http://arxiv.org/abs/1804.03196}{{\ttfamily arXiv:1804.03196 [hep-th]}}.

\bibitem{Sharapov:2019pdu}
A.~Sharapov, E.~Skvortsov, and T.~Tran, ``{Towards massless sector of
  tensionless strings on AdS$_5$},'' {\em Phys. Lett.} {\bfseries B800} (2020)
  135094,
\href{http://arxiv.org/abs/1908.00050}{{\ttfamily arXiv:1908.00050 [hep-th]}}.

\bibitem{Grigoriev:1999qz}
M.~A. Grigoriev and P.~H. Damgaard, ``{Superfield BRST charge and the master
  action},'' \href{http://dx.doi.org/10.1016/S0370-2693(00)00050-2}{{\em Phys.
  Lett.} {\bfseries B474} (2000) 323--330},
\href{http://arxiv.org/abs/hep-th/9911092}{{\ttfamily arXiv:hep-th/9911092
  [hep-th]}}.

\bibitem{Cattaneo2001OnTA}
A.~S. Cattaneo and G.~Felder, ``{On the AKSZ Formulation of the Poisson Sigma
  Model},'' {\em Letters in Mathematical Physics} {\bfseries 56} (2001)
  163--179, \href{http://arxiv.org/abs/math/0102108}{{\ttfamily
  arXiv:math/0102108 [math.QA]}}.

\bibitem{Ikeda:2012pv}
N.~Ikeda, \href{http://dx.doi.org/10.1142/9789813144613\_0003}{``{Lectures on
  AKSZ Sigma Models for Physicists},''} in {\em {Workshop on Strings, Membranes
  and Topological Field Theory}}, pp.~79--169.
\newblock WSPC, 2017.
\newblock \href{http://arxiv.org/abs/1204.3714}{{\ttfamily arXiv:1204.3714
  [hep-th]}}.

\bibitem{FIORENZA_2012}
D.~Fiorenza, C.~L. Rogers, and U.~Schreiber, ``{A Higher Chern--Weil
  Derivations of AKSZ $\sigma$-Models},'' {\em Int. J. Geom. Methods Mod.
  Phys.} {\bfseries 10} no.~01, (Nov, 2012) 1250078,
  \href{http://arxiv.org/abs/1108.4378}{{\ttfamily arXiv:1108.4378 [math-ph]}}.

\bibitem{henneaux1994quantization}
M.~Henneaux and C.~Teitelboim, {\em Quantization of Gauge Systems}.
\newblock Princeton paperbacks. Princeton University Press, 1994.

\bibitem{Grigoriev:2020xec}
M.~Grigoriev and A.~Kotov, ``{Presymplectic AKSZ formulation of Einstein
  gravity},'' \href{http://arxiv.org/abs/2008.11690}{{\ttfamily
  arXiv:2008.11690 [hep-th]}}.

\bibitem{Shaynkman:2000ts}
O.~V. Shaynkman and M.~A. Vasiliev, ``Scalar field in any dimension from the
  higher spin gauge theory perspective,'' {\em Theor. Math. Phys.} {\bfseries
  123} (2000) 683--700,
\href{http://arxiv.org/abs/hep-th/0003123}{{\ttfamily hep-th/0003123}}.

\bibitem{Khavkine2013PresymplecticCA}
I.~Khavkine, ``Presymplectic current and the inverse problem of the calculus of
  variations,'' {\em Journal of Mathematical Physics} {\bfseries 54} (2013)
  111502, \href{http://arxiv.org/abs/1210.0802}{{\ttfamily arXiv:1210.0802
  [math-ph]}}.

\bibitem{Olver}
{Peter J. Olver}, {\em {Applications of Lie Groups to Differential Equations}}.
\newblock Graduate Texts in Mathematics 107. Springer US, 2nd~ed., 1986.

\bibitem{BigDick}
L.~A. Dickey, \href{http://dx.doi.org/10.1142/1109}{{\em Soliton Equations and
  Hamiltonian Systems}}.
\newblock World Scientific, 1991.

\bibitem{Sharapov2016VariationalTG}
A.~A. Sharapov, ``{Variational Tricomplex, Global Symmetries and Conservation
  Laws of Gauge Systems},'' {\em SIGMA} {\bfseries 12} (2016) 098,
  \href{http://arxiv.org/abs/1607.01626}{{\ttfamily arXiv:1607.01626
  [math-ph]}}.

\bibitem{Witten1988Nu}
E.~{Witten}, ``{2 + 1 dimensional gravity as an exactly soluble system},''
  \href{http://dx.doi.org/10.1016/0550-3213(88)90143-5}{{\em Nucl. Phys.}
  {\bfseries B311} no.~1, (Dec., 1988) 46--78}.

\bibitem{Sharapov:2017yde}
A.~A. Sharapov and E.~D. Skvortsov, ``{Formal higher-spin theories and
  Kontsevich--Shoikhet--Tsygan formality},'' {\em Nucl. Phys.} {\bfseries B921}
  (2017) 538--584,
\href{http://arxiv.org/abs/1702.08218}{{\ttfamily arXiv:1702.08218 [hep-th]}}.

\bibitem{actions}
{Undisclosed set of authors} {\em , to appear} .

\bibitem{Vasiliev:1980as}
M.~A. Vasiliev, ``{`Gauge' form of description of massless fields with
  arbitrary spin},''
{\em Sov. J. Nucl. Phys.} {\bfseries 32} (1980) 439.

\bibitem{MacDowell:1977jt}
S.~W. MacDowell and F.~Mansouri, ``Unified geometric theory of gravity and
  supergravity,''
{\em Phys. Rev. Lett.} {\bfseries 38} (1977) 739.

\bibitem{Witten:2003ya}
E.~Witten, ``{SL(2,Z) action on three-dimensional conformal field theories with
  Abelian symmetry},'' in {\em {From fields to strings: Circumnavigating
  theoretical physics. Ian Kogan memorial collection (3 volume set)}},
  pp.~1173--1200.
\newblock 2003.
\newblock
\href{http://arxiv.org/abs/hep-th/0307041}{{\ttfamily arXiv:hep-th/0307041
  [hep-th]}}.
\newblock

\bibitem{Leigh:2003ez}
R.~G. Leigh and A.~C. Petkou, ``{SL(2,Z) action on three-dimensional CFTs and
  holography},'' \href{http://dx.doi.org/10.1088/1126-6708/2003/12/020}{{\em
  JHEP} {\bfseries 12} (2003) 020},
\href{http://arxiv.org/abs/hep-th/0309177}{{\ttfamily arXiv:hep-th/0309177
  [hep-th]}}.

\bibitem{Giombi:2013yva}
S.~Giombi, I.~R. Klebanov, S.~S. Pufu, B.~R. Safdi, and G.~Tarnopolsky, ``{AdS
  Description of Induced Higher-Spin Gauge Theory},'' {\em JHEP} {\bfseries 10}
  (2013) 016,
\href{http://arxiv.org/abs/1306.5242}{{\ttfamily arXiv:1306.5242 [hep-th]}}.

\bibitem{Joung:2012hz}
E.~Joung, L.~Lopez, and M.~Taronna, ``{Generating functions of
  (partially-)massless higher-spin cubic interactions},''
  \href{http://dx.doi.org/10.1007/JHEP01(2013)168}{{\em JHEP} {\bfseries 01}
  (2013) 168},
\href{http://arxiv.org/abs/1211.5912}{{\ttfamily arXiv:1211.5912 [hep-th]}}.

\bibitem{Francia:2016weg}
D.~Francia, G.~L. Monaco, and K.~Mkrtchyan, ``{Cubic interactions of
  Maxwell-like higher spins},''
  \href{http://dx.doi.org/10.1007/JHEP04(2017)068}{{\em JHEP} {\bfseries 04}
  (2017) 068},
\href{http://arxiv.org/abs/1611.00292}{{\ttfamily arXiv:1611.00292 [hep-th]}}.

\bibitem{Metsaev:2018xip}
R.~R. Metsaev, ``{Light-cone gauge cubic interaction vertices for massless
  fields in AdS(4)},''
  \href{http://dx.doi.org/10.1016/j.nuclphysb.2018.09.021}{{\em Nucl. Phys.}
  {\bfseries B936} (2018) 320--351},
\href{http://arxiv.org/abs/1807.07542}{{\ttfamily arXiv:1807.07542 [hep-th]}}.

\bibitem{Sezgin:2005pv}
E.~Sezgin and P.~Sundell, ``{An Exact solution of 4-D higher-spin gauge
  theory},'' {\em Nucl.Phys.} {\bfseries B762} (2007) 1--37,
\href{http://arxiv.org/abs/hep-th/0508158}{{\ttfamily arXiv:hep-th/0508158
  [hep-th]}}.

\bibitem{Sezgin:2011hq}
E.~Sezgin and P.~Sundell, ``{Geometry and Observables in Vasiliev's Higher Spin
  Gravity},'' {\em JHEP} {\bfseries 07} (2012) 121,
\href{http://arxiv.org/abs/1103.2360}{{\ttfamily arXiv:1103.2360 [hep-th]}}.

\bibitem{Aragone:1979hx}
C.~Aragone and S.~Deser, ``{Consistency Problems of Hypergravity},''
{\em Phys. Lett.} {\bfseries B86} (1979) 161.

\bibitem{Aminneborg:1996iz}
S.~Aminneborg, I.~Bengtsson, S.~Holst, and P.~Peldan, ``{Making anti-de Sitter
  black holes},'' \href{http://dx.doi.org/10.1088/0264-9381/13/10/010}{{\em
  Class. Quant. Grav.} {\bfseries 13} (1996) 2707--2714},
\href{http://arxiv.org/abs/gr-qc/9604005}{{\ttfamily arXiv:gr-qc/9604005
  [gr-qc]}}.

\bibitem{Aros:2019pgj}
R.~Aros, C.~Iazeolla, P.~Sundell, and Y.~Yin, ``{Higher spin fluctuations on
  spinless 4D BTZ black hole},''
  \href{http://dx.doi.org/10.1007/JHEP08(2019)171}{{\em JHEP} {\bfseries 08}
  (2019) 171},
\href{http://arxiv.org/abs/1903.01399}{{\ttfamily arXiv:1903.01399 [hep-th]}}.

\bibitem{Ammon:2012wc}
M.~Ammon, M.~Gutperle, P.~Kraus, and E.~Perlmutter, ``{Black holes in three
  dimensional higher spin gravity: A review},''
  \href{http://dx.doi.org/10.1088/1751-8113/46/21/214001}{{\em J. Phys.}
  {\bfseries A46} (2013) 214001},
\href{http://arxiv.org/abs/1208.5182}{{\ttfamily arXiv:1208.5182 [hep-th]}}.

\bibitem{Gelfond:2006be}
O.~Gelfond, E.~Skvortsov, and M.~Vasiliev, ``{Higher spin conformal currents in
  Minkowski space},'' {\em Theor.Math.Phys.} {\bfseries 154} (2008) 294--302,
\href{http://arxiv.org/abs/hep-th/0601106}{{\ttfamily arXiv:hep-th/0601106
  [hep-th]}}.

\bibitem{Gelfond:2014pja}
O.~Gelfond and M.~Vasiliev, ``{Conserved Higher-Spin Charges in $AdS_4$},''
  {\em Phys. Lett.} {\bfseries B754} (2014) 187--194,
  \href{http://arxiv.org/abs/1412.7147}{{\ttfamily arXiv:1412.7147 [hep-th]}}.

\bibitem{universe3040078}
P.~Smirnov and M.~Vasiliev, ``{Gauge Non-Invariant Higher-Spin Currents in
  $AdS_4$},'' \href{http://dx.doi.org/10.3390/universe3040078}{{\em Universe}
  {\bfseries 3} no.~4, 78, (2017) },
  \href{http://arxiv.org/abs/1312.6638}{{\ttfamily arXiv:1312.6638 [hep-th]}}.

\bibitem{sharapov2020characteristic}
A.~Sharapov and E.~Skvortsov, ``{Characteristic Cohomology and Observables in
  Higher Spin Gravity},'' \href{http://arxiv.org/abs/2006.13986}{{\ttfamily
  arXiv:2006.13986 [hep-th]}}.

\bibitem{2002JMP43283V}
I.~{Vaisman}, ``{Fedosov quantization on symplectic ringed spaces},''
  \href{http://dx.doi.org/10.1063/1.1427411}{{\em Journal of Mathematical
  Physics} {\bfseries 43} no.~1, (Jan., 2002) 283--298},
  \href{http://arxiv.org/abs/math/0106070}{{\ttfamily arXiv:math/0106070
  [math.SG]}}.

\bibitem{2020TMP204.1079G}
N.~D. {Gorev}, B.~M. {Elfimov}, and A.~A. {Sharapov}, ``{Deformation
  quantization of framed presymplectic manifolds},''
  \href{http://dx.doi.org/10.1134/S0040577920080085}{{\em Theoretical and
  Mathematical Physics} {\bfseries 204} no.~2, (Aug., 2020) 1079--1092}.

\bibitem{88c5409ac2d1498fa6234cbfe818edba}
K.~Rejzner and E.~Hawkins, ``{The Star Product in Interacting Quantum Field
  Theory},'' \href{http://dx.doi.org/10.1007/s11005-020-01262-4}{{\em Letters
  in Mathematical Physics} (Feb., 2020) 1--57}.

\bibitem{Iazeolla:2017dxc}
C.~Iazeolla, E.~Sezgin, and P.~Sundell, ``{On Exact Solutions and Perturbative
  Schemes in Higher Spin Theory},''
  \href{http://dx.doi.org/10.3390/universe4010005}{{\em Universe} {\bfseries 4}
  no.~1, (2018) 5},
\href{http://arxiv.org/abs/1711.03550}{{\ttfamily arXiv:1711.03550 [hep-th]}}.

\bibitem{Anninos:2020hfj}
D.~Anninos, F.~Denef, Y.~T.~A. Law, and Z.~Sun, ``{Quantum de Sitter horizon
  entropy from quasicanonical bulk, edge, sphere and topological string
  partition functions},''
\href{http://arxiv.org/abs/2009.12464}{{\ttfamily arXiv:2009.12464 [hep-th]}}.

\bibitem{Loday}
J.-L. Loday, {\em {Cyclic Homology}}.
\newblock Springer, 1998.

\bibitem{Co}
A.~Connes, {\em {Noncommutative Geometry}}.
\newblock Academic Press, 1995.

\bibitem{Feigin1987}
B.~L. Feigin and B.~L. Tsygan, {\em Additive K-theory}, pp.~67--209.
\newblock Springer Berlin Heidelberg, Berlin, Heidelberg, 1987.

\bibitem{MacLane}
S.~MacLane, {\em {Homology}}.
\newblock Springer, 1995.

\bibitem{Connes:85}
A.~Connes, ``{Non commutative differential geometry },'' {\em Publ. Math. IHES}
  {\bfseries 62} (1985) 41--144.

\bibitem{Kassel1986}
C.~Kassel, ``{A K\"unneth Formula for the Cyclic Cohomology of
  $\mathbb{Z}/2$-Graded Algebras},'' {\em Mathematische Annalen} {\bfseries
  275} (1986) 683--699.

\bibitem{Tsygan_1983}
B.~L. Tsygan, ``{The homology of matrix Lie algebras over rings and the
  Hochschild homology},'' {\em Russian Mathematical Surveys} {\bfseries 38}
  no.~2, (Apr, 1983) 198--199.

\bibitem{LQ}
J.-L. Loday and D.~Quillen, ``{Cyclic homology and the Lie algebra homology of
  matrices},'' {\em Commentarii Mathematici Helvetici} {\bfseries 59} no.~1,
  (Jan, 1984) 565--594.

\bibitem{coutinho_1995}
S.~C. Coutinho, \href{http://dx.doi.org/10.1017/CBO9780511623653}{{\em A Primer
  of Algebraic D-Modules}}.
\newblock London Mathematical Society Student Texts. Cambridge University
  Press, 1995.

\bibitem{AFLS}
J.~Alev, M.~Farinati, T.~Lambre, and A.~Solotar, ``{Homologie des invariants
  d'une algèbre de Weyl sous l'action d'un groupe fini},'' {\em Journal of
  Algebra} {\bfseries 232} no.~2, (2000) 564--577.

\bibitem{Pinczon}
G.~Pinczon, ``{On Two Theorems about Symplectic Reflection Algebras},'' {\em
  Lett. Math. Phys.} {\bfseries 82} (2007) 237--253.

\bibitem{FFS}
B.~Shoikhet, G.~Felder, and B.~Feigin, ``{Hochschild cohomology of the Weyl
  algebra and traces in deformation quantization},'' {\em Duke Mathematical
  Journal} {\bfseries 127} no.~3, (2005) 487--517.

\bibitem{Shoikhet:2000gw}
B.~Shoikhet, ``{A proof of the Tsygan formality conjecture for chains},'' {\em
  Advances in Mathematics} {\bfseries 179} no.~1, (2003) 7 -- 37.

\bibitem{COQUEREAUX1995333}
R.~Coquereaux and E.~Ragoucy, ``{Currents on Grassmann algebras},'' {\em
  Journal of Geometry and Physics} {\bfseries 15} no.~4, (1995) 333 -- 352,
  \href{http://arxiv.org/abs/hep-th/9310147}{{\ttfamily arXiv:hep-th/9310147}}.

\bibitem{grensing2004berezin}
G.~Grensing and M.~Nitschmann, ``Berezin integration over anticommuting
  variables and cyclic cohomology,''
  \href{http://arxiv.org/abs/hep-th/0401231}{{\ttfamily arXiv:hep-th/0401231
  [hep-th]}}.

\bibitem{mackenzie_2005}
K.~C.~H. Mackenzie, \href{http://dx.doi.org/10.1017/CBO9781107325883}{{\em
  General Theory of Lie Groupoids and Lie Algebroids}}.
\newblock London Mathematical Society Lecture Note Series. Cambridge University
  Press, 2005.

\bibitem{Kast}
R.~Coquereaux, A.~Jadczyk, and D.~Kastler, ``{Differential and integral
  geometry of Grassmann algebras},'' {\em Reviews in Mathematical Physics}
  {\bfseries 03} no.~01, (1991) 63--99.

\end{thebibliography}
\end{document}